\documentclass[12pt]{article}
\usepackage{amsfonts,amsmath,graphicx,amssymb,geometry,arydshln,textcomp,amsthm,framed,color,appendix,lscape}
\usepackage[bookmarksopen=true]{hyperref}
\usepackage[round, authoryear]{natbib}
\usepackage[nodisplayskipstretch]{setspace}
\usepackage{ragged2e}
\usepackage{xcolor,xspace,colortbl,rotating}
\usepackage{tabulary}
\usepackage{booktabs}
\usepackage{multirow}
\usepackage{bm}
\usepackage[ruled,linesnumbered]{algorithm2e}
\usepackage{cleveref}
\usepackage{threeparttable}
\usepackage[subrefformat=parens,labelformat=parens]{subfig}
\usepackage{etoc}
\usepackage{enumerate}
\usepackage{ltablex,caption}
\usepackage{bookmark}
\DeclareGraphicsExtensions{.pdf,.svg,.eps,.ps,.png,.jpg,.jpeg}
\theoremstyle{definition}
\newtheorem{theorem}{Theorem}

\newtheorem{assumption}{Assumption}

\newtheorem {proposition}{Proposition}
\newtheorem {remark}{Remark}

\crefname{subequation}{subequation}{subequations}
\Crefname{subequation}{Assumption}{assumptions}
\crefname{assumption}{Assumption}{Assumptions}
\hypersetup{colorlinks,linkcolor={blue},citecolor={blue},urlcolor={blue}}

\makeatletter
\newcommand\subsubsubsection{\@startsection{paragraph}{4}{\z@}{-2.5ex\@plus -1ex \@minus -.25ex}{1.25ex \@plus .25ex}{\normalfont\normalsize\bfseries}}
\newcommand\subsubsubsubsection{\@startsection{subparagraph}{5}{\z@}{-2.5ex\@plus -1ex \@minus -.25ex}{1.25ex \@plus .25ex}{\normalfont\normalsize\bfseries}}
\makeatother

\begin{document}

\author{Xiao Huang\thanks{Department of Economics, Finance, and Quantitative Analysis, Coles College of Business, Kennesaw State University, GA 30144, USA. Email: xhuang3@kennesaw.edu.}}
\title{\Large Lassoed Boosting and Linear Prediction in the Equities Market}
\date{ \today}
\maketitle

\doublespace

\begin{abstract}
	We consider a two-stage estimation method for linear regression. First, it uses the lasso in \cite{tibshirani1996lasso} to screen variables and, second, re-estimates the coefficients using the least-squares boosting method in \cite{friedman2001gbm} on every set of selected variables. Based on the large-scale simulation experiment in \cite{hastie2017extended}, lassoed boosting performs as well as the relaxed lasso in \cite{meinshausen2007relaxed} and, under certain scenarios, can yield a sparser model. Applied to predicting equity returns, lassoed boosting gives the smallest mean-squared prediction error compared to several other methods. 
	%An application in predicting equity returns also shows that lassoed boosting can give the smallest mean squared prediction error among several methods under consideration. 

\end{abstract}

\bigskip


\bigskip


\newpage  

\normalsize

\doublespace
%\linenumbers

\section{Introduction} \label{intro}

To analyze consumer behavior, predict sales, and track price movement, business and economic researchers must routinely sift through massive amount of data to select relevant variables. In an influential paper, \cite{tibshirani1996lasso} proposed a shrinkage method called lasso to simplify estimation, and it has become a critical tool in high-dimensional analysis. Extensions include the elastic net in \cite{zou2005elasticnet} and the group lasso in \cite{yuan2006grouplasso}. \cite{fanandli2001SCAD,zhang2010MCplus,mazumderetal2011} also discuss nonconvex penalty function approaches.  \cite{buhlmann2011highdimenstats, hastie2015slsparcity} provide thorough expositions of the lasso and related methods.

With so many variable-selection methods available, data analysts will profit from some general advice on their use. \cite{hastie2017extended} recently conducted a large-scale simulation to compare the performance of (a) the lasso; (b) forward stepwise selection, which generates models by sequentially adding the regressor that most improves the fit; (c) best subset selection of regressors for each model size; (d) the relaxed lasso in \cite{meinshausen2007relaxed}, which emerged as the overall winner with good accuracy and sparsity recovery. The paper also commented on many other competitors of the lasso, such as the Dantzig selector and square-root lasso, and concluded that their performance is either close to the lasso or somewhere between the lasso and the best subset.

This paper investigates whether we can design a new estimator that is as simple as the relaxed lasso but even more effective under certain scenarios. We propose one example: \textit{lassoed boosting}. The relaxed lasso uses linear interpolation between every lasso solution and the corresponding least-squares (LS) solution to create additional coefficient paths; the linear interpolation forces coefficients to grow proportionally toward an LS solution. In lassoed boosting, we use the lasso in the first stage to screen variables and, for each subset of variables, we use LS-boost (\cite{friedman2001gbm}) to grow coefficients in the second stage.  We hypothesize that, for some data, a good solution may appear outside the grid of (proportional) solutions generated by the relaxed lasso; using boosting to rebuild coefficients allows us to explore possibly better solutions. Our method complements the use of the relaxed lasso in practice. 

Both lassoed boosting and the relaxed lasso can be connected to a strand of literature on refitting strategies (see, e.g., \cite{chzhenetal2019lassorefit}
and references therein). The simple idea of combining the lasso with LS-boost comes with some obvious benefits. Using a large iteration number in the second stage, LS-boost should (a) mitigate the overshrinkage problem of the lasso; (b) remove the proportional constraints when spawning solutions so that coefficient paths can grow freely; and (c) give the estimation procedure a second chance, increasing the likelihood of finding a sparser model. Moreover, tuning both the lasso and LS-boost procedures should lead to better solutions.

This paper discusses the method of lassoed boosting and its good performance in the simulation experiment in \cite{hastie2017extended} and an application. Based on the results in \cite{freundetal2017boosting} (hereafter FGM), we also discuss the convergence rate of lassoed boosting estimator. We apply lassoed boosting to predict equity returns and compare the results to several other methods.
%we discuss the convergence property of LS-boost and the faster rate of lassoed boosting under certain scenarios. 
%Third, based on the idea of integrated gradients in \cite{sundararajan2017attribution}, we use path integrated gradients to study the difference in parameter attribution between the lasso and LS-boost. We show that the lasso and LS-boost in general exhibit different parameter attribution patterns, providing a new perspective on the comparison of these two methods. 
An R package \texttt{lboost} that implements our method can be found at \url{https://github.com/xhuang20/lboost}.

The rest of the paper is organized as follows. Section 2 discusses the convergence property of LS-boost and lassoed boosting. Section 3 introduces several other two-stage methods. Section 4 discusses the simulation experiment, and Section 5, the application to equity-returns prediction. Section 6 concludes. The online supplement contains all proofs, additional discussions, and figures.

\section{Lassoed boosting}
We begin by defining the lasso and LS-boost procedures. Consider $n$ observations $\{(x_i,y_i)\}_1^n$,  where $x_i = (x_{i1},\ldots, x_{ip})$ is the $1 \times p$ row vector of variables and $y_i$ is the $i$th response variable. In matrix notation, define the $n \times 1$ vector $\boldsymbol{\mathrm{y} }$ and the $n \times p$ matrix $\boldsymbol{\mathrm{X}}$ with the $j$th column $\boldsymbol{\mathrm{x}}_j$ and the $i$th row $x_i$. Let $\beta^*$ be the true coefficient vector and $u_i \sim (0, \sigma^2)$. Let $\lVert \cdot \rVert_1$ and $\lVert \cdot \rVert_2$ be the $\ell_1$ and $\ell_2$ norms, respectively. Consider the linear regression model
\begin{equation} \label{eq:lrm}
    \boldsymbol{\mathrm{y} } = \boldsymbol{\mathrm{X}} \beta^* + \boldsymbol{\mathrm{u}}.
	%y_i = \beta_0 + \sum_{j=1}^{p} x_{ij} \beta_j + \varepsilon_i,
\end{equation}
The LS solution $\hat{\beta}_{\text{LS}}$ is obtained by minimizing the following loss function:
\begin{equation} \label{eq:ls loss}
	 L_n(\beta) = \frac{1}{2n} \lVert \boldsymbol{\mathrm{y} } - \boldsymbol{\mathrm{X}} \beta \rVert_{2}^{2}.
\end{equation}

The lasso estimate, $\hat{\beta}^{\lambda}$, results from minimizing
\begin{equation} \label{eq:lasso loss}
	\frac{1}{2n} \lVert \boldsymbol{\mathrm{y} } - \boldsymbol{\mathrm{X}} \beta \rVert_{2}^{2} + \lambda \lVert \beta \rVert_1
\end{equation}
for some tuning parameter $\lambda > 0$. A sequence of $\lambda$s, $\left\{\lambda_q\right\}_0^Q$, is used to tune the coefficient solutions with $\lambda_0 = \max_j |\frac{1}{n}\langle\boldsymbol{\mathrm{x}}_j,\boldsymbol{\mathrm{y} }\rangle| > \lambda_1 > \cdots > \lambda_Q$. Let $\mathcal{A}_q$ be the active set of selected variables at step $q$ and $\hat{\beta}^{\lambda_q}$ be the coefficient estimate.  When $\lambda = \lambda_0$, no variable is selected and  $\mathcal{A}_0$ is empty; the size of the active set increases as more variables are included in $\mathcal{A}_q$ when $\lambda$ decreases. When $\lambda=0$, $\mathcal{A}_q$ includes all variables  and \cref{eq:lasso loss} reduces to \cref{eq:lrm}.

The LS-boost algorithm works by choosing the variable $x_{j_k}$ that best fits the current residual $\hat{\boldsymbol{\mathrm{u}} }^{k-1}$ at each iteration step $k$ and then updating the $j_k$th regression coefficient. Choose a learning rate $0<\varepsilon<1$. Initialize $\hat{\beta}^0 = 0$ and $\hat{\boldsymbol{\mathrm{u}}}^0 = \boldsymbol{\mathrm{y}}$. For each iteration $k \geq 1$,

\medskip
Step 1. Select the variable $\boldsymbol{\mathrm{x}}_{j_k}$ with 
\begin{equation*}
	j_k \in \operatorname*{argmin}_{1 \leq j \leq p} \sum_{i=1}^{n}(\hat{u}_i^{k-1} - \hat{\beta}_j x_{ij})^2 \text{ with } \hat{\beta}_j = \frac{\sum_{i=1}^{n} \hat{u}_i^{k-1}x_{ij}}{\sum_{i=1}^{n}x_{ij}^2}.
\end{equation*}

Step 2. Update $\hat{\beta}^k$ and $\hat{\boldsymbol{\mathrm{u}}}^k$ by
\begin{align*}
	\hat{\beta}^k_{j_k} &= \hat{\beta}^{k-1}_{j_k} + \varepsilon \hat{\beta}_{j_k}, \enspace\hat{\beta}^k_{j} = \hat{\beta}^{k-1}_{j} \text{ for } j \neq j_k, \enspace \text{and } \hat{\boldsymbol{\mathrm{u}}}^k = \hat{\boldsymbol{\mathrm{u}}}^{k-1} - \varepsilon \cdot \boldsymbol{\mathrm{x}}_{j_k} \hat{\beta}_{j_k}.
\end{align*}

Iterating between Step 1 and Step 2 until we reach a prespecified stopping criterion gives the solution paths. LS-boost can sometimes generate coefficient paths similar to those of the lasso, but they are two different methods in general.

\subsection{The algorithm and its implementation}
\begin{figure}[htp]
	\centering
	\subfloat[Relaxed lasso solution paths]{\label{fig:1a}\includegraphics[width=0.48\linewidth,keepaspectratio]{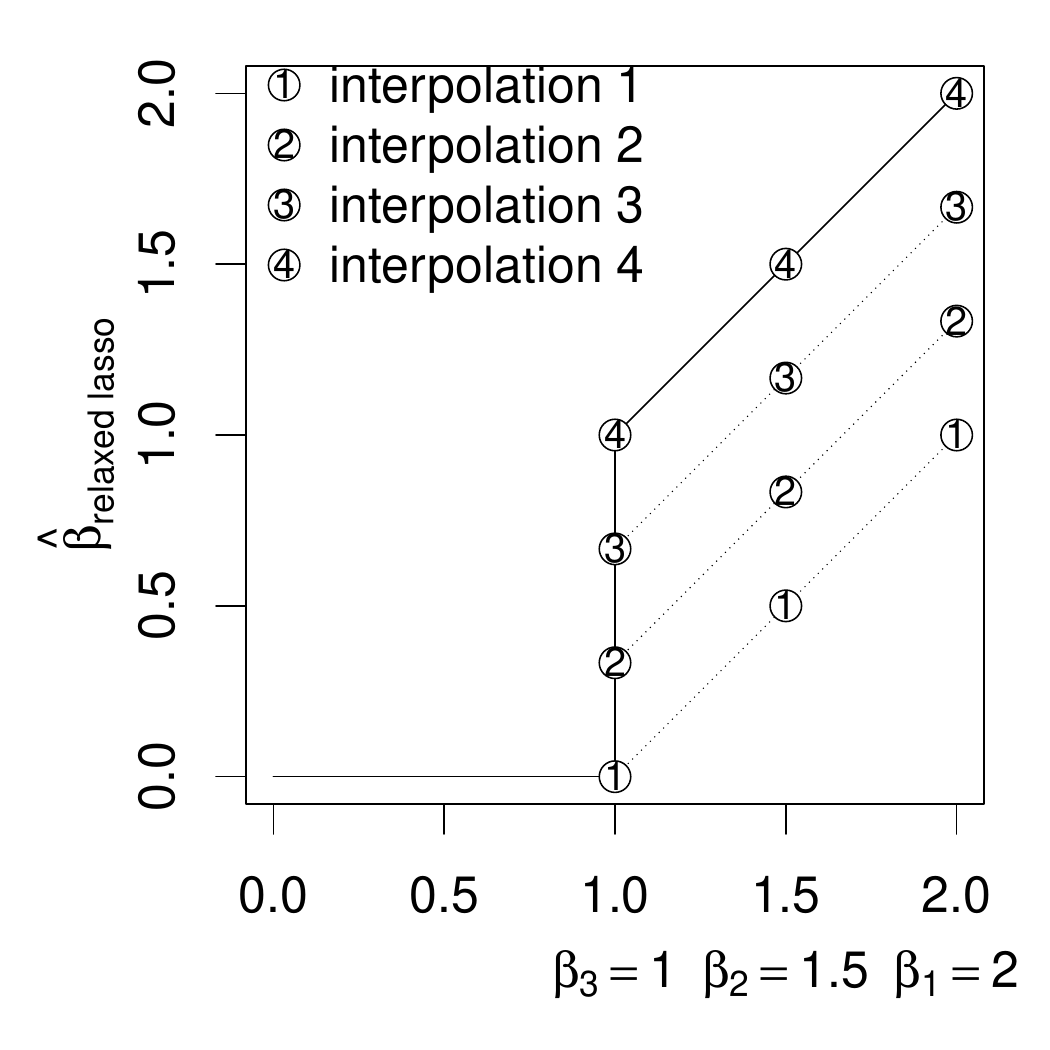} }%
	\subfloat[Lassoed boosting solution paths]{\label{fig:1b}\includegraphics[width=0.48\linewidth,keepaspectratio]{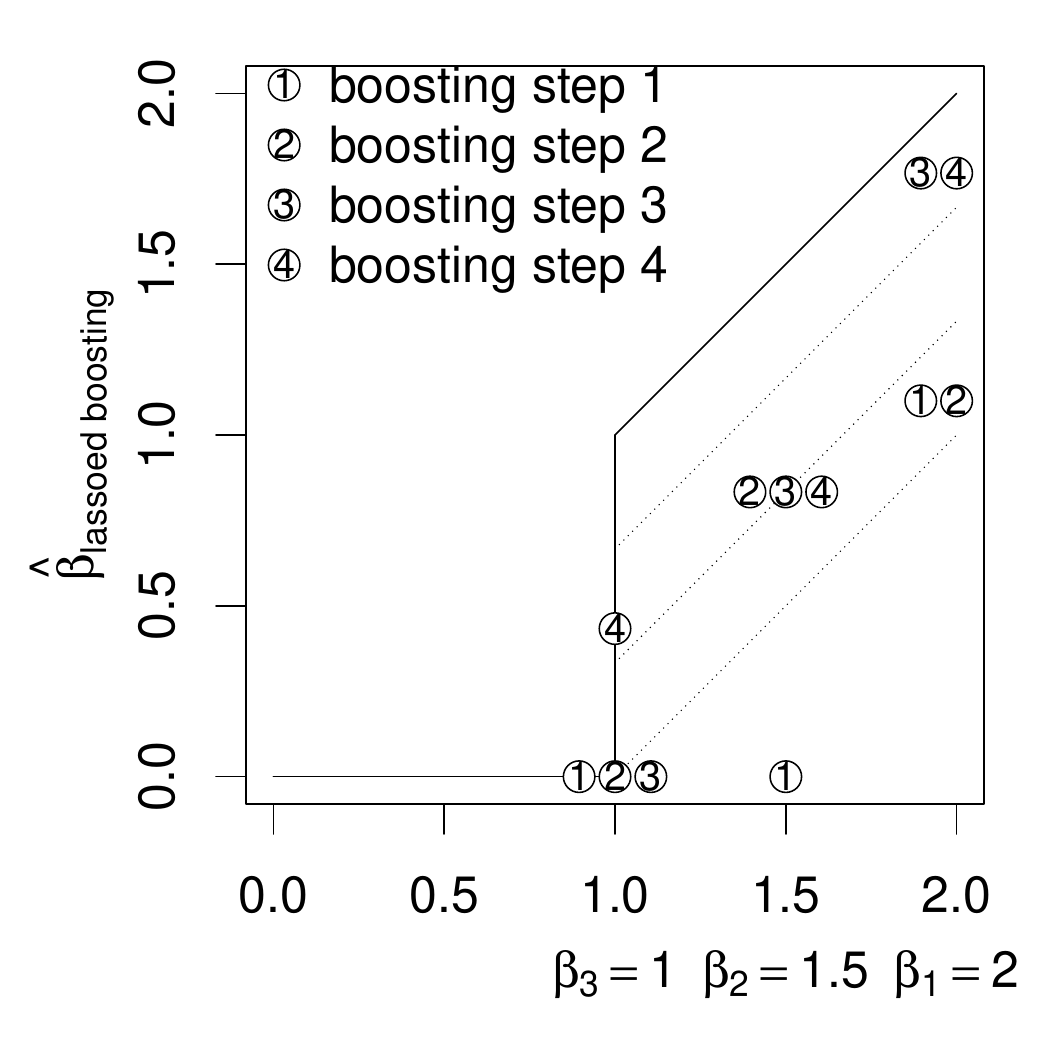} }\\
	\subfloat[Step 2 in relaxed lasso]{\label{fig:1c}\includegraphics[width=0.48\linewidth,keepaspectratio]{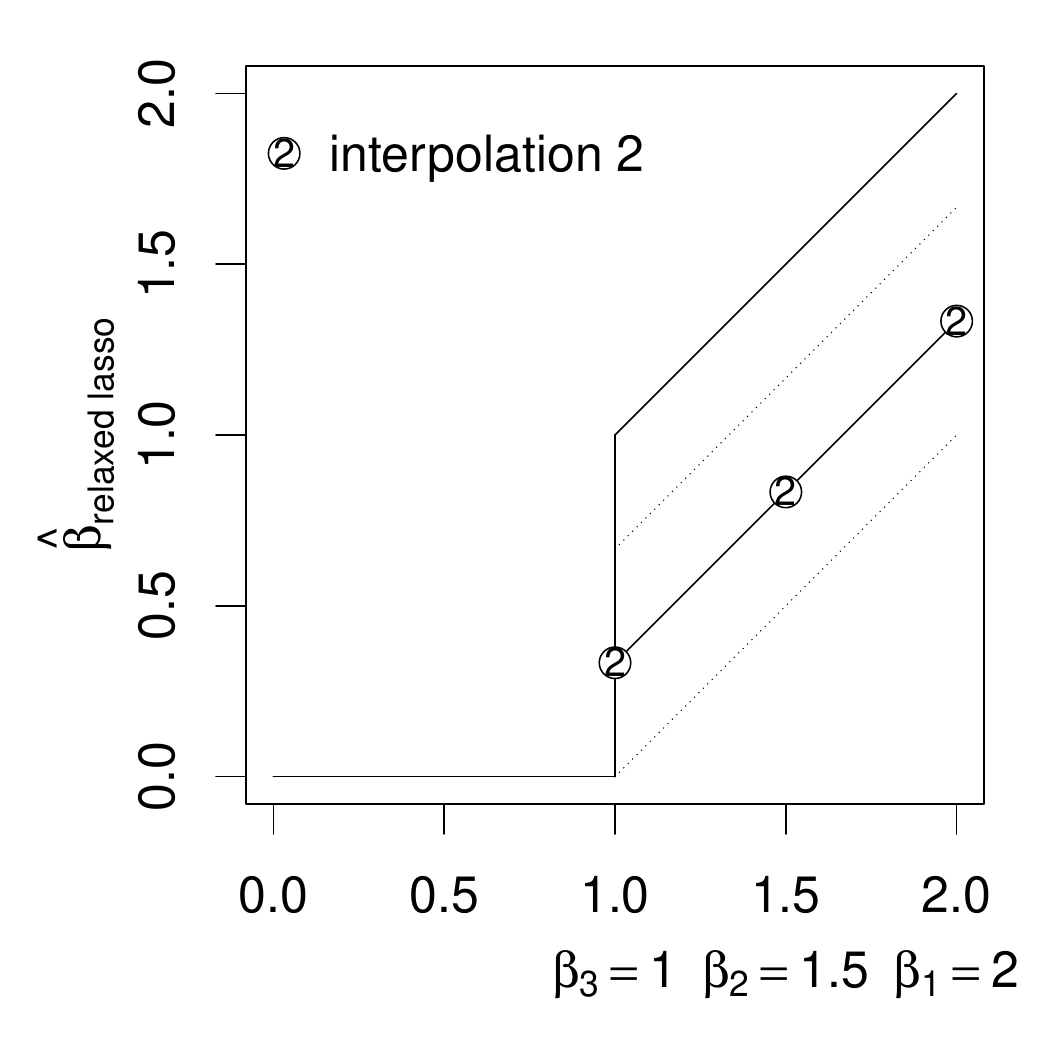} }%
	\subfloat[Step 2 in lassoed boosting]{\label{fig:1d}\includegraphics[width=0.48\linewidth,keepaspectratio]{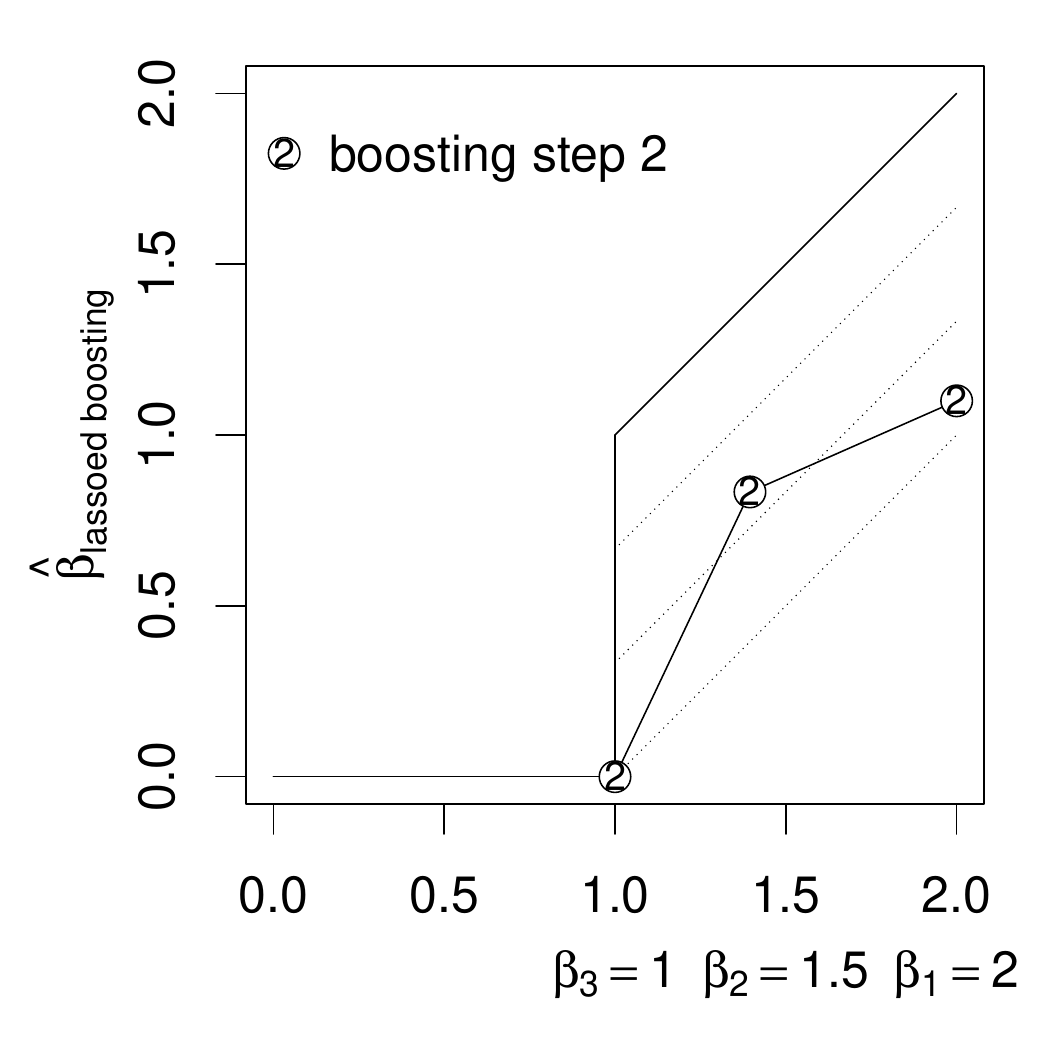} }%
	\caption{\Cref{fig:1a} gives an example of the relaxed lasso solution paths for three parameters with linear interpolation weights $(0,0.33,0.66,1)$. \Cref{fig:1b} shows a boosting solution path for the same three parameters with four steps. \Cref{fig:1c,fig:1d} select the coefficient solutions of the second interpolation and the second boosting step, respectively.}%
	\label{fig:par_paths}%
\end{figure}

Lassoed boosting works by rebuilding coefficient paths for variables in each $\mathcal{A}_q$ using LS-boost.
\begin{center}
	{\SetAlgoNoLine%
		\begin{algorithm}[H]
			\caption{Lassoed boosting} 
			\label{alg:lassoed boosting} 
			\DontPrintSemicolon
			\SetKwFor{For}{for}{do}{end~for}
			\SetKw{KwAssume}{Assume}
			\KwAssume{a sequence of $Q$ tuning parameters $\left\{\lambda_q\right\}_1^Q$ for the lasso problem.}\\
			\For{$q = 1$ to $Q$}{
				Use the lasso to obtain an active set of variables, $\mathcal{A}_q$, for each $\lambda_q$.\\
				Use LS-boost to compute the coefficient path for each variable in $\mathcal{A}_q$.
			}
			\Return $Q$ sets of coefficient paths for validation or cross-validation.
		\end{algorithm}
	}
\end{center}
A few remarks are in order.
\begin{remark} \label{rmk:path difference}
	Both the relaxed lasso and lassoed boosting use the lasso in the first stage. Afterward, the relaxed lasso takes the lasso solution $\hat{\beta}^{\lambda_q}$, along with the full LS solution $\hat{\beta}_{\text{LS}}^{\lambda_q}$ for variables in $\mathcal{A}_q$ and a sequence of weights such as $\{0,0.33,0.66,1.0\}$, to generate the interpolated coefficient path
	\begin{equation} \label{eq:relaxed lasso}
		\hat{\beta}_{\text{relax}}^{\lambda_q} = \text{weight} \times \hat{\beta}_{\text{lasso}}^{\lambda_q} + (1 - \text{weight}) \times \hat{\beta}_{\text{LS}}^{\lambda_q}.
	\end{equation}
	As long as the lasso solution paths are monotonic, \cref{eq:relaxed lasso} creates a sequence of solution paths that grows proportionately toward the LS solution $\hat{\beta}_{\text{LS}}^{\lambda_q}$, and the computation cost is close to zero. Lassoed boosting does not use the lasso solution $\hat{\beta}^{\lambda_q}_{\text{lasso}}$ but only the variables in $\mathcal{A}_q$ to start LS-boost, so the generated solution paths will differ from those of the relaxed lasso. \Cref{fig:par_paths} illustrates this difference in the estimates for $\beta_1, \beta_2$, and $\beta_3$. \Cref{fig:1c,fig:1d} compare interpolation step 2 and boosting step 2 for the two methods. The boosting estimates exhibit no proportional increase. 
\end{remark}

\begin{remark} \label{rmk:sparsity}
	Our simulation and application indicate that lassoed boosting sometimes yields a sparser model (see \Cref{fig:1c,fig:1d}). Starting with an active set $\mathcal{A}_q =(\boldsymbol{\mathrm{x}}_1,\boldsymbol{\mathrm{x}}_2,\boldsymbol{\mathrm{x}}_3)$, the relaxed lasso pulls the lasso solution, marked \raisebox{.5pt}{\textcircled{\raisebox{-.9pt} {1}}}, proportionately toward the LS solution, marked \raisebox{.5pt}{\textcircled{\raisebox{-.9pt} {4}}} in \Cref{fig:1a}, and all three $\beta$s increase in the second interpolation in \Cref{fig:1c}. With boosting, coefficients are updated one at a time and $\beta_3$ is not updated in step 2 in \Cref{fig:1d} despite the fact that $\boldsymbol{\mathrm{x}}_3$ is already included in the active set. Hence, given the same set of variables in $\mathcal{A}_q$, LS-boost might generate sparser solution paths than the relaxed lasso does.
\end{remark}

\begin{remark} \label{rmk:stopping}
	The solution path of the relaxed lasso always includes the LS solution for a given active set, which is not the case for LS-boost. After four steps, the relaxed lasso reaches the LS solution in \Cref{fig:1a}, while the solution of LS-boost, marked \raisebox{.5pt}{\textcircled{\raisebox{-.9pt} {4}}} in \Cref{fig:1b}, does not. Using an information criterion such as the corrected AIC in \cite{hurvich1998smoothing} for early stopping is a common practice in boosting to avoid overfitting. It is easy to verify that, for many data, a typical boosting solution stops short of the LS solution. One can increase the iteration number, but in practice there is no guarantee that the solution will be close to the LS solution even when $n > p$.
\end{remark}

%See Figure~\subref*{fig:1a} for details. \Cref{fig:1a} is cool.

\subsection{Convergence Results}

In this section, we discuss the asymptotic convergence result for lassoed boosting. Let  $\hat{\beta}^{\lambda_q, k}$ be the boosting solution at step $k$ for the variables in the active set $\mathcal{A}_q$ associated with $\lambda_q$. Let $K_{\mathcal{A}_q} = |\mathcal{A}_q|$ be the number of elements in $\mathcal{A}_q$. The active set for the true model is $\mathcal{A}=\{1,\ldots,s\}$ and $K_{\mathcal{A}} = s$. We make the following assumptions for  \Cref{prop:asymptotic rate,prop:two-stage refit}.

\begin{assumption} \label{assumption:general}
	The data are generated according to \cref{eq:lrm} with $u_i \sim (0,\sigma^2)$. $\boldsymbol{\mathrm{X}}$ is deterministic.
\end{assumption}

\begin{assumption} \label{assumption:sparsity}
	The parameter vector is $s$-sparse so $\beta^* =(\beta_1,\ldots,\beta_s,0,\ldots,0)^T$ and $s < p$.
\end{assumption}

We also implicitly assume $\log(p)/n \rightarrow 0$ in the proof of \Cref{prop:asymptotic rate,prop:two-stage refit} in the supplement (Section S.1), but this assumption is not needed for \Cref{thm:prediction convergence}.
\subsubsection{The asymptotic rate}

%Define the expected loss function for $\beta^{\lambda_q, k}$ 
%\begin{equation} \label{eq:expected loss}
%	L(\hat{\beta}^{\lambda_q,k}) = E(Y - X^T \hat{\beta}^{\lambda_q,k})^2 - \sigma^2.
%\end{equation}

%\Cref{prop:asymptotic rate} gives the asymptotic convergence rate of lassoed boosting.
The following proposition shows that lassoed boosting shares the same loss-function convergence rate with the relaxed lasso in Theorem 6 in \cite{meinshausen2007relaxed}.

\begin{proposition} \label{prop:asymptotic rate}
	Under \Cref{assumption:general,assumption:sparsity}, as $n \rightarrow \infty$, we have
	\begin{equation*}
		\inf_{\lambda_q, k \in [1,\infty]} L_n(\hat{\beta}^{\lambda_q,k}) = O_p(n^{-1}). 
	\end{equation*}
\end{proposition}
\noindent The proof is given in Section S.1. The convergence rate for lassoed boosting is faster than the lasso's in Theorem 5 in \cite{meinshausen2007relaxed}.

%Rigorously speaking, because we use Theorem 11.3 in \cite{hastie2015slsparcity} in the proof of \Cref{prop:asymptotic rate}, we need to borrow all assumptions in that theorem and the result in \Cref{prop:asymptotic rate} holds with high probability.

The relaxed lasso and lassoed boosting share the same fast convergence rate because both of their solution paths include the LS solution when the lasso correctly identifies the variables. This observation suggests that any lasso-based two-stage method that includes the LS solution in the second step will also enjoy the rate in \Cref{prop:asymptotic rate}. \Cref{prop:two-stage refit} summarizes this result. Let $\hat{\beta}_{\text{two-stage}}^{\lambda_q, \mathcal{K}}$ be a two-stage estimator that uses either the lasso solution or the active set $\mathcal{A}_q$ to generate solution paths that include the full LS solution for each $\mathcal{A}_q$, and $\mathcal{K}$ is the vector of all tuning parameters in the second stage.
\begin{proposition} \label{prop:two-stage refit}
	Under \Cref{assumption:general,assumption:sparsity}, as $n \rightarrow \infty$,
	\begin{equation*}
	\inf_{\lambda_q, \mathcal{K}} L_n(\hat{\beta}_{\text{two-stage}}^{\lambda_q, \mathcal{K}}) = O_p(n^{-1}). 
	\end{equation*}
\end{proposition}
\noindent See Section S.1 for the proof. \Cref{prop:two-stage refit} indicates that the relaxed lasso and lassoed boosting are two examples of a large class of two-stage estimators.

\subsubsection{Linear convergence of predictions}

We begin our discussion of some linear convergence results for LS-boost and extend it to lassoed boosting on an active set $\mathcal{A}_q$. Let $\hat{\beta}^{k}$ be the LS-boost solution at step $k$; $\hat{\beta}^{k}_{\text{LS}}$ a possibly non-unique least squares solution at step $k$; and $\hat{\beta}_{\text{LS}}$ the full LS solution. Note that $\hat{\beta}_{\text{LS}}$ is non-unique when $p > n$. Theorem 2.1 in FGM gives the linear convergence result for $\lVert \boldsymbol{\mathrm{X}} \hat{\beta}^k - \boldsymbol{\mathrm{X}} \hat{\beta}_{\text{LS}} \rVert_2$. First, we investigate the convergence result for $\lVert \boldsymbol{\mathrm{X}} \hat{\beta}^k - \boldsymbol{\mathrm{X}}\beta^* \rVert_2$. 

\begin{figure}[htp]
	\centering
	\subfloat[Distinct solution regions]{\label{fig:2a}\includegraphics[width=0.33\linewidth,keepaspectratio]{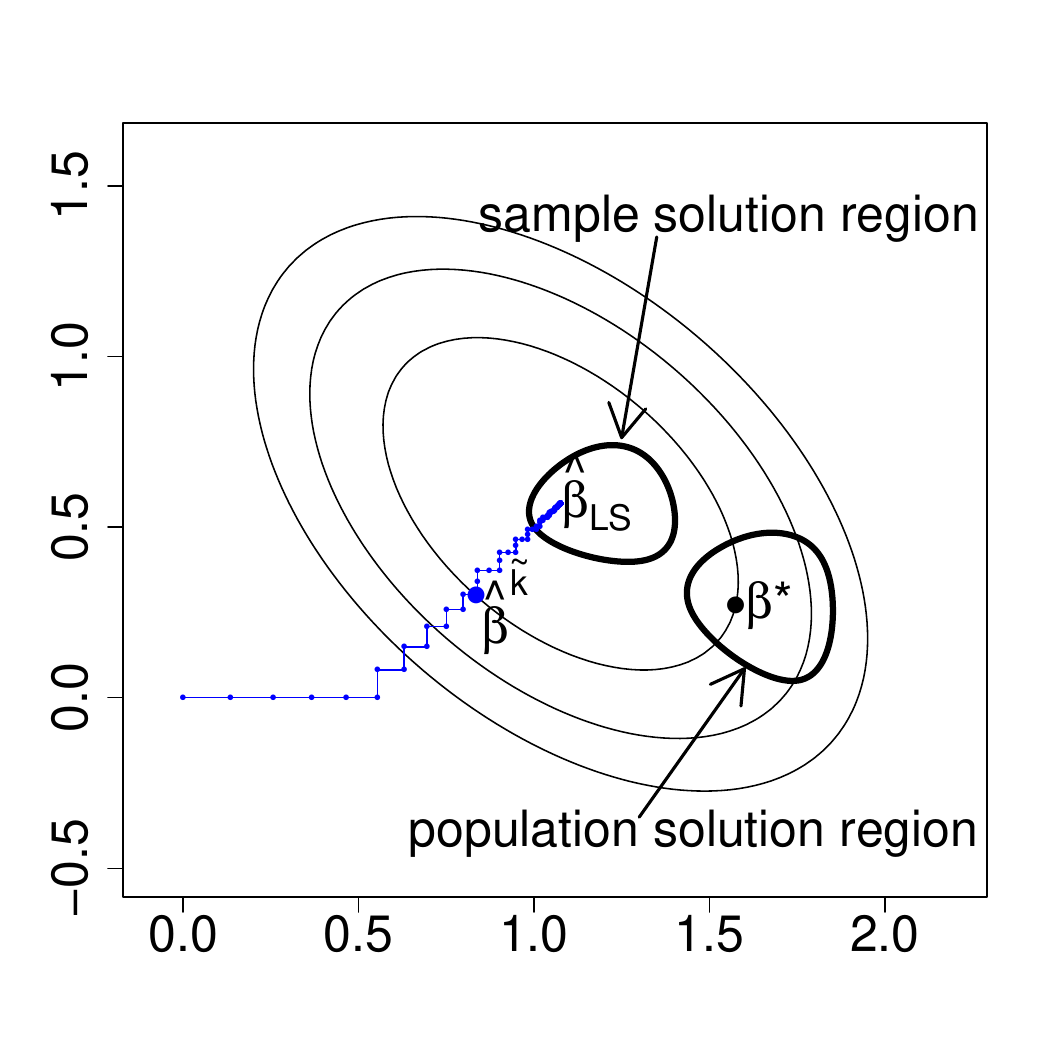} }%
	\subfloat[Overlapping solution regions]{\label{fig:2b}\includegraphics[width=0.33\linewidth,keepaspectratio]{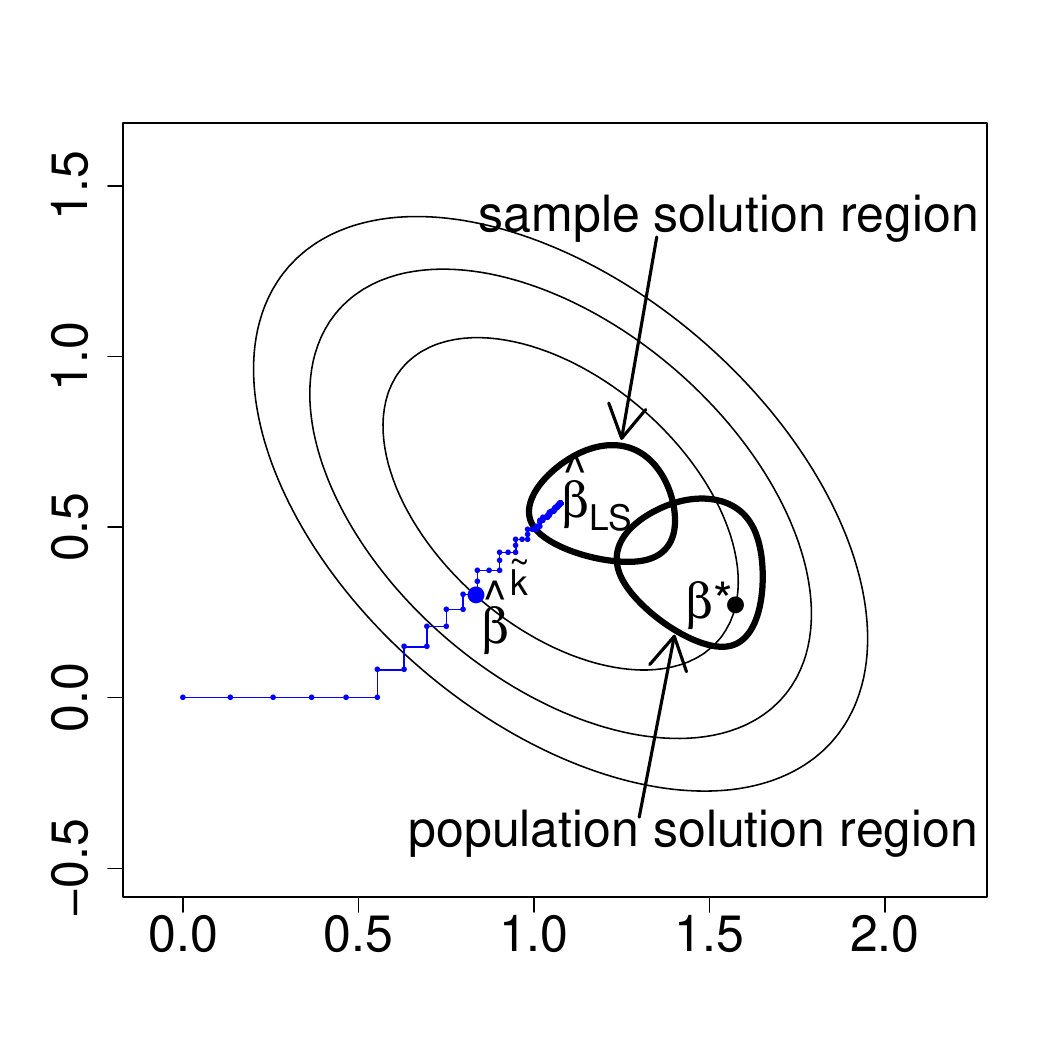} }%
	\subfloat[$\hat{\beta}_{LS}$ and $\beta^*$ have the same LS loss]{\label{fig:2c}\includegraphics[width=0.33\linewidth,keepaspectratio]{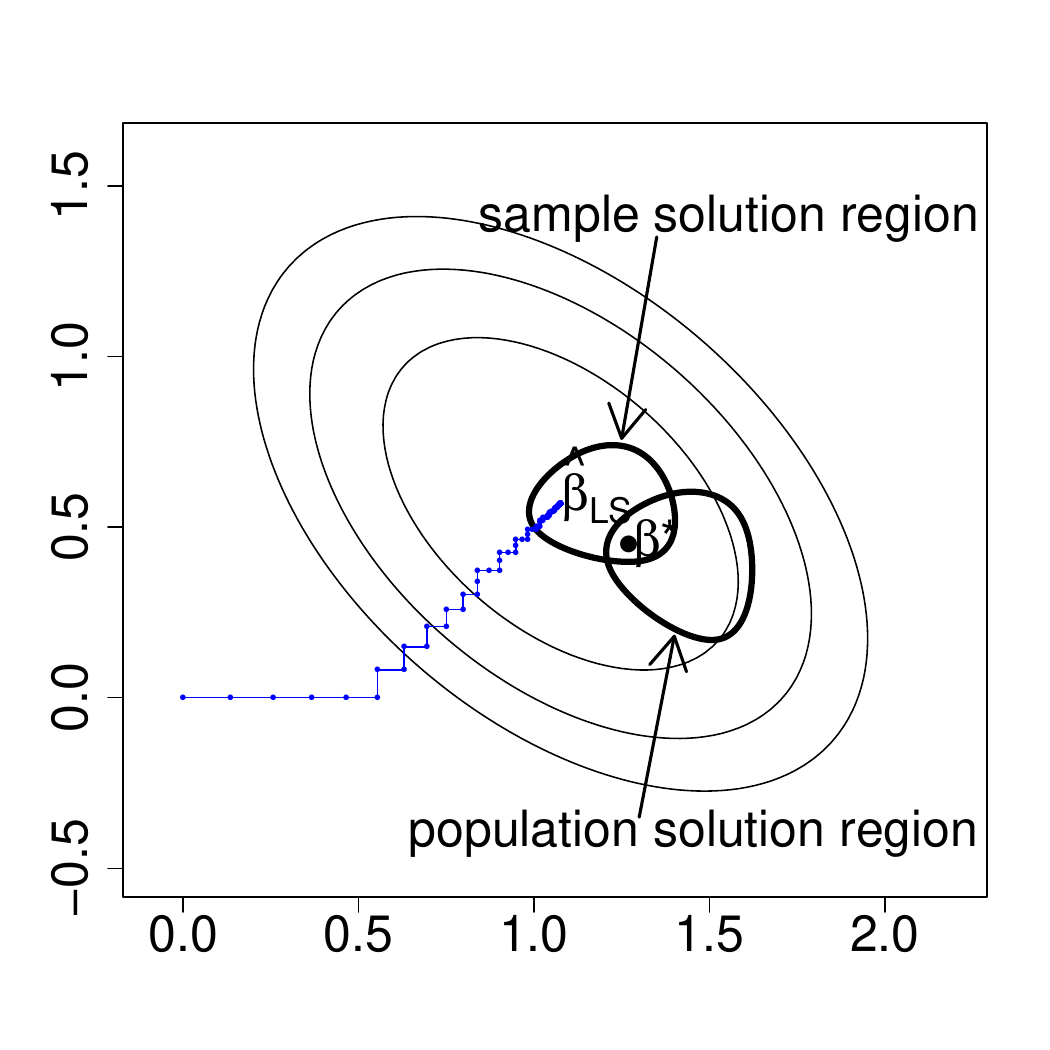} }%
	\caption{Assume there are two unidentified elements in $\beta^*$. The blue line is an LS-boost solution path starting from a zero vector and $\hat{\beta}^{\tilde{k}}$ is the boosting solution at step $\tilde{k}$. \Cref{fig:2a,fig:2b} show $L_n(\hat{\beta}_{LS}) < L_n(\beta^*)$, and in \Cref{fig:1c}, $L_n(\hat{\beta}_{LS}) = L_n(\beta^*)$.}%
	\label{fig:contours}%
\end{figure}

Without any identification assumption, both $\hat{\beta}_{\text{LS}}$ and $\beta^*$ are underidentified, as shown in \Cref{fig:2a}. The boosting solution converges to $\hat{\beta}_{\text{LS}}$, which is one of many solutions in the flat sample solution region, or the irregular shape at the center of the contour plot. The linear model in \cref{eq:lrm} is also underidentified at the population level, leading to a flat population solution region in \Cref{fig:2a}. In general, the two regions do not completely overlap. \Cref{fig:2b,fig:2c} show where they might intersect, and in \Cref{fig:2c}, $L_n(\hat{\beta}_{LS}) = L_n(\beta^*)$.

Let $\lambda_{\text{pmin}}(\boldsymbol{\mathrm{X}}^T\boldsymbol{\mathrm{X}})$ be the smallest nonzero eigenvalue of $\boldsymbol{\mathrm{X}}^T\boldsymbol{\mathrm{X}}$ and define 
\begin{equation} \label{eq:gamma}
	\gamma:= \left(1 - \frac{\varepsilon(2 - \varepsilon) \lambda_{\text{pmin}}(\boldsymbol{\mathrm{X}}^T\boldsymbol{\mathrm{X}})}{4p}\right).
\end{equation}
FGM show that $0.75 \leq \gamma < 1$. Let $\nabla L_n(\beta)$ be the gradient vector at $\beta$.
\begin{theorem} \label{thm:prediction convergence}
	Under \cref{assumption:general}, for $k \geq 0$, LS-boost has the following prediction bound:
	\begin{equation} \label{eq:prediction bound}
		\lVert \boldsymbol{\mathrm{X}} \hat{\beta}^k - \boldsymbol{\mathrm{X}}\beta^* \rVert_2 \leq \lVert \boldsymbol{\mathrm{X}} \hat{\beta}^k_\text{LS} \rVert_2 \gamma^{k/2} + \sqrt{2n \lVert \nabla L_n(\beta^*) \rVert_2 \cdot \lVert \hat{\beta}_\text{LS} - \beta^* \rVert_2}.
	\end{equation}
\end{theorem}
\noindent A proof is given in the online supplement. Compared to Theorem 2.1 in FGM, \cref{eq:prediction bound} has an extra term that relates to the gradient vector and the $\ell_2$ error of $\hat{\beta}_\text{LS}$. Without additional assumptions, this extra term will not disappear as $k \rightarrow \infty$. 

\begin{remark}
	In the special case when $\beta^*$ is located inside the sample solution region (see \Cref{fig:2c}), $\lVert \nabla L_n(\beta^*) \rVert_2 = 0$ and \cref{eq:prediction bound} reduces to the result in Theorem 2.1 in FGM. This result holds even when $\lVert \hat{\beta}_\text{LS} - \beta^* \rVert_2 > 0$.
\end{remark}

%Consider the prediction at a new data point, $\boldsymbol{\mathrm{X}}_\text{new}$, independent of the sample $\{\boldsymbol{\mathrm{y}},\boldsymbol{\mathrm{X}}\}$. Assume $\boldsymbol{\mathrm{X}}_\text{new}$ is a $p \times 1$ vector. The proof of \Cref{thm:prediction convergence} shows a similar result holds at $\boldsymbol{\mathrm{X}}_\text{new}$.
%\begin{equation} \label{eq:new prediction bound}
%	\lVert \boldsymbol{\mathrm{X}}_\text{new}^T \hat{\beta}^k - \boldsymbol{\mathrm{X}}_\text{new}^T \beta^* \rVert_2 \leq \lVert \boldsymbol{\mathrm{X}}_\text{new}^T \hat{\beta}^k_\text{LS} \rVert_2 \gamma^{k/2} + \sqrt{2 \lVert \nabla L_n(\beta^*) \rVert_2 \cdot \lVert \hat{\beta}_\text{LS} - \beta^* \rVert_2}.
%\end{equation}

Clearly, \cref{eq:prediction bound} indicates that, in a finite sample case when $n \not\to \infty$, LS-boost prediction will not recover the true sparse regression function, $\boldsymbol{\mathrm{X}} \beta^*$. Theorem 12.2 in \cite{buhlmann2011highdimenstats} and Theorem 1 in \cite{buhlmann2006boostinghigh} show that, as both $n \to \infty$ and $k \to \infty$, $\lVert \boldsymbol{\mathrm{X}} \hat{\beta}^k - \boldsymbol{\mathrm{X}}^T \beta^* \rVert_2^2/n = o_p(1)$. This result does not contradict the nonasymptotic result in \cref{eq:prediction bound}. A heuristic argument follows.

\begin{remark}
	As $n \to \infty$ and sample data get closer to population, the sample solution region will converge to the population solution region at the center. We expect $ \nabla L_n(\beta^*) \to 0$ in \cref{eq:prediction bound}, so the second term in \cref{eq:prediction bound} will disappear asymptotically, and we will have $\lVert \boldsymbol{\mathrm{X}} \hat{\beta}^k - \boldsymbol{\mathrm{X}}\beta^* \rVert_2^2/n \rightarrow o_p(1)$ when $k \rightarrow \infty$. Section S.2 provides a more detailed explanation.
\end{remark}

\begin{remark}
	Both Theorem 12.2 of \cite{buhlmann2011highdimenstats} and \Cref{thm:prediction convergence} present a prediction convergence result. How do they compare?  \Cref{thm:prediction convergence} uses an exponential function to the base of $\gamma$ to characterize the convergence of $\lVert \boldsymbol{\mathrm{X}} \hat{\beta}^k - \boldsymbol{\mathrm{X}}\beta^* \rVert_2^2/n$ as $k\rightarrow \infty$, while Theorem 12.2 in \cite{buhlmann2011highdimenstats} relies on a power function of $k$ to achieve the same goal. See Section S.2 for a more detailed discussion.
\end{remark}

Next, we discuss the faster convergence rate of lassoed boosting. Let $|\mathcal{A}_q| = p_q$. Consider two active sets $\mathcal{A}_{q_1}$ and $\mathcal{A}_{q_2}$ that are generated on the lasso solution path. We focus on the specific case when $p_{q_1} < p_{q_2} < n$ with either $n < p$ or $n \ge p$ and discuss why it is worth considering.

The linear convergence rate $\gamma$ plays a critical role in determining the speed of convergence in \Cref{thm:prediction convergence} and Theorem 2.1 in FGM. Figure 4 in FGM shows a general pattern: $\gamma$ decreases, and $\lambda_{\text{pmin}}(\boldsymbol{\mathrm{X}}^T\boldsymbol{\mathrm{X}})$ increases as $p$ increases. However, the two relationships are not strictly monotonic, suggesting that the convergence will be faster as $p$ increases, in agreement with result (ii) in Theorem 2.1 of FGM
\begin{equation} \label{eq:FGM beta convergence}
	\lVert \hat{\beta}^k - \hat{\beta}_\text{LS}^k \rVert_2 \leq \frac{\lVert \boldsymbol{\mathrm{X}} \hat{\beta}_\text{LS} \rVert_2}{\sqrt{\lambda_{\text{pmin}}(\boldsymbol{\mathrm{X}}^T\boldsymbol{\mathrm{X}})}}\gamma^{k/2}.
\end{equation}
A similar conclusion can be drawn for the convergence result in \Cref{thm:prediction convergence}. We would expect the opposite: the convergence for the estimator and the prediction will slow down when $p$ increases and more variables add ``noise" and competition to the variable selection process.

We provide an alternative explanation to complement the results in Figure 4 in FGM. Notice it represents cases when $p > n$ with $n=50$ and $p\geq 73$. The same figures will give a different pattern when $p < n$. In lassoed boosting, we apply LS-boost sequentially to variables in $\left\{\mathcal{A}_q\right\}_{q=1}^{Q}$. If the lasso does a good job, and the model really is sparse, we expect that in the early stage of variable selection, some of the $\mathcal{A}_q$ will include the correct variables and $|\mathcal{A}_q|$ will be much smaller than $n$ ($\ll n$). For a sparse model, those active sets with $|\mathcal{A}_q| \ll n$ are arguably the most interesting since the boosting solutions spawned on these sets will be more likely to mimic the true and sparse elements in $\beta^*$. LS-boost on each $\mathcal{A}_q$ can be viewed as separate exercises. We can add the subscript $\mathcal{A}_q$ to results in \Cref{thm:prediction convergence} and \cref{eq:FGM beta convergence}. Consider \cref{eq:FGM beta convergence} for the active set $\mathcal{A}_q$ with $|\mathcal{A}_q| \ll n $.
\begin{equation} \label{eq:FGM beta convergence on active set}
\lVert \hat{\beta}^k_{\mathcal{A}_q} - \hat{\beta}_{\text{LS},\mathcal{A}_q}^k \rVert_2 \leq \frac{\lVert \boldsymbol{\mathrm{X}}_{\mathcal{A}_q} \hat{\beta}_{\text{LS},\mathcal{A}_q} \rVert_2 }{\sqrt{\lambda_{\text{pmin}}(\boldsymbol{\mathrm{X}}_{\mathcal{A}_q}^T\boldsymbol{\mathrm{X}}_{\mathcal{A}_q})}}\gamma^{k/2}_{\mathcal{A}_q},
\end{equation}
where all quantities are restricted to the active set $\mathcal{A}_q$ and 
\begin{equation} \label{eq:gamma on active set}
\gamma_{\mathcal{A}_q}:= \left(1 - \frac{\varepsilon(2 - \varepsilon) \lambda_{\text{pmin}}(\boldsymbol{\mathrm{X}}_{\mathcal{A}_q}^T\boldsymbol{\mathrm{X}}_{\mathcal{A}_q})}{4p_q}\right).
\end{equation}
Since $|\mathcal{A}_q| < n$, all parameters can be identified. 

\begin{figure}[htp]
	\centering
	\subfloat[plot of $\gamma$ as $|\mathcal{A}_q|$ increases]{\label{fig:3a}\includegraphics[width=0.48\linewidth,keepaspectratio,scale=1]{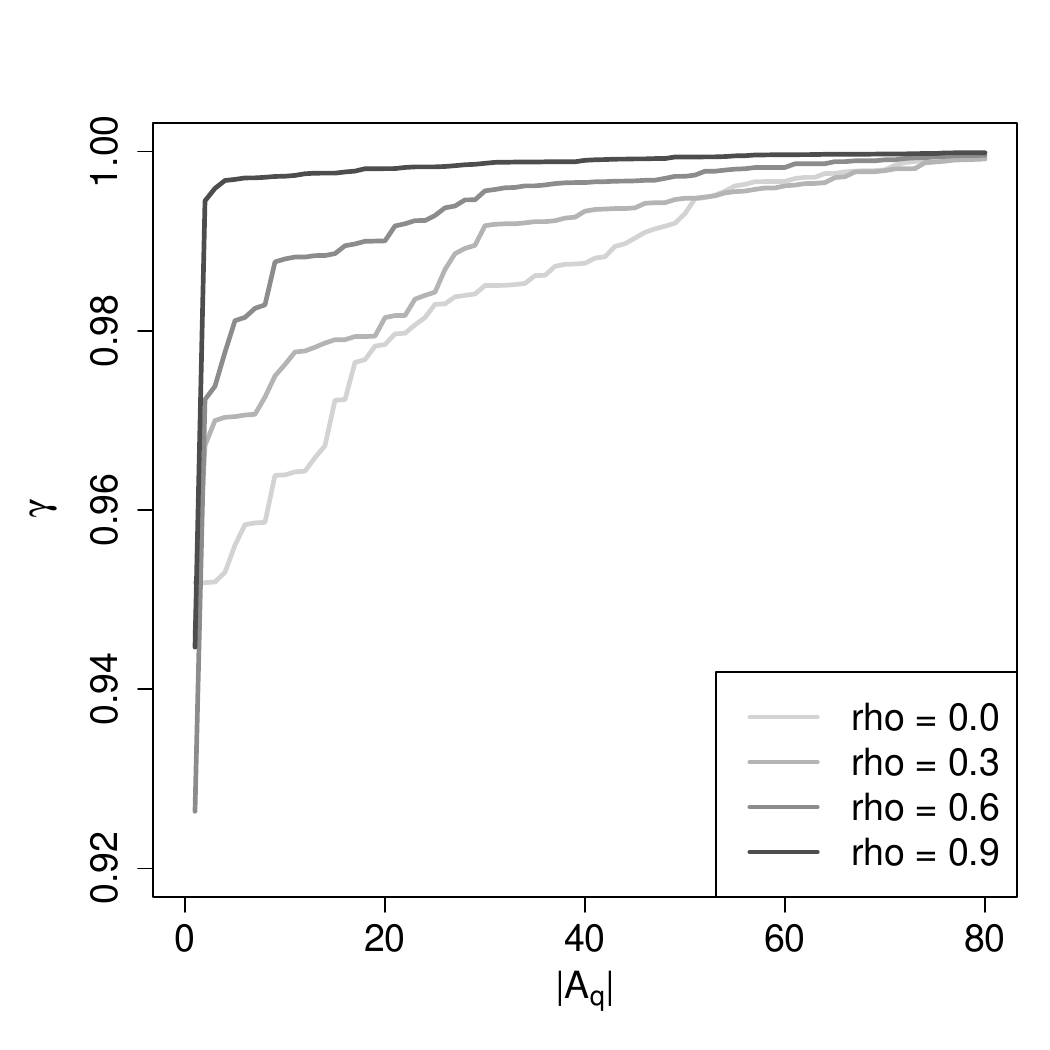} }%
	\subfloat[plot of $\lambda_{\text{pmin}}(\boldsymbol{\mathrm{X}}_{\mathcal{A}_q}^T\boldsymbol{\mathrm{X}}_{\mathcal{A}_q})$ as $|\mathcal{A}_q|$ increases]{\label{fig:3b}\includegraphics[width=0.48\linewidth,keepaspectratio,scale=1]{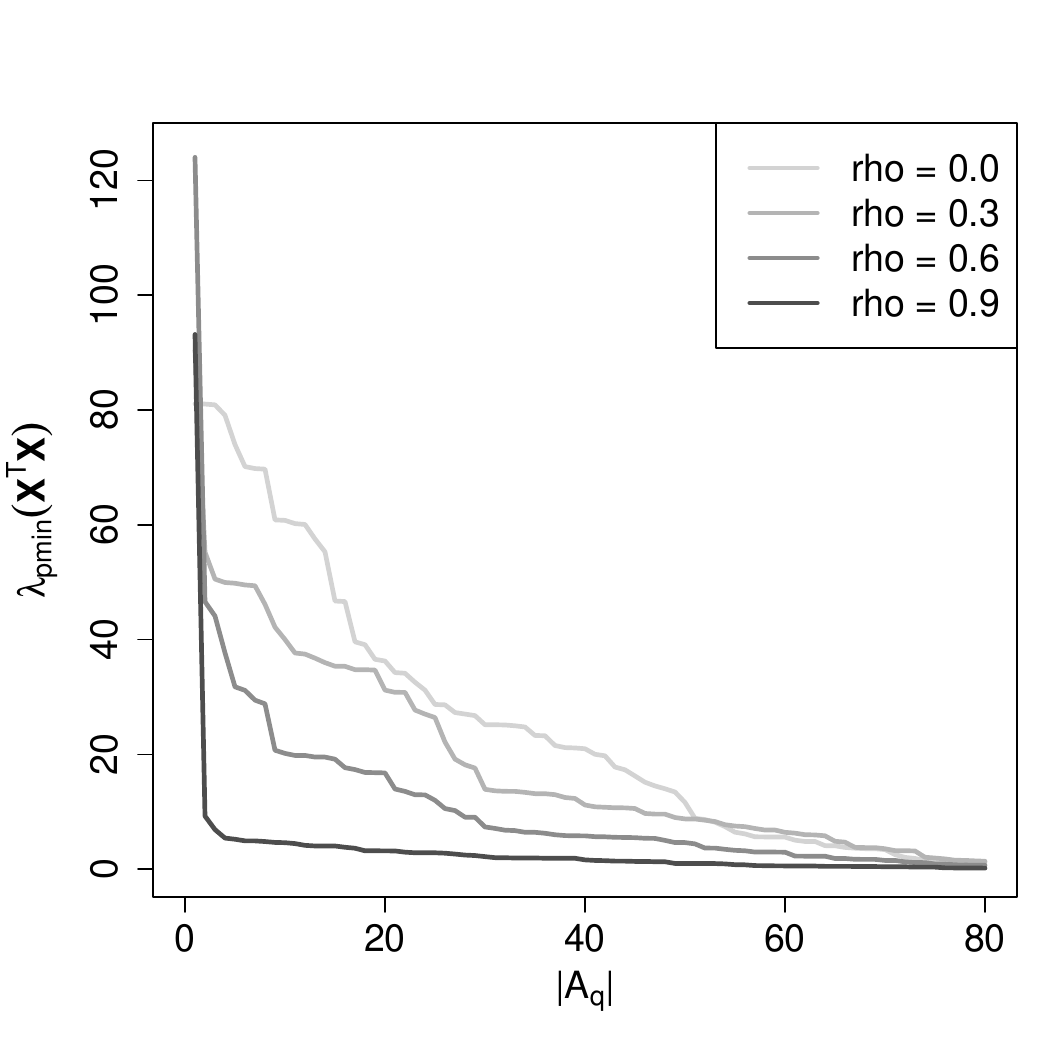} }
	\caption{Simulation results for $\gamma_{\mathcal{A}_q}$ and $\lambda_{\text{pmin}}(\boldsymbol{\mathrm{X}}_{\mathcal{A}_q}^T\boldsymbol{\mathrm{X}}_{\mathcal{A}_q})$ with $n=100$ and $|\mathcal{A}_q| \leq 80$. The symbol rho refers to correlation among variables. \Cref{fig:3a} shows $\gamma_{\mathcal{A}_q}$ is an increasing function of $|\mathcal{A}_q|$. \Cref{fig:3b} shows $\lambda_{\text{pmin}}(\boldsymbol{\mathrm{X}}_{\mathcal{A}_q}^T\boldsymbol{\mathrm{X}}_{\mathcal{A}_q})$ is a decreasing function of $|\mathcal{A}_q|$.}%
	\label{fig:rate}%
\end{figure}

Next, we show a simulation result that describes $\gamma_{\mathcal{A}_q}$ and $\lambda_{\text{pmin}}(\boldsymbol{\mathrm{X}}_{\mathcal{A}_q}^T\boldsymbol{\mathrm{X}}_{\mathcal{A}_q})$ as a function of $|\mathcal{A}_q|$. \Cref{fig:rate} describes the relationship of the linear convergence rate and minimum eigenvalue with $p_q(=|\mathcal{A}_q|)$ when $p_q < n$. It reverses the patterns in Figure 4 in FGM. We stress that both figures are correct, but \Cref{fig:rate} helps to explain the convergence rate when $p_q < n$. As the lasso penalty parameter $\lambda_q$ decreases, and $p_q$ increases, the minimum eigenvalue of $\boldsymbol{\mathrm{X}}_{\mathcal{A}_q}^T\boldsymbol{\mathrm{X}}_{\mathcal{A}_q}$ will decrease monotonically and remain positive. Such decrease in minimum eigenvalue as matrix dimension increases is a standard result in matrix theory (see, e.g., Theorem 4.3.8 in \cite{hornjohnson1985matrixanalysis}). \Cref{fig:3b} and \cref{eq:gamma on active set} imply \Cref{fig:3a}. Hence, for two active sets $\mathcal{A}_{q_1}$ and $\mathcal{A}_{q_2}$ with $p_{q_1} < p_{q_2} < n$, we have $\gamma_{\mathcal{A}_{q_1}} < \gamma_{\mathcal{A}_{q_2}}$ and the convergence rate for $\hat{\beta}^k_{\mathcal{A}_q}$ in \cref{eq:FGM beta convergence on active set} is faster when boosting on $\mathcal{A}_{q_1}$ than on $\mathcal{A}_{q_2}$. This result applies to both \cref{eq:FGM beta convergence on active set} and \Cref{thm:prediction convergence}  after we replace $\boldsymbol{\mathrm{X}}^T\boldsymbol{\mathrm{X}}$ with $\boldsymbol{\mathrm{X}}_{\mathcal{A}_q}^T\boldsymbol{\mathrm{X}}_{\mathcal{A}_q}$ in lassoed boosting. It is beneficial,  particularly in the early stages of the lasso; when $\mathcal{A}_q$ includes correct variables and $p_q < n$, the parameters are identified; convergence is faster; and the convergence rate for prediction in \Cref{thm:prediction convergence} is also faster. When $|\mathcal{A}_q| > n$, parameters are underidentified and $\hat{\beta}_{\text{LS},\mathcal{A}_q}^k $ in \cref{eq:FGM beta convergence on active set} may differ for each $k$ and $\mathcal{A}_q$.

\section{Additional examples of two-stage procedure}

In this section, we give several additional examples of two-stage estimators.
\subsection{Lassoed forward stagewise regression}
The forward stagewise regression algorithm in \cite{hastie2009elements} is a popular method for building coefficients and the regression function in small steps. In the $k$th step, it identifies the predictor $\boldsymbol{\mathrm{x}}_{j_k}$ that best correlates with the current residual $\boldsymbol{\mathrm{r}}_k$ and makes the following update:
\begin{align} \label{alg:forward stagewise}
	&\hat{\beta}_{j_k}^{k+1} = \hat{\beta}_{j_k}^{k} + \varepsilon \cdot \text{sign}(\boldsymbol{\mathrm{r}}_k^T\boldsymbol{\mathrm{x}}_{j_k}), \hat{\beta}_{j}^{k+1} = \hat{\beta}_{j}^{k} \text{ for } j \neq j_k,\text{ and }\boldsymbol{\mathrm{r}}_{k+1} = \boldsymbol{\mathrm{r}}_k - \varepsilon \cdot \text{sign}(\boldsymbol{\mathrm{r}}_k^T\boldsymbol{\mathrm{x}}_{j_k}) \boldsymbol{\mathrm{x}}_{j_k}. 
\end{align}
 This algorithm can also be implemented on lasso-generated active sets. We call it lassoed forward stagewise regression.
 
Like \Cref{thm:prediction convergence}, \Cref{thm:prediction convergence for forward} in the online supplement gives a convergence result for the prediction function of the estimator in \Cref{alg:lassoed forward}. In the case of $p_q < n$, applying \Cref{thm:prediction convergence for forward} to the active set $\mathcal{A}_q$ and using the same argument as for lassoed boosting, we conclude that the convergence rate for lassoed forward stagewise regression is faster.
 \begin{center}
 	{\SetAlgoNoLine%
 		\begin{algorithm}[H]
 			\caption{Lassoed forward stagewise regression} 
 			\label{alg:lassoed forward} 
 			\DontPrintSemicolon
 			\SetKwFor{For}{for}{do}{end~for}
 			\SetKw{KwAssume}{Assume}
 			\KwAssume{a sequence of $Q$ tuning parameters $\left\{\lambda_q\right\}_1^Q$ for the lasso problem.}\\
 			\For{$q = 1$ to $Q$}{
 				Use the lasso to obtain an active set of predictors, $\mathcal{A}_q$, for each $\lambda_q$.\\
 				Use forward stagewise regression to compute the coefficient path for each predictor in $\mathcal{A}_q$.
 			}
 			\Return $Q$ sets of coefficient paths for validation (or cross-validation) purposes.
 		\end{algorithm}
 	}
 \end{center}
%Mimicking Theorem 3.1 in FGM and \Cref{thm:prediction convergence}, we give a prediction convergence result for \Cref{alg:lassoed forward}.
%\begin{theorem} \label{thm:prediction convergence for forward}
%	Let $k \geq 0$ be the number of iterations. Under \cref{assumption:general}, there exists an $i \in \left\{0,\cdots,k\right\}$ so that the following bound hold:
%	\begin{equation} \label{eq:prediction bound forward}
%	\lVert \boldsymbol{\mathrm{X}} \hat{\beta}^i - \boldsymbol{\mathrm{X}}\beta^* \rVert_2 \leq \frac{\sqrt{p}}{\sqrt{\lambda}(\boldsymbol{\mathrm{X}}^T\boldsymbol{\mathrm{X}})} \left[\frac{\lVert \boldsymbol{\mathrm{X}}\hat{\beta}_\text{LS} \rVert_2^2}{\varepsilon(k+1)} + \epsilon\right] + \sqrt{2n \lVert \nabla L_n(\beta^*) \rVert_2 \cdot \lVert \hat{\beta}_\text{LS} - \beta^* \rVert_2}.
%	\end{equation}
%\end{theorem}

%The proof is similar to that of \Cref{thm:prediction convergence} and is given in the online supplement. In the case of $p_q < n$, applying \Cref{thm:prediction convergence for forward} to the active set $\mathcal{A}_q$ and using the same argument for lassoed boosting, we conclude that lassoed forward stagewise regression has a faster convergence rate compared to forward stagewise regression.

\subsection{Twiced lasso} 

 \citet[p.~91]{hastie2009elements} write, ``one can use the lasso to select the set of non-zero predictors, and then apply the lasso again, but using only the selected predictors from the first step. This is known as the \textit{relaxed lasso} (Meinshausen, 2007).'' The idea of using the lasso twice refers to the ``Simple Algorithm" in \citet[p.~377]{meinshausen2007relaxed}, where after obtaining the active set $\mathcal{A}_q$, the lasso is applied again on $\mathcal{A}_q$ with the penalty parameter on $[0,\lambda_q]$. \Cref{alg:twiced lasso} describes the procedure of lassoing twice, and we call it twiced lasso.
 %We provide a trivial extension of the relaxed lasso to facilitate its comparison with lassoed boosting and call it twiced lasso. 
 \begin{center}
 	{\SetAlgoNoLine%
 		\begin{algorithm}[H]
 			\caption{Twiced lasso} 
 			\label{alg:twiced lasso} 
 			\DontPrintSemicolon
 			\SetKwFor{For}{for}{do}{end~for}
 			\SetKw{KwAssume}{Assume}
 			\KwAssume{a sequence of $Q$ tuning parameters $\left\{\lambda_q\right\}_1^Q$ for the lasso problem}\\
 			\For{$q = 1$ to $Q$}{
 				Use the lasso to obtain an active set of variables, $\mathcal{A}_q$, for each $\lambda_q$.\\
 				Apply the lasso again on $\mathcal{A}_q$ to compute the coefficient path for each predictor in $\mathcal{A}_q$ with $\left\{\lambda_1, \cdots, \lambda_Q\right\}$.
 			}
 			\Return $Q$ sets of coefficient paths for validation (or cross-validation) purposes.
 		\end{algorithm}
 	}
 \end{center}
 The key difference between twiced lasso and the relaxed lasso is, in step 4 of \Cref{alg:twiced lasso}, the former always starts at $\lambda_1$ while the latter starts at $\lambda_{q+1}$. Obviously, starting from $\lambda_1$ for every $\mathcal{A}_q$ creates computational redundancy. However, the formulation in \Cref{alg:twiced lasso} allows direct comparison with lassoed boosting in \Cref{alg:lassoed boosting} and helps to explain the change in the convergence rate.
 
 Adapt the restricted eigenvalues condition in equation (11.10) in \cite{hastie2015slsparcity} to the active set $\mathcal{A}_q$ and we have, as a constant $\tilde{\lambda} > 0$,
 \begin{equation} \label{eq:restricted eigenvalues}
 	\frac{\nu^T \left(\boldsymbol{\mathrm{X}}_{\mathcal{A}_q}^T\boldsymbol{\mathrm{X}}_{\mathcal{A}_q}/n\right) \nu}{\lVert \nu \rVert_2^2} \geq \tilde{\lambda} \text{ for all nonzero } \nu \text{ in a constrained set } \mathcal{C}.
 \end{equation}
The constrained set $\mathcal{C}$ defines directions along which parameters can be identified. The discussion on lassoed boosting also applies here. When the lasso does a good job selecting variables in step 3 of \Cref{alg:twiced lasso}, $\mathcal{A}_q$ contains the correct variables, and $p_q < n$. In this case, the least-squares loss is strictly convex ($\nabla^2 L_n(\beta) = \boldsymbol{\mathrm{X}}_{\mathcal{A}_q}^T\boldsymbol{\mathrm{X}}_{\mathcal{A}_q}/n $ is positive definite) and \cref{eq:restricted eigenvalues} holds for all $\nu \in \mathbb{R}^{p_q}$. Hence, $\tilde{\lambda} = \lambda_{\text{pmin}}(\boldsymbol{\mathrm{X}}_{\mathcal{A}_q}^T\boldsymbol{\mathrm{X}}_{\mathcal{A}_q}/n)$,  and as \Cref{fig:3b} suggests, $\tilde{\lambda}$ decreases as $p_q$ increases. Theorem 11.1 in \cite{hastie2015slsparcity} implies
\begin{equation} \label{eq:lasso beta convergence}
	\lVert \hat{\beta}_{\mathcal{A}_q} - \beta^*_{\mathcal{A}_q} \rVert_2 \leq \frac{3}{\lambda_{\text{pmin}}(\boldsymbol{\mathrm{X}}_{\mathcal{A}_q}^T\boldsymbol{\mathrm{X}}_{\mathcal{A}_q}/n)} \sqrt{\frac{p_q}{n}} \sqrt{n} \lambda_n,
\end{equation}
where $\lambda_n$ is the tuning parameter such that $\lambda_n \geq 2 \lVert \boldsymbol{\mathrm{X}}_{\mathcal{A}_q}^T \boldsymbol{\mathrm{u}} \rVert_{\infty}/n > 0$. A smaller $p_q$ will tighten the bound in \cref{eq:lasso beta convergence}, implying a faster convergence rate. 

\begin{remark}
	In sum, in the twiced lasso, the first-stage screening can pare down the number of variables so that the convergence rate for the second-stage lasso estimator will be faster when $p_q < n$.
\end{remark}

%In \cref{eq:lasso beta convergence} when $p_q < s$, $\beta^*_{\mathcal{A}_q}$ is not necessarily equal to the corresponding elements in $\beta^*$ and its value varies, depending on $\boldsymbol{\mathrm{X}}_{\mathcal{A}_q}$. 
When $s \leq p_q < n$, and $\mathcal{A}_q$ correctly includes the nonzero variables, the non-zero elements of $\beta^*_{\mathcal{A}_q}$ are equal to those in $\beta^*$. 

\subsection{Twiced boosting}

Another extension uses boosting twice. Step 1 in \Cref{alg:twiced boosting} uses LS-boost to screen variables and organizes them into sequentially increasing sets. Step 2 uses LS-boost again to grow the coefficient paths on each active set. The iteration stopping criterion in step 1 is when the active set includes all variables, and we can use the standard AIC-type criterion to stop boosting in the second stage.
\begin{center}
	{\SetAlgoNoLine%
		\begin{algorithm}[H]
			\caption{Twiced boosting} 
			\label{alg:twiced boosting} 
			\DontPrintSemicolon
			\SetKwFor{For}{for}{do}{end~for}
			\SetKw{KwRun}{Run}
			\KwRun{a first round of LS-boost to obtain a sequence of $Q$ different active sets of variables, $\mathcal{A}_q$ with $q=1,\cdots,Q$} --- not the same $\mathcal{A}_q$ from the lasso procedure.\\
			%\hfill \textbackslash\textbackslash not the same $\mathcal{A}_q$ from the lasso \\
			\For{$q = 1$ to $Q$}{
				Run LS-boost on each active set of variables, $\mathcal{A}_q$, and obtain coefficient paths for all predictors in $\mathcal{A}_q$.
			}
			\Return $Q$ sets of coefficient paths for validation (or cross-validation) purposes
		\end{algorithm}
	}
\end{center}

\Cref{alg:twiced boosting} differs from the twin boosting algorithm in \cite{buhlmann2010twinboosting}. Twin boosting uses the estimate in the first stage to guide variable selection in the second stage, similar to the idea of the adaptive lasso in \cite{zou2006adaptivelasso}.

\section{Monte Carlo simulation}

Our Monte Carlo simulations study the performance of five estimators: forward stepwise, lasso, lassoed boosting, relaxed lasso, and twiced lasso (see online supplement). We do not consider the best subset selection method in \cite{bertsimas2016bestsubset} due to its computation cost. Our simulation design is the same as that in \cite{hastie2017extended} except that we use 50, rather than 10, equally spaced values between $0$ and $1$ as weights for the relaxed lasso.

\subsection{Simulation setup}

The data-generating process (DGP) is determined by a combination of $4$ beta types, $4$ sample-size and sparsity patterns, $3$ error correlation values, and $10$ signal-to-noise (SNR) levels, for a total of $480(=4\times4\times3\times10)$ DGPs.

\textit{Beta types}. We consider four coefficient settings for the sparse coefficient $\beta^*$.
\begin{enumerate}[align=left]
	\setlength{\itemsep}{0pt}
	\setlength{\parskip}{0pt}
	\item[beta-type 1:] $\beta^*$ has $s$ elements equal to 1, equally spaced between $1$ and $p$, and the rest equal to $0$;
	\item[beta-type 2:] The first $s$ elements of $\beta^*$ equal 1, and the rest equal $0$;
	\item[beta-type 3:] The first $s$ elements of $\beta^*$ equal $s$ interpolated values between $10$ and $0.5$, and the rest equal $0$;
	\item[beta-type 5:] The first $s$ elements of $\beta^*$ equal 1. The rest decay to $0$, $\beta^*_i = 0.5^{i-s}$ for $i=s+1,\cdots,p$.
\end{enumerate}
The first three beta-types are considered in \cite{bertsimas2016bestsubset}; beta-type 5 is added in \cite{hastie2017extended}. We do not consider beta-type 4 in \cite{bertsimas2016bestsubset} due to its similar result to beta-type 3.

\textit{Size and sparsity}. We consider the following data size and sparsity combinations:
\[
\begin{array}{ll}
	\bullet \text{ low: }n=100, p = 10, s = 5 &\bullet \text{ medium: }n = 500, p = 100, s = 5  \\
	\bullet \text{ high-5: }n = 50, p = 1000, s = 5 & \bullet \text{ high-10: }n = 100, p = 1000, s = 10
\end{array}
\]

\textit{Error correlation}. We try $\rho = 0, 0.3$ and $0.7$ in the DGP.

\textit{SNR levels}. The SNR level takes $10$ values: $v=(0.05,0.09,0.14,0.25,0.42,0.71,1.22,2.07,3.52,6.00)$.

Let $\hat{\beta}$ be the estimated coefficient from one of the five methods. 

\textit{Evaluation metrics}. We apply the following four evaluation metrics on the test data:
\begin{itemize}
	\item Relative risk:	\begin{equation*}
		\text{RR}(\hat{\beta}) = \frac{E(x_0^T\hat{\beta}-x_0^T\beta^*)^2}{E(x_0^T\beta^*)^2} = \frac{(\hat{\beta}-\beta^*)^T\Sigma (\hat{\beta}-\beta^*)}{{\beta^*}^T\Sigma \beta^*}.
	\end{equation*}
	$\text{RR}(\hat{\beta}) = 0$ if $\hat{\beta} = \beta^*$; $\text{RR}(\hat{\beta}) = 1$ if $\hat{\beta} = 0$ (the null model.)
	\item Relative test error:
	\begin{equation*}
		\text{RTE}(\hat{\beta}) = \frac{E(y_0-x_0^T\beta)^2}{\sigma^2} = \frac{(\hat{\beta}-\beta^*)^T\Sigma (\hat{\beta}-\beta^*) + \sigma^2}{\sigma^2}.
	\end{equation*}
	$\text{RTE}(\hat{\beta}) = 1$ if $\hat{\beta} = \beta^*$; $\text{RTE}(\hat{\beta}) = \text{SNR} + 1$ if $\hat{\beta} = 0$.
	\item Proportion of variance explained:
	\begin{equation*}
		\text{PVE}(\hat{\beta}) = 1 - \frac{E(y_0-x_0^T\beta)^2}{\text{Var}(y_0)} = 1 -  \frac{(\hat{\beta}-\beta^*)^T\Sigma (\hat{\beta}-\beta^*) + \sigma^2}{{\beta^*}^T\Sigma \beta^* + \sigma^2}.
	\end{equation*}
	$\text{PVE}(\hat{\beta}) = \text{SNR}/(1+\text{SNR})$ if $\hat{\beta} = \beta^*$; $\text{RTE}(\hat{\beta}) = 0$ if $\hat{\beta} = 0$.
	\item Number of nonzeros: $\text{NNZ}(\hat{\beta}) = \lVert \hat{\beta} \rVert_0 = \sum_{j=1}^{p} 1\{\hat{\beta}_j \neq 0 \}$. In figures that plot $\lVert \hat{\beta} \rVert_0$, we also print the number (averaged over 10 replications) of correctly identified coefficients for each method at each SNR level, which allows us to infer the true- and false- positive rates in variable recovery.
\end{itemize}
The simulation study has the following steps:
\begin{description}
	\setlength{\itemsep}{0pt}
	\setlength{\parskip}{0pt}
	\item[Step 1 Data simulation] Choose a beta-type for $\beta^*$. Draw rows of $\boldsymbol{\mathrm{X}}$ i.i.d. from $N(0,\Sigma_{p \times p})$, where the $ij$th element of $\Sigma$ is $\rho^{|i-j|}$ with $\rho = 0,0.35$ or $0.7.$ Draw $\boldsymbol{\mathrm{y}}$ from $N(\boldsymbol{\mathrm{X}}\beta^*, \sigma^2 \boldsymbol{\mathrm{I}}_n)$ and $\sigma^2 = {\beta^*}^T\Sigma \beta^*/v$.
	\item[Step 2 Model selection] Run each method on $(\boldsymbol{\mathrm{X}}, \boldsymbol{\mathrm{y} })$ with a sequence of tuning parameters; select a tuning parameter by minimizing prediction error on a validation set the same size as $(\boldsymbol{\mathrm{X}}, \boldsymbol{\mathrm{y} })$ and generated independently.
	\item[Step 3 Model evaluation] Record four metrics.
	\item[Step 4 Average] Repeat steps 1 to 3 ten times and compute the average of all metrics.
\end{description}
%The simulation has a total of $480$ DGPs. with the following variations:
%\begin{table}[h!]
%	\begin{center}
%		%\caption{Your first table.}
%		\label{tab:DGP}
%		\begin{tabular}{c|c|c|c|c} % <-- Alignments: 1st column left, 2nd middle and 3rd right, with vertical lines in between
%			beta types & size and sparsity & error correlations & SNR levels & Total combinations \\
%			\hline
%			4 & 4 & 3 & 10 & 480
%		\end{tabular}
%	\end{center}
%\end{table}

%\begin{itemize}
%	\item low: $n=100, p = 10, s = 5$;
%	\item medium: $n = 500, p = 100, s = 5$;
%	\item high-5: $n = 50, p = 1000, s = 5$;
%	\item high-10: $n = 100, p = 1000, s = 10$.
%\end{itemize}
%n total, we have $ 4 (\text{beta-type})\times 4 (\text{size and sparsity}) \times 10 (\text{SNR}) \times 3 (\rho) = 480$ DGPs. 

Parameter tuning again follows that in \cite{hastie2017extended}. We use the R package \texttt{glmnet} to generate the lasso solutions and to select variables for lassoed boosting. We use $50$ values of $\lambda$ for the lasso at the low setting and $100$ values of $\lambda$ in the other three settings. We also use $50$ values of equally spaced weights between $0$ and $1$ for the relaxed lasso. The forward stepwise procedure is tuned up to $50$ steps. We use the R package $mboost$ to apply LS-boost to each active set of variables. Early stopping in boosting typically prevents an LS solution, while the relaxed lasso can always reach a full LS solution. To reduce the difference between lassoed-boosting and relaxed-lasso solutions, we use the corrected AIC in \cite{hurvich1998smoothing} to obtain the iteration number and then double it as the final stopping criterion in LS-boost. In the interval between $1$ and the final stopping number, we use $50$ equally spaced steps and extract the corresponding LS-boost coefficients on each $\mathcal{A}_q$. The learning rate is set to $0.01$.

Like \cite{hastie2017extended}, for all methods, we use a training set to generate solution paths, a validation set to select the best solution, and a test set to compute the final prediction. Hence, our results require no direct tuning of the lasso penalty parameter $\lambda$.

Lassoed boosting is more computationally expensive than the relaxed lasso. By the time the lasso and the relaxed lasso finishes, lassoed boosting just gets started at generating solutions for each of the active set. We can employ parallelization and warm starts to improve the speed and reduce redundancy in computation. 

\subsection{Simulation results}
This discussion focuses on cases for beta-type 2 and $\rho = 0.35$ based on validation tuning. The supplement includes the complete set of results for both validation and oracle tuning, where parameters are chosen to minimize the average risk over $10$ replications. We also skip the results for the twiced lasso because they resemble those for the relaxed lasso, but they are reported in all figures in the supplement. 

\Cref{fig:low,fig:med,fig:hi5,fig:hi10} plot the average of relative risk (RR), relative test error (RTE), proportion of variance explained (PVE), and number of nonzero (NNZ) coefficients for each of the data size-and-sparsity combinations over $10$ replications. The dotted line in the RTE plots is the RTE of the null model; the dotted line in the PVE plots is the perfect score SNR/(1 + SNR); the horizontal dotted line in the number of nonzeros plots is the value of $s$. On top of each NNZ plot, we print the average number of correctly identified variables over 10 replications for each method at each of the 10 SNR levels, allowing us to infer the true positive rate.

In the RR plots across \Cref{fig:low,fig:med,fig:hi5,fig:hi10}, the performance of lassoed boosting and the relaxed lasso are much the same, except for a small difference in \Cref{fig:hi5}. In \Cref{fig:low,fig:med} with small and medium settings, lassoed boosting and the relaxed lasso outperform forward stepwise and the lasso when SNR is low; the performance of all four methods starts to converge when SNR increases. In \Cref{fig:hi5,fig:hi10} with large $p$, forward stepwise and the lasso seem to outperform the other two methods when SNR is small. However, some low SNR results must be interpreted with caution. For example, in the NNZ plot, the lasso, lassoed boosting, and the relaxed lasso can barely identify any correct variable, and the RR score is larger than $1$. Forward stepwise sticks to the null model at very low SNR values and obtains a null score of $RR=1$. 

In all RTE plots in \Cref{fig:low,fig:med,fig:hi5,fig:hi10}, lassoed boosting and the relaxed lasso perform almost identically, and significantly outperform the other two methods when SNR is relatively high. The only exception appears in \Cref{fig:med}, where the RTE of forward stepwise is the smallest of the four when SNR is relatively large, though the numerical difference is small compared to that of lassoed boosting and the relaxed lasso.

In the PVE plots in \Cref{fig:low,fig:med}, all four methods give good results that seems to improve as SNR increases. This result is no surprise since all methods better identify the correct variables at higher SNR. In \Cref{fig:hi5,fig:hi10}, lassoed boosting and relaxed lasso perform similarly and better than the other options at most of the SNR levels.

In the NNZ plots across all four figures, lassoed boosting and the relaxed lasso show similar capability in sparsity recovery.  In \Cref{fig:low}, when SNR is relatively high, all methods recover the true variables; the relaxed lasso achieves slightly sparser models than lassoed boosting. However, when SNR decreases, lassoed boosting sometimes achieves sparser models while identifying, on average, the same number of nonzero parameters. The NNZ plot in \Cref{fig:hi5} seems to suggest that the relaxed lasso performs better when SNR is low, which is not generally the case (see the third row in  \Cref{fig:betatype13nzs} for instances when lassoed boosting yields sparser models at low SNR levels). Again, in \Cref{fig:hi5}, almost no nonzero parameter is recovered when SNR is low, making any comparison dubious.

 The first three metrics seems to suggest that lassoed boosting and the relaxed lasso share similar coefficient paths, but the NNZ plots in \Cref{fig:low,fig:med,fig:hi5,fig:hi10} demonstrate that they differ. Several examples of beta-types 1 and 3 in \Cref{fig:betatype13nzs} from the supplement show that lassoed boosting produces sparser models in quite a few cases, and while the lasso can recover the true parameters, its false inclusion rate is high. Furthermore, from the NNZ plots, we conclude that all methods may fail to recover variables correctly when SNR is very low. 

Overall, we can roughly conclude that the performance of lassoed boosting is comparable to that of the relaxed lasso in the first three metrics, and it can produce a sparser model under certain scenarios, which happens more likely when the SNR is moderately large in validation tuning (see \Cref{fig:betatype13nzs} for several examples). We can also easily spot multiple occasions when the relaxed lasso gives sparser models. 

\begin{figure}[htp]
	\centering
	\includegraphics{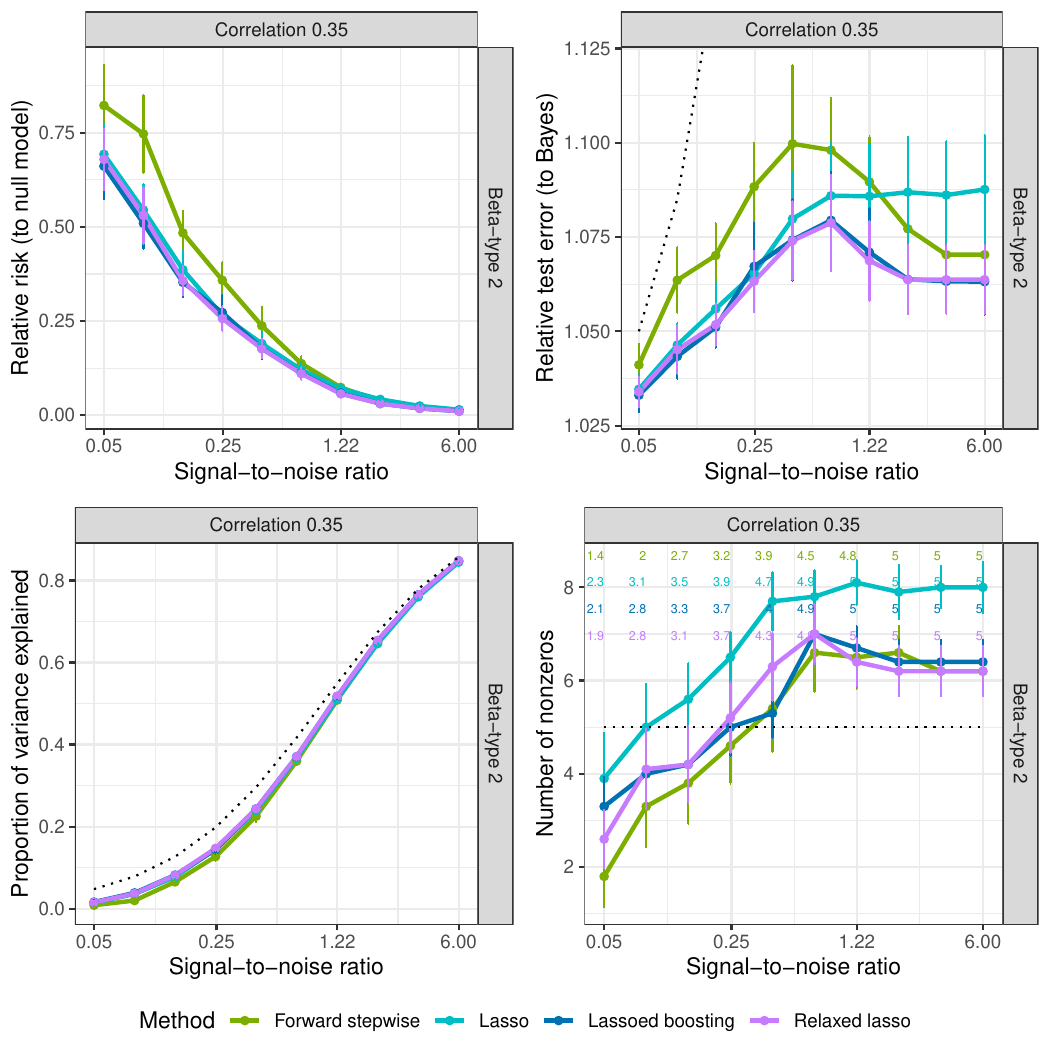}
	\caption{Curves of relative risk, relative test error, proportion of variance explained, and number of nonzeros as a function of SNR in the low setting with $n=100, p=10,$ and $s =5$. The numbers at the top of the nonzeros figure are the average variables correctly identified by each method at each of the $10$ SNR values. The order of the numbers, from row $1$ to row $4$, matches the order of the four legend labels from left to right.}%
	\label{fig:low}%
\end{figure}

\begin{figure}[htp]
	\centering
	\includegraphics{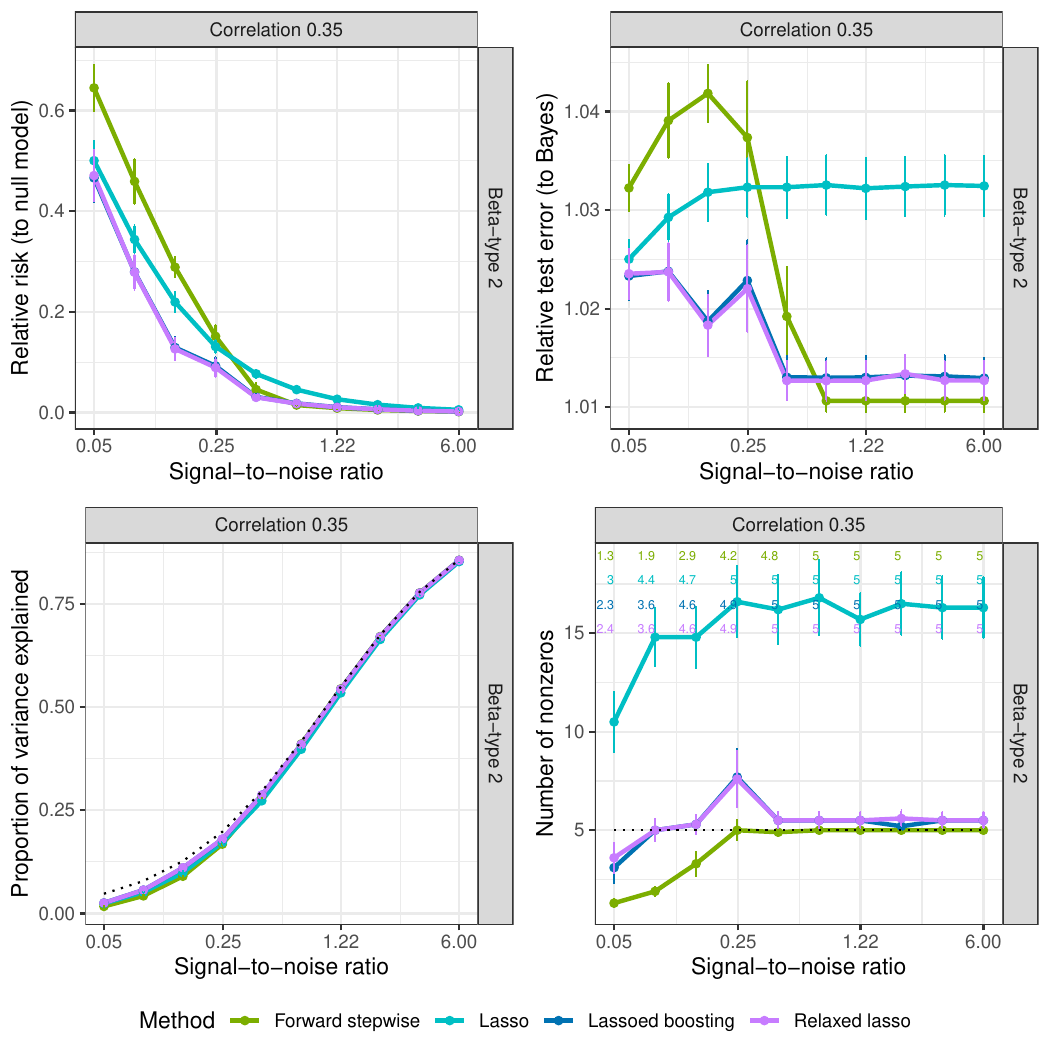}
	\caption{Curves of relative risk, relative test error, proportion of variance explained, and number of nonzeros as a function of SNR in the medium setting with $n=500, p=100,$ and $s =5$. The numbers at the top of the nonzeros figure are the average variables correctly identified by each method at each of the $10$ SNR values. The order of the numbers, from row $1$ to row $4$, matches the order of the four legend labels from left to right.}%
	\label{fig:med}%
\end{figure}

\begin{figure}[htp]
	\centering
	\includegraphics{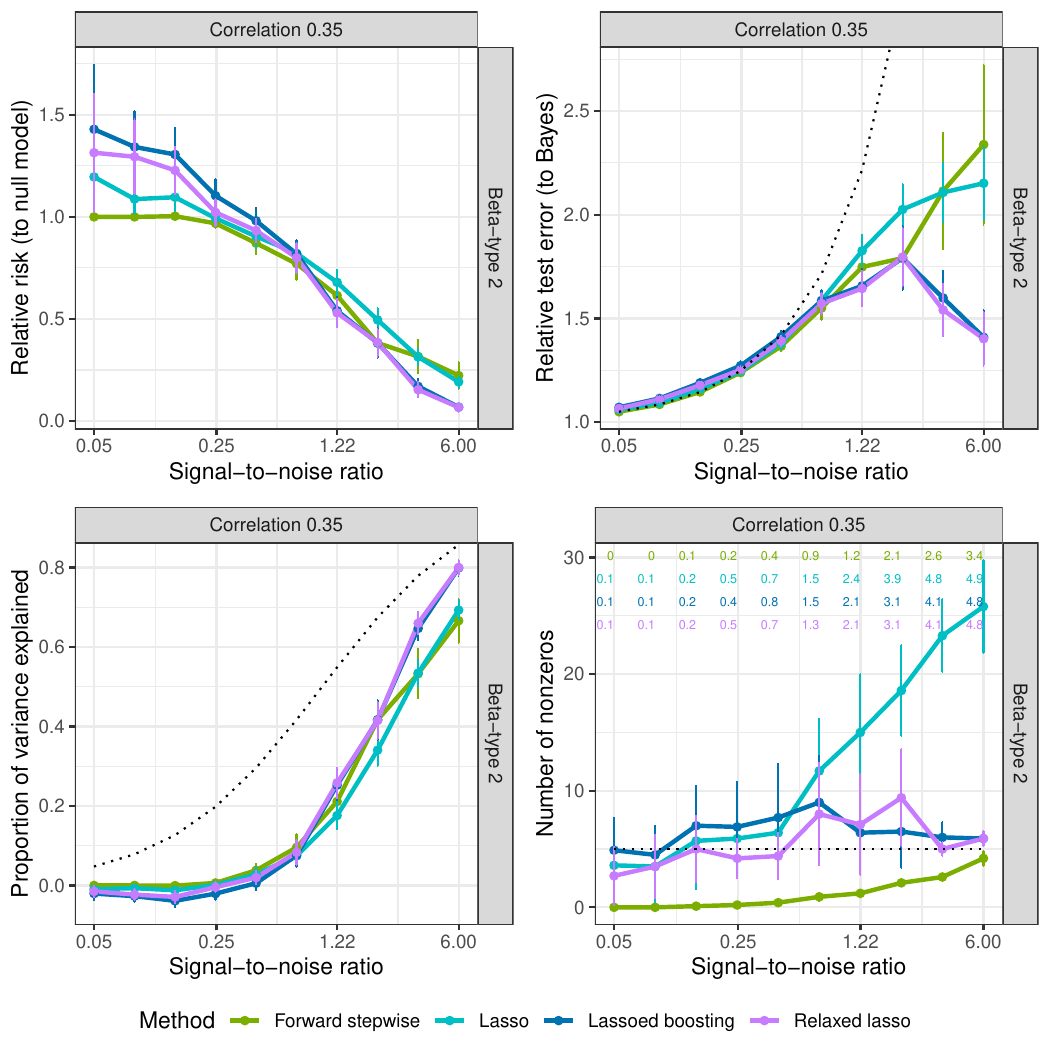}
	\caption{Curves of relative risk, relative test error, proportion of variance explained, and number of nonzeros as a function of SNR in the high-5 setting with $n=50, p=1000,$ and $s =5$. The numbers at the top of the nonzeros figure are the average variables correctly identified by each method at each of the $10$ SNR values. The order of the numbers, from row $1$ to row $4$, matches the order of the four legend labels from left to right.}%
	\label{fig:hi5}%
\end{figure}

\begin{figure}[htp]
	\centering
	\includegraphics{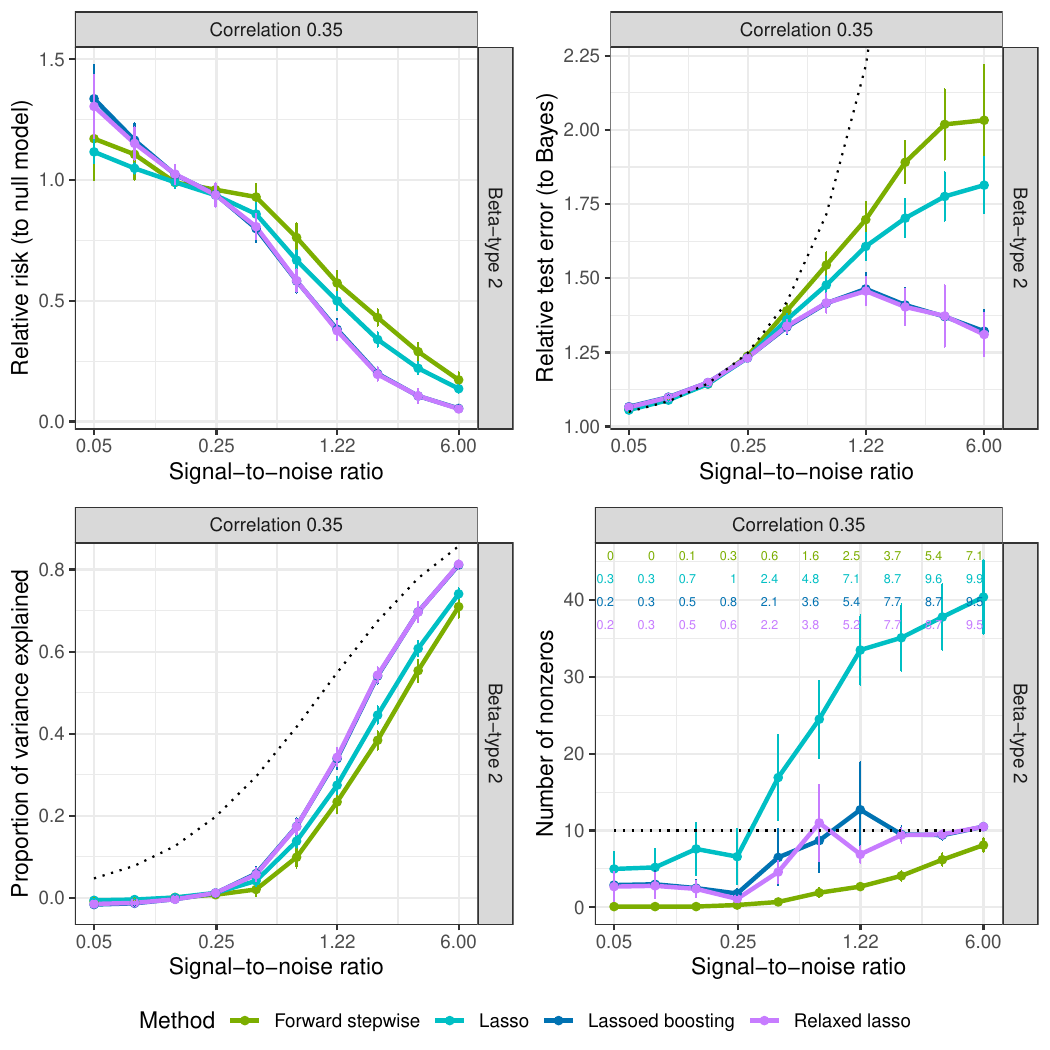}
	\caption{Curves of relative risk, relative test error, proportion of variance explained, and number of nonzeros as a function of SNR in the high-10 setting with $n=100, p=1000,$ and $s =10$. The numbers at the top of the nonzeros figure are the average variables correctly identified by each method at each of the $10$ SNR values. The order of the numbers, from row $1$ to row $4$, matches the order of the four legend labels from left to right.}%
	\label{fig:hi10}%
\end{figure}

\begin{figure}[htp]
	\centering
	\includegraphics{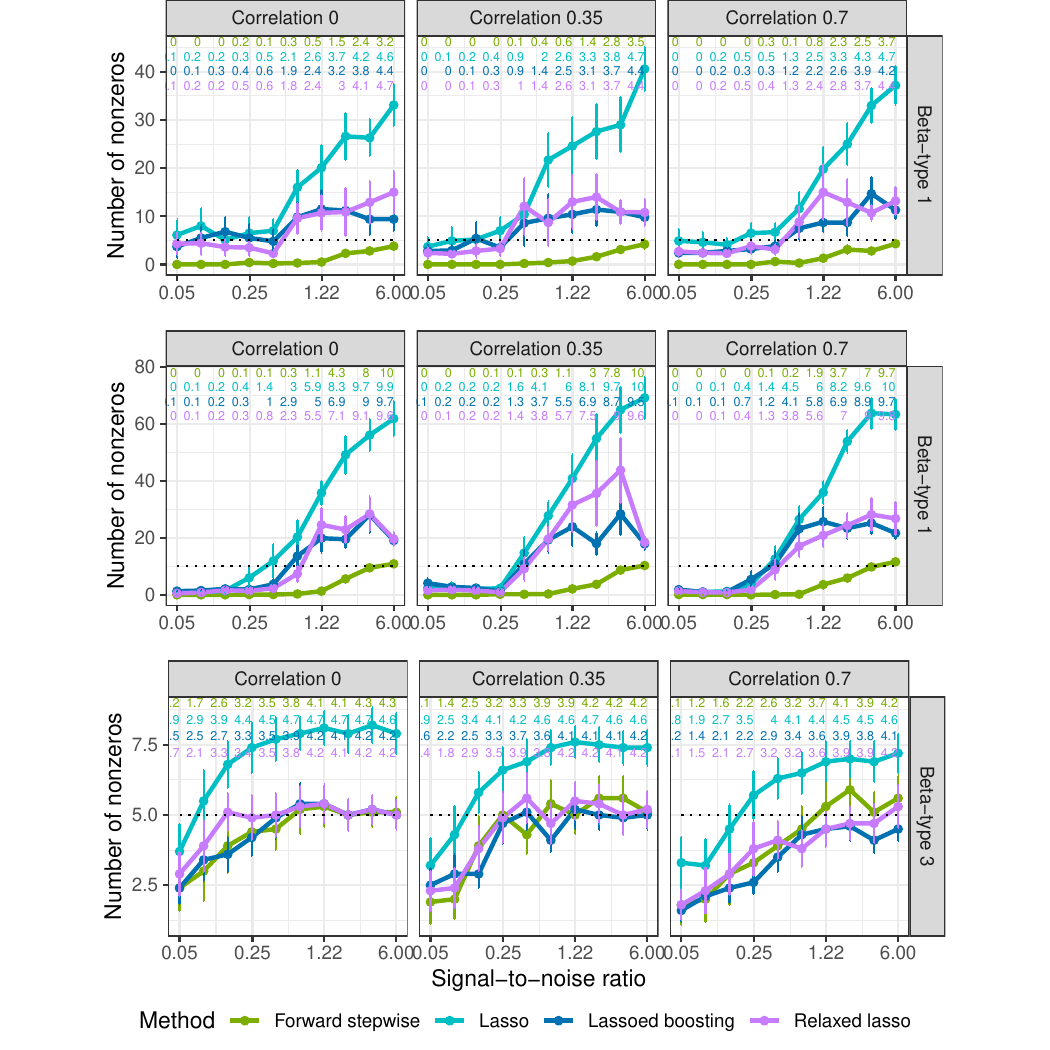}
	\caption{Top row: NNZ plots for beta-type 1 in the high-5 setting with $n=50, p = 1000, s=5$. Middle row: NNZ plots for beta-type 1 in the high-10 setting with $n=100, p = 1000, s=10$. Bottom row: NNZ plots for beta-type 3 in the low setting with $n=100, p=10, s= 5$.The numbers at the top of each figure are the average variables correctly identified by each method at each of the $10$ SNR values. The order of the numbers, from row $1$ to row $4$, matches the order of the four legend labels from left to right.}%
	\label{fig:betatype13nzs}%
\end{figure}

\section{An application in equities return prediction}

In this section, we compare the prediction accuracy of various methods to lassoed boosting's. Sections S.6 and S.7 also discuss the use of path integrated gradient to compare the path differences between the LS-boost and the lasso.

 \cite{Greenetal2017stockreturns} use $94$ variables from the databases CRSP, Compustat and I/B/E/S to study the determinants of average monthly US stock returns in a series of Fama-MacBeth regressions. We update their data to 2018 and use data from 2010 to 2018 for a total of $376544$ firm-month observations in $108$ months. The number of stocks (firms) in each month varies from $3269$ to $3883$. We add $8$ variables that were removed in \cite{Greenetal2017stockreturns} due to collinearity concerns. The dummy variable \textit{ipo} is removed since it varies little during certain time periods. In total, we have $101$ ($94+8-1$) variables (see \Cref{tab:variable list} in \Cref{supp:variable def}). For example, we have $3883$ observations for January 2010. $\boldsymbol{\mathrm{y} }$ becomes a $3883 \times 1$ vector of one-month-ahead, cross-sectional stock returns, and $\boldsymbol{\mathrm{X} }$ includes all variables plus a column of ones for the intercept. Missing values are replaced by the mean of the variable. We use $50\%$, $25\%$, and $25\%$ of the $3883$ observations for training, validation, and testing, respectively, and record the mean-squared prediction error (MSPE) for the test set of that month. After repeating this exercise for the remaining $107$ months, we report the mean and median of the $108$ MSPEs.

The goal of our exercise is to find a good linear model to predict stock returns, rather than to identify what determines the cross-section of expected stock returns, which is addressed in the large finance literature on anomalies. Active arbitrage in the market also implies that many analysts may have exploited all the variables under consideration, so we do not expect any linear model to exhibit broad predictive power. The data ignore many macro variables and such popular variables as the Fama-French factors and the \textit{q}-factors.

\Cref{tab:prediction} reports the prediction results from different methods. The tuning procedure in simulation is also used for the lasso, forward stepwise (FS), twiced lasso (TLasso), relaxed lasso (RLasso), and lassoed bossing ($\text{LB}_1$) in \Cref{tab:prediction}. $\text{LB}_2$ reports lassoed boosting with a learning rate equal to $0.001$ and a stopping criterion equal to the corrected AIC. GHZ represents a linear regression with $12$ variables identified in \cite{Greenetal2017stockreturns} as significant in explaining the cross-section of stock returns but only for the non-microcap return data before $2003$, and only two are significant for data after $2003$. We use the $12$-variable regression model for simple benchmarking purposes.

\begin{table}[htp] \centering
	\begin{center}
		\caption{Average MSPE and model size of the different methods} 
		\label{tab:prediction} 
		\begin{threeparttable}
			\begin{tabular}{llllllll}  
				\toprule
				& Lasso & FS & TLasso &RLasso& $\text{LB}_1$ & $\text{LB}_2$ &  GHZ \\
				\midrule
				Mean & $0.020057$ & $0.023027$&$0.020057$&$0.020046$& $0.020027$ &$\mathbf{0.020015}$&$0.020455$\\
				Median & $0.018206$&$0.021190$&$0.018212$&$0.018188$&$0.018221$ &$\mathbf{0.018144}$&$0.018484$\\
				\# of tun. par.  & $100$& $50$&$100 \times 50$&$100 \times 50$&$100 \times 50$ &$100 \times 50$& \text{N/A}\\
				avg. model size&$22.14$ & $6.15$&$12.44$&$12.37$&$12.47$ &$9.22$& $12$\\	
				\bottomrule
			\end{tabular}
			\begin{tablenotes}[flushleft]
				\setlength\labelsep{0pt}
				\item[] \textit{Notes}: FS, TLasso, RLasso, and LB refer to forward stepwise, twiced lasso, relaxed lasso, and lassoed boosting, respectively. GHZ refers to a linear regression model based on the $12$ variables identified in \cite{Greenetal2017stockreturns} as significant in explaining the cross-section of expected returns for non-microcap stocks.  Rows 1 and 2 report the mean and median of MSPE for the $108$ monthly test sets from January, 2010 to December, 2018. Row 3 reports the number of tuning parameters used, where the forward stepwise (FS) method is tuned over $50$ steps. Row 4 reports the average number of selected variables over $108$ test sets for each method. For the GHZ method, the number of variables (model size) is fixed at $12$.
			\end{tablenotes}
		\end{threeparttable}
	\end{center} 
\end{table}
%\vspace*{-0.8cm}

The differences in prediction among the methods are small. The average MSPE of $\text{LB}_1$ is smaller than that of the lasso and the relaxed lasso but its median MSPE is larger. The relaxed lasso outperforms the lasso with a smaller mean MSPE and a smaller median MSPE. The results of the twiced lasso closely follow those of the lasso and the relaxed lasso. When tuned with a learning rate of $0.001$, lassoed boosting starts to perform even better. $\text{LB}_2$ has the smallest mean MSPE and median MSPE among all the methods. Note that the lasso and lasso-based methods all outperform the FS method and the $12$-variable GHZ model.

The numerical difference between the relaxed lasso and lassoed boosting ($\text{LB}_2$) in \Cref{tab:prediction} is also very small. The difference in mean MSPE is $0.000031$. Does such a small difference matter? We note that MSPE is defined as
\begin{equation} \label{eq:MSPE}
	\text{MSPE} = \frac{1}{n_T} \sum_{i=1}^{n_T} (y_i - x_i \hat{\beta})^2,
\end{equation}
where $n_T$ is the size of the test set. MSPE measures the average squared distance between prediction and the true value. By taking the square root of the MSPE, we obtain the average absolute distance, which is $0.0056$ ($=0.56\%$), which translates to $56$ cents of difference when trading $100$ dollars. A practitioner can better assess its economic significance. Based on the $108$ MSPEs, a paired one-sided \textit{t}-test for the null hypothesis $\text{MSPE}_{\text{RLasso}} < \text{MSPE}_{\text{LB}_2} $ gives a \textit{p}-value of $0.007$. Hence, we reject the null and conclude that the MSPE of $\text{LB}_2$ is smaller than that of the relaxed lasso.

\begin{figure}[htp]
	\centering
	\subfloat{\label{fig:nnzboostingVSlasso}\includegraphics[width=0.49\linewidth,keepaspectratio,scale=1]{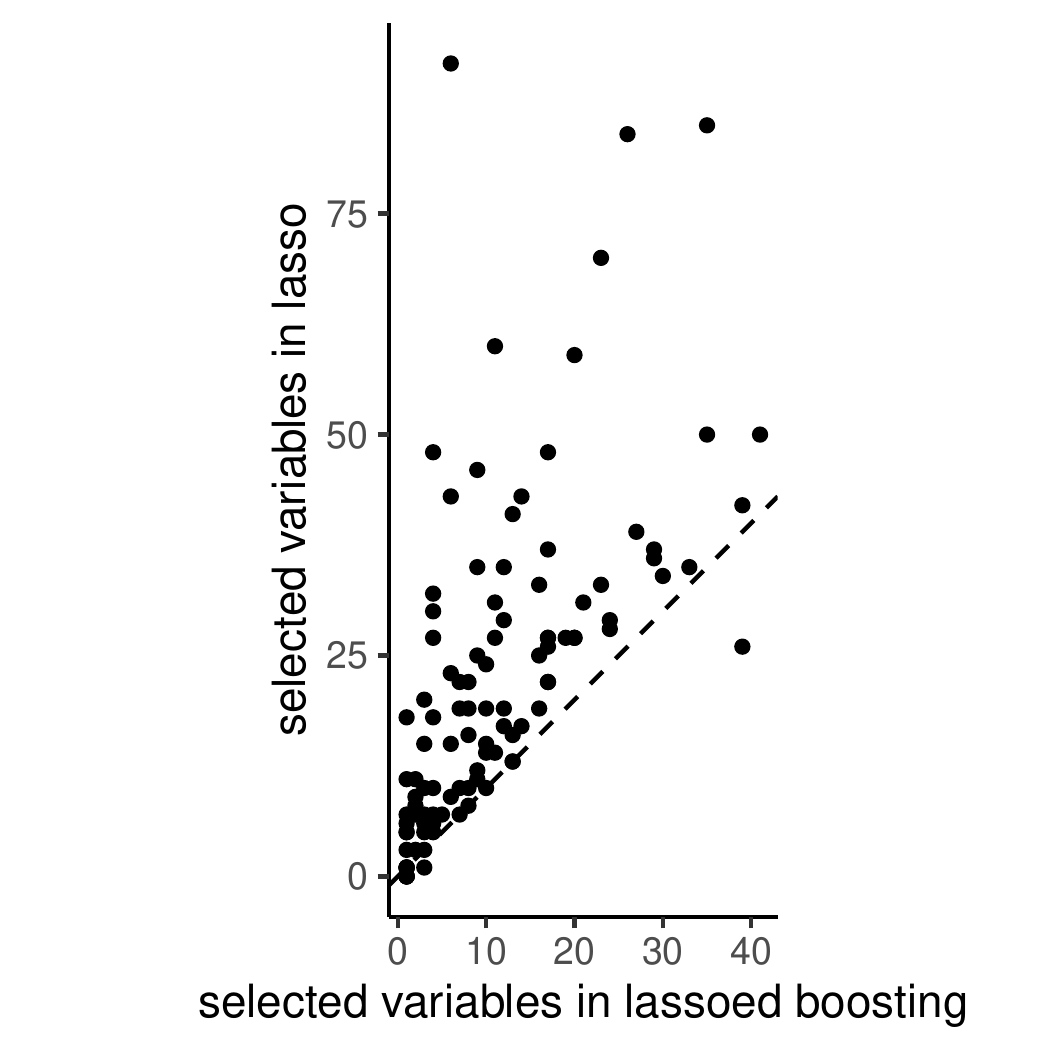} }%
	\subfloat{\label{fig:nnzboostingVSrelaxed}\includegraphics[width=0.49\linewidth,keepaspectratio,scale=1]{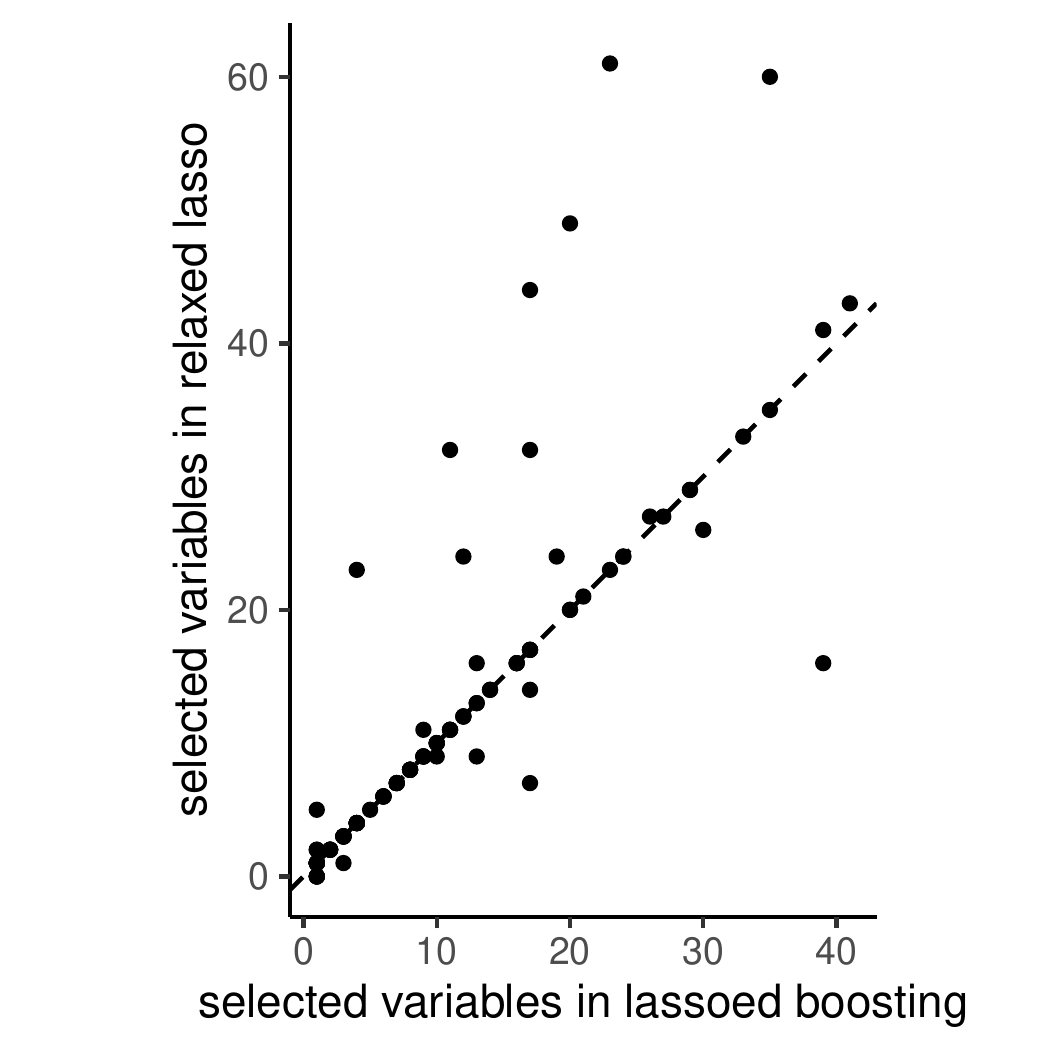} }
	\caption{Each dot in the left figure represents the number of selected nonzeros in lassoed boosting and lasso for each of the $108$ test sets. The dotted line is $45\text{\textdegree}$ line. Out of $108$ test sets, compared to the lasso, lassoed boosting has a smaller, bigger, and the equal model size in 91, 6, and 11 cases, respectively. The right figure compares the same differences between lassoed boosting and the relaxed lasso, where the three numbers are 25, 14, and 69. }%
	\label{fig:application_nnz}%
\end{figure}

The last row of \Cref{tab:prediction} reports the average number of selected variables across the $108$ test sets. On average, the number for $\text{LB}_2$ is less than half that for the lasso. Compared to the relaxed lasso, lassoed boosting also offers a sparser solution on average. \Cref{fig:application_nnz} plots the number of nonzeros identified in lassoed boosting ($\text{LB}_2$) vs. the lasso (left) and the relaxed lasso (right). The dotted line is $45\text{\textdegree}$ line. Out of $108$ test sets, lassoed boosting yields a sparser model in $91$ cases; the lasso in only $6$. Compared to the relaxed lasso, the number of nonzeros identified by lassoed boosting is smaller in $25$ cases, larger in $14$ cases, and the same in $69$ cases. Overall, our empirical results are consistent with simulation: lassoed boosting can yield sparser model in certain cases. The large variation in model size also indicates model instability, though a stable model like GHZ underperforms the unstable models.

Combining the information in \Cref{tab:prediction} and \Cref{fig:application_nnz}, we conclude that, based on this application, lassoed boosting can deliver better prediction with a more compact model.

\section{Conclusions}

This paper finds that lassoed boosting, a refitting strategy based on the lasso, has good finite-sample properties in both our simulation experiment and an application. We also introduce the idea of using path integrated gradient to study the difference between the lasso and LS-boost in parameter attribution. 

Our application uses one data set and the simulation compares only five variable-selection methods. It takes one line of R code to connect our method to the \texttt{bestsubset} package in \cite{hastie2017extended} and to reproduce the simulation results (see the github page for instructions.) We invite readers to try lassoed boosting in more simulations and applications to further study its properties.

Our work can be extended in several directions. First, lassoed boosting should be compared to other methods in a classification exercise such as credit rating and default analysis. Second, a valid \textit{p}-value should be attached to the estimates. Third, \Cref{prop:two-stage refit} suggests that both the relaxed lasso and lassoed boosting may be examples of many other two-stage approaches. An analyst should explore other possibilities to find a better method.

\section*{Acknowledgments}

I am grateful to two referees for their insightful comments and suggestions. The computation and research support from the Office of Research and Coles College of Business at Kennesaw State University are greatly acknowledged.

\newpage
\spacing{1.45}
\bibliographystyle{ecca}
\bibliography{reference}

\newpage
\setcounter{page}{1}
\spacing{1.42}

\begin{appendices}
{
	%\renewcommand{\appendixtocname}{Supplement}
	%\crefalias{section}{supp}
	
	\centering \title{\large\MakeUppercase{Supplementary Material to ``Lassoed Boosting and Linear Prediction in the Equities Market"}\footnote{Email: xhuang3@kennesaw.edu}\\[10pt]} }
	\begin{center}
		\large
		\author{\textsc{Xiao Huang}}\\
		\date{\today}
	\end{center}
	\maketitle
    
\bigskip    
    
This supplement contains all proofs, additional discussions and figures.

\setstretch{2}
\bigskip

\setstretch{1.3}
\localtableofcontents
\setstretch{2}

\newpage %\spacing{1.35}
\setstretch{1.35}
\setcounter{section}{19}
\setcounter{equation}{0}
\renewcommand{\theequation}{S.\arabic{equation}}
\subsection{Proofs} \label{supp:asymptotic}
\begin{proof}[Proof of \Cref{prop:asymptotic rate}]
	 From Theorem 11.3 in \cite{hastie2015slsparcity}, we know there exists a $\lambda$ such that the lasso is variable selection consistent as $\log(p)/n \rightarrow 0$ when $n \rightarrow \infty$. Let $\lambda_*$ be a penalty value so that $\mathcal{A}_{\lambda_*} = \mathcal{A}$. Given the correctly identified active set $\mathcal{A}_{\lambda_*}$, the LS-boost solution at step $k=\infty$ is equal to the LS solution on the active set $\mathcal{A}_{\lambda_*}$, $\hat{\beta}^{\lambda_*,\infty} = \hat{\beta}_{LS}^{\lambda_*}$. Since $\inf_{\lambda, s \in [1,\infty]} L_n(\hat{\beta}^{\lambda,k}) \leq L_n(\hat{\beta}_{LS}^{\lambda_*})$, we have, for $c > 0$,
	\begin{align*}
		P(\inf_{\lambda, k \in [1,\infty]} L_n(\hat{\beta}^{\lambda,k}) > c n^{-1}) &\leq P(L_n(\hat{\beta}^{\lambda_*}_{LS}) > c n^{-1}) \rightarrow 0 \text{ as } n \rightarrow \infty,
	\end{align*}
	where the last result follows the standard convergence property of the LS estimator.
\end{proof}

\begin{proof}[Proof of \Cref{prop:two-stage refit}]
	The proof is similar to the proof of \Cref{prop:asymptotic rate}. Using Theorem 11.3 in \cite{hastie2015slsparcity}, as $n \rightarrow \infty$, gives a $\lambda_*$ so that $\mathcal{A}_{\lambda_*} = \mathcal{A}$ with high probability. Since $\inf_{\lambda, \mathcal{K}} L_n(\hat{\beta}^{\lambda,\mathcal{K}}) \leq L_n(\hat{\beta}_{LS}^{\lambda_*})$ as the tuning vector $\mathcal{K}$ can generate the case of a full LS solution, we have, for $c > 0$,
	\begin{align*}
	P(\inf_{\lambda,\mathcal{K}} L_n(\hat{\beta}^{\lambda,\mathcal{K}}) > c n^{-1}) &\leq P(L_n(\hat{\beta}^{\lambda_*}_{LS}) > c n^{-1}) \rightarrow 0 \text{ as } n \rightarrow \infty,
	\end{align*}
	where the last result again follows the standard convergence property of the LS estimator.
\end{proof}

\begin{proof}[Proof of \Cref{thm:prediction convergence}]
	By the triangular inequality, we have
	\begin{equation} \label{eq:triangular prediction}
	\lVert \boldsymbol{\mathrm{X}} \hat{\beta}^k - \boldsymbol{\mathrm{X}} \beta^* \rVert_2 \leq \lVert \boldsymbol{\mathrm{X}} \hat{\beta}^k - \boldsymbol{\mathrm{X}} \hat{\beta}_\text{LS} \rVert_2 + \lVert \boldsymbol{\mathrm{X}} \hat{\beta}_\text{LS} - \boldsymbol{\mathrm{X}} \beta^* \rVert_2,
	\end{equation}
	where the bound for the term $\lVert \boldsymbol{\mathrm{X}} \hat{\beta}^k - \boldsymbol{\mathrm{X}} \hat{\beta}_\text{LS} \rVert_2$ is given in Theorem 2.1 in \cite{freundetal2017boosting}. Consider the second term $\lVert \boldsymbol{\mathrm{X}} \hat{\beta}_\text{LS} - \boldsymbol{\mathrm{X}} \beta^* \rVert_2$. Recall $L_n(\beta^*) \geq L_n(\hat{\beta}_\text{LS})$.
	\begin{align*}
		2n(L_n(\beta^*) - L_n(\hat{\beta}_\text{LS})) &= \lVert \boldsymbol{\mathrm{y}} - \boldsymbol{\mathrm{X}} \beta^* \rVert^2_{2} - \lVert \boldsymbol{\mathrm{y}} - \boldsymbol{\mathrm{X}} \hat{\beta}_\text{LS} \rVert^2_{2}\\
		&= -2 \boldsymbol{\mathrm{y}}^T\boldsymbol{\mathrm{X}} \beta^* + \lVert \boldsymbol{\mathrm{X}} \beta^* \rVert^2_2 + 2 \boldsymbol{\mathrm{y}}^T\boldsymbol{\mathrm{X}} \hat{\beta}_\text{LS} - \lVert \boldsymbol{\mathrm{X}} \hat{\beta}_\text{LS} \rVert^2_2\\
		&= -2 \hat{\beta}_\text{LS} \boldsymbol{\mathrm{X}}^T\boldsymbol{\mathrm{X}} \beta^* + \lVert \boldsymbol{\mathrm{X}} \beta^* \rVert^2_2 + 2 \hat{\beta}_\text{LS} \boldsymbol{\mathrm{X}}^T\boldsymbol{\mathrm{X}} \hat{\beta}_\text{LS} - \lVert \boldsymbol{\mathrm{X}} \hat{\beta}_\text{LS} \rVert^2_2\\
		&= \lVert \boldsymbol{\mathrm{X}} \hat{\beta}_\text{LS} - \boldsymbol{\mathrm{X}} \beta^* \rVert^2_2,
	\end{align*}
	where we use the f.o.c. of \cref{eq:ls loss}, $\boldsymbol{\mathrm{y}}^T\boldsymbol{\mathrm{X}} = \hat{\beta}_\text{LS} \boldsymbol{\mathrm{X}}^T\boldsymbol{\mathrm{X}}$, in the third equality. Hence, we have
	\begin{equation} \label{eq:prediction and loss}
		\lVert \boldsymbol{\mathrm{X}} \hat{\beta}_\text{LS} - \boldsymbol{\mathrm{X}} \beta^* \rVert_2 = \sqrt{2n(L_n(\beta^*) - L_n(\hat{\beta}_\text{LS}))}.
	\end{equation}
	By the convexity of $L_n(\cdot)$, we have
	
	\begin{align*} \label{eq:loss convexity}
		L_n(\hat{\beta}_\text{LS}) &\geq L_n(\beta^*) + \nabla L_n(\beta^*)^T (\hat{\beta}_\text{LS} - \beta^*)\\
		&\geq L_n(\beta^*) - \lVert \nabla L_n(\beta^*) \rVert_2 \cdot \lVert \hat{\beta}_\text{LS} - \beta^* \rVert_2.
	\end{align*}
	Rearranging the last inequality gives
	\begin{equation} \label{eq:loss inequality}
		L_n(\beta^*) - L_n(\hat{\beta}_\text{LS}) \leq \lVert \nabla L_n(\beta^*) \rVert_2 \cdot \lVert \hat{\beta}_\text{LS} - \beta^* \rVert_2.
	\end{equation}
	Substituting Theorem 2.1 (iii) in FGM and \cref{eq:prediction and loss,eq:loss inequality} into \cref{eq:triangular prediction} completes the proof.
\end{proof}

Next, similar to \Cref{thm:prediction convergence}, we give a prediction convergence result for \Cref{alg:lassoed forward}.
\begin{theorem} \label{thm:prediction convergence for forward}
	Let $k \geq 0$ be the number of iterations. Under \cref{assumption:general}, there exists an $i \in \left\{0,\cdots,k\right\}$ so that the following bound hold:
	\begin{equation} \label{eq:prediction bound forward}
		\lVert \boldsymbol{\mathrm{X}} \hat{\beta}^i - \boldsymbol{\mathrm{X}}\beta^* \rVert_2 \leq \frac{\sqrt{p}}{\sqrt{\lambda}(\boldsymbol{\mathrm{X}}^T\boldsymbol{\mathrm{X}})} \left[\frac{\lVert \boldsymbol{\mathrm{X}}\hat{\beta}_\text{LS} \rVert_2^2}{\varepsilon(k+1)} + \epsilon\right] + \sqrt{2n \lVert \nabla L_n(\beta^*) \rVert_2 \cdot \lVert \hat{\beta}_\text{LS} - \beta^* \rVert_2}.
	\end{equation}
\end{theorem}
\begin{proof} [Proof of \Cref{thm:prediction convergence for forward}]
	Applying the triangular inequality, we have
	\begin{equation} \label{eq:triangular prediction forward}
	\lVert \boldsymbol{\mathrm{X}} \hat{\beta}^i - \boldsymbol{\mathrm{X}} \beta^* \rVert_2 \leq \lVert \boldsymbol{\mathrm{X}} \hat{\beta}^i - \boldsymbol{\mathrm{X}} \hat{\beta}_\text{LS} \rVert_2 + \lVert \boldsymbol{\mathrm{X}} \hat{\beta}_\text{LS} - \boldsymbol{\mathrm{X}} \beta^* \rVert_2.
	\end{equation}
	Bound for the first term on the r.h.s. follows Theorem 3.1 (iii) in FGM, while bound for the second term on the r.h.s. follows the same result in the proof of \Cref{thm:prediction convergence}. Combining the two results gives the bound in \Cref{thm:prediction convergence for forward}.
\end{proof}

\subsection{Comparison of convergence results in \Cref{thm:prediction convergence} and Theorem 12.2 in \cite{buhlmann2011highdimenstats}} \label{supp:compare convergence}

The convergence rate in Theorem 12.2 in \cite{buhlmann2011highdimenstats} is obtained after choosing an iteration number $k$ (``$m$" in the book's notation) to minimize the upper bound. We can consider the pre-optimized rate result in the first equation on page 426 of \cite{buhlmann2011highdimenstats} to gain some insight. For ease of reference, we reproduce the equation below.
\begin{equation} \label{eq:rate in buhlmann}
	\frac{1}{n} \lVert \boldsymbol{\mathrm{X}} \hat{\beta}^k - \boldsymbol{\mathrm{X}} \beta^* \rVert_{2}^2 = \lVert \hat{R}^k f^0 \rVert_{n}^2 \leq \max\{2 k^{-\frac{D_\kappa}{2+D_\kappa}},\kappa^{-1}(1-\kappa/2)^{-1} 2 \Delta_n(\lVert \beta^0_n \rVert_1 + k \gamma_n) + k \Delta_n  \},
\end{equation}
where we replace ``$m$" in the original equation with $k$ to denote the boosting iteration number, and
\begin{align*}
f^0 &= \boldsymbol{\mathrm{X}} \beta^0_n \text{ (and } \beta^0_n = \beta^* \text{ in this paper),}\\
\hat{R}^k f^0 &= \boldsymbol{\mathrm{X}} \beta_n^0 - \boldsymbol{\mathrm{X}} \hat{\beta}^k, \quad \kappa \in (0,\frac{1}{2}), \quad D_\kappa = (1-\kappa)(1-\kappa/2), \\
	\Delta_n &= \max_{j=1,\cdots,p} \frac{1}{n}\sum_{i=1}^{n}|u_i x_{ij}|= O_p(\sqrt{\log(p)/n}),\\
	\gamma_n &= (1+\sigma^2) + o_p(1) = O_p(1).
\end{align*}
Under the assumption of $\lVert \beta^0_n \rVert_1 = o\left(\sqrt{\frac{n}{\log(p)}}\right)$ and $k = o(\sqrt{n/\log(p)})$ in the same theorem, the second term in \cref{eq:rate in buhlmann} is dominated by $\kappa^{-1}(1-\kappa/2)^{-1} 2 \Delta_n \lVert \beta^0_n \rVert_1$, which has order $ \Delta_n \lVert \beta^0_n \rVert_1$. Although it is $o_p(1)$, it is not a function of $k$ and we cannot compare it directly to the result in \Cref{thm:prediction convergence}. For this reason, let us consider the first term in \cref{eq:rate in buhlmann}. Given the value of $\kappa$, $\frac{D_\kappa}{2+D_\kappa}$ varies on the interval $(0.158, 0.333)$. Hence $2 k^{-\frac{D_\kappa}{2+D_\kappa}}$ is a power function of $k$ with its power on the interval of $(-0.333, -0.158)$. 

From \Cref{thm:prediction convergence,eq:triangular prediction}, we have
\begin{align}
		\frac{1}{n} \lVert \boldsymbol{\mathrm{X}} \hat{\beta}^k - \boldsymbol{\mathrm{X}}\beta^*  \rVert_2^2 & \leq \frac{1}{n} \lVert \boldsymbol{\mathrm{X}} \hat{\beta}^k_{\text{LS}} \rVert_2^2 \gamma^k + \frac{1}{n}\lVert \boldsymbol{\mathrm{X}} \hat{\beta}_\text{LS} - \boldsymbol{\mathrm{X}} \beta^* \rVert_2\nonumber\\
		& \leq \frac{1}{n} \lVert \boldsymbol{\mathrm{y}} \rVert_2^2 \gamma^k + 
		\frac{1}{n}\lVert \boldsymbol{\mathrm{X}} \hat{\beta}_\text{LS} - \boldsymbol{\mathrm{X}} \beta^* \rVert_2, \label{eq:thm 7 result}
\end{align} 
where we can assume $\frac{1}{n} \lVert \boldsymbol{\mathrm{y}} \rVert_2^2 = O_p(1)$. Hence the first term on the r.h.s. of \cref{eq:thm 7 result} is an exponential function to the base of $\gamma$ with $\gamma \in [0.75, 1)$. Thus, we conclude that while Theorem 2.12. in \cite{buhlmann2011highdimenstats} uses a power function to describe the convergence of the prediction in LS-boost as the procedure iterates, \Cref{thm:prediction convergence} (and Theorem 2.1 in FGM) uses an exponential function to characterize the convergence.

With additional assumptions, we can further comment on the behavior of the second term on the r.h.s. of \cref{eq:thm 7 result}. Assume the error term $u_i$ belongs to the class of sub-Gaussian distribution with variance proxy $\sigma^2$, $u_i \sim \text{subG}(\sigma^2)$.
If $p \leq n$, Theorem 2.2 in \cite{rigollet2019hds} shows the second term is $O_p(1/n)$; if $p > n$ and $\beta^*$ is $s$-sparse, Corollary 2.8 in \cite{rigollet2019hds} indicates that the second term is $O_p(\log(p)/n)$. These results hold with high probability. To summarize, under the additional assumption of $O_p(\log(p)/n) \rightarrow 0$ as $n,p \rightarrow \infty$, the second term in \cref{eq:thm 7 result} disappears asymptotically, which allows us to focus on the exponential function, $\gamma^k$, to study the convergence of predictions in LS-boost.

\subsection{Validation tuning figures in simulation}
\subsubsection{Low setting: $n=100, p=10, s=5$}
\subsubsubsection{Relative risk (to null model)}
    \includegraphics[scale=0.79]{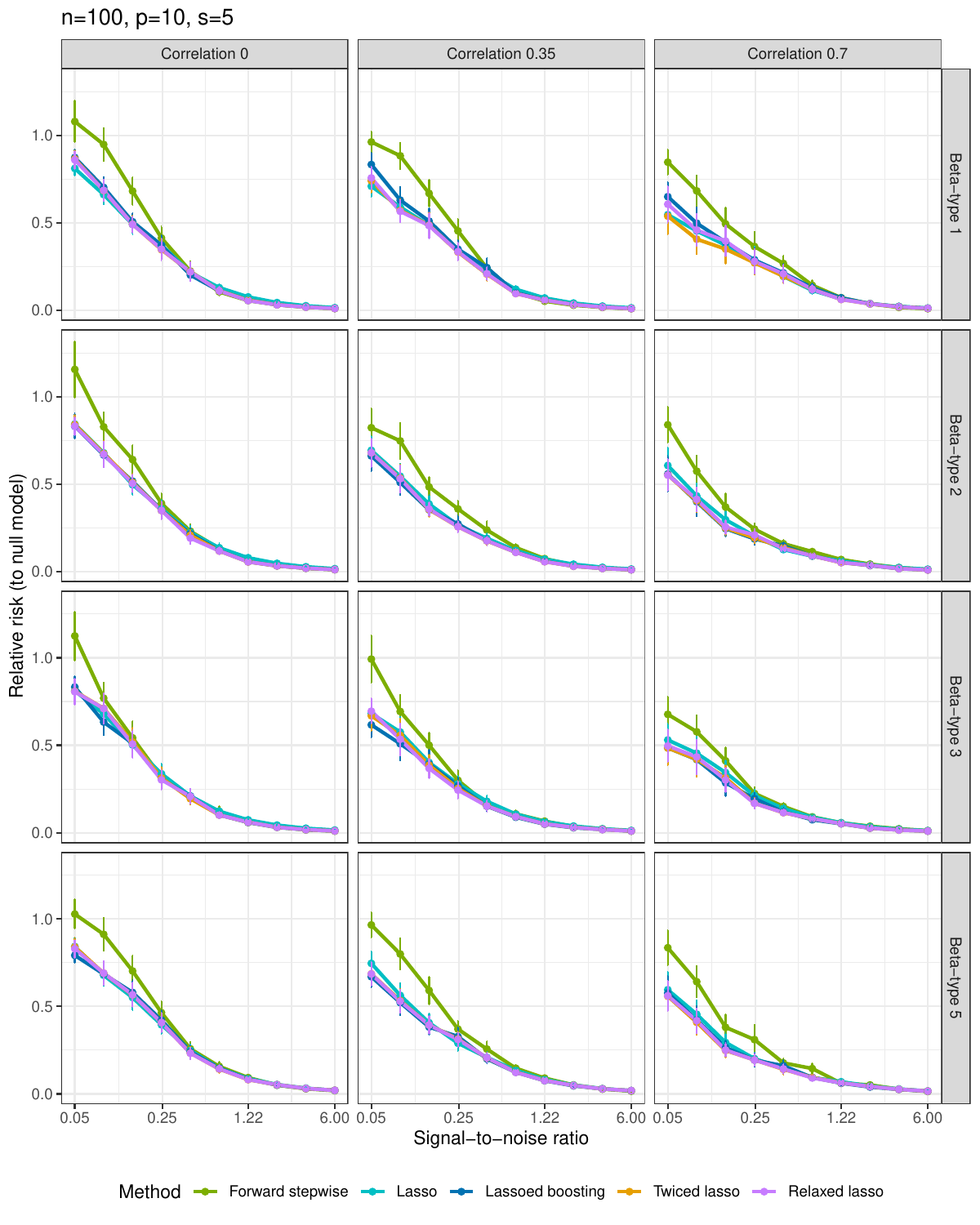}
\subsubsubsection{Relative test error (to Bayes)}
    \includegraphics[scale=0.82]{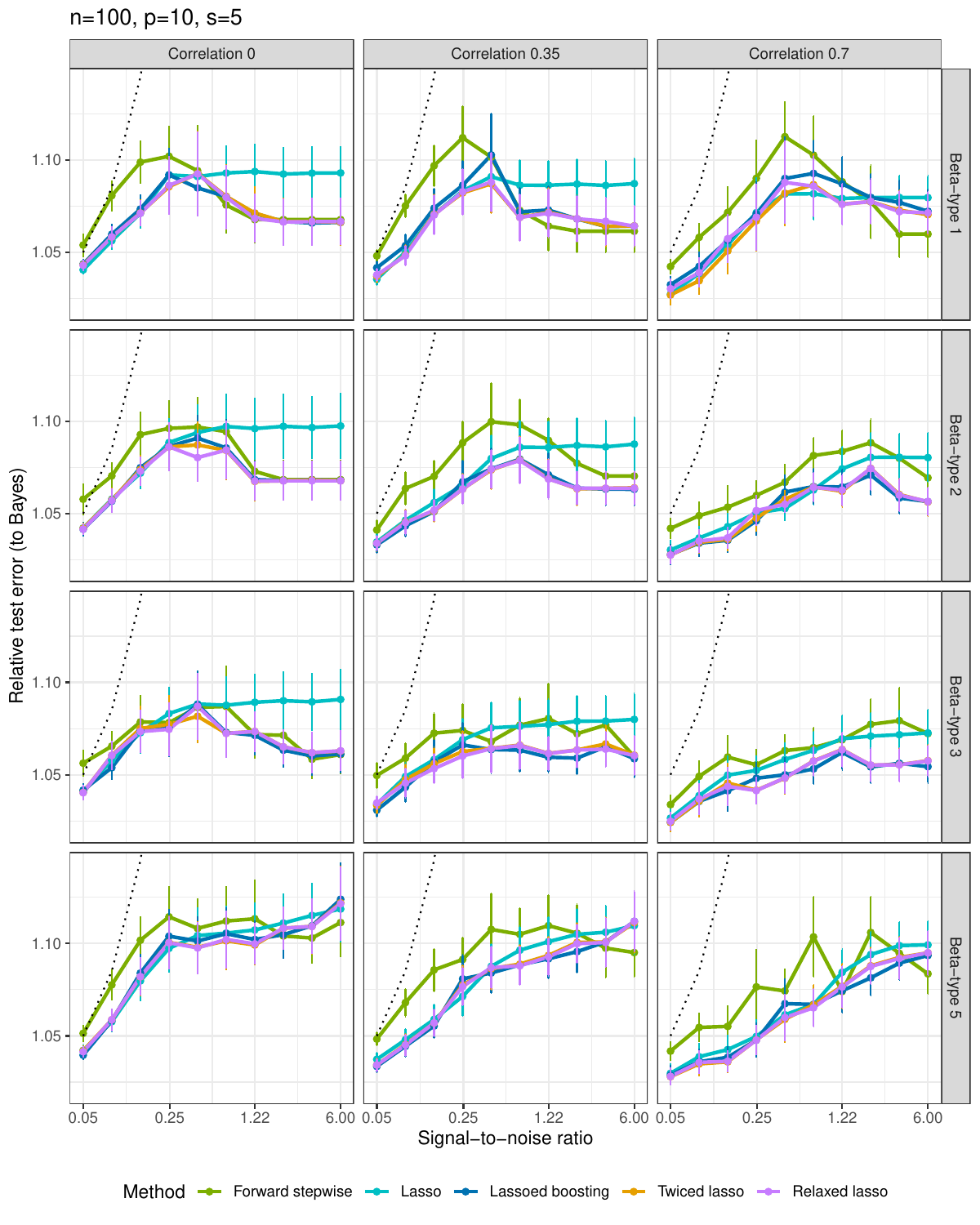}
\subsubsubsection{Proportion of variance explained}
    \includegraphics[scale=0.82]{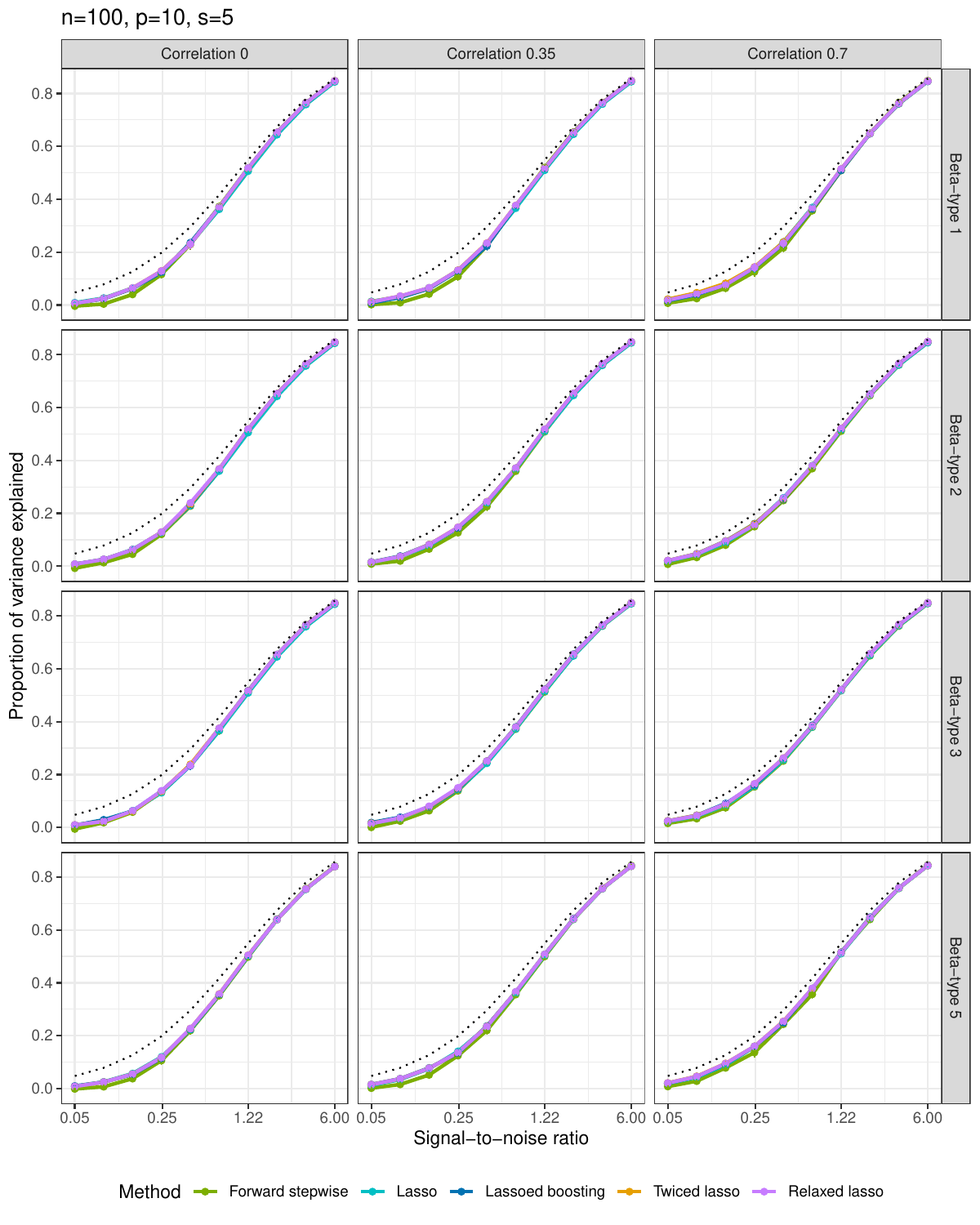}
\subsubsubsection{Number of nonzero coefficients}
    \includegraphics[scale=0.82]{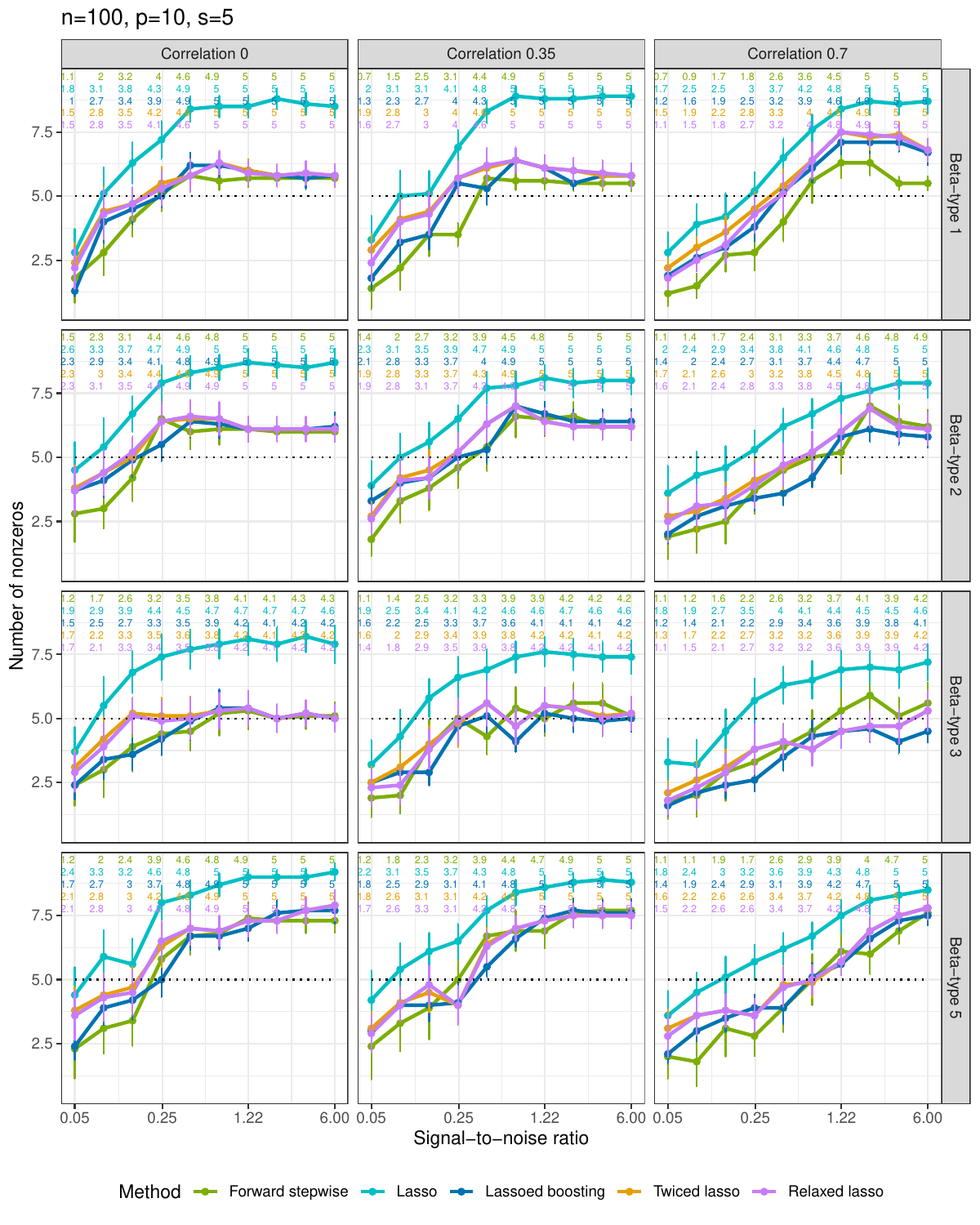}
    
\subsubsection{Medium setting: $n=500, p=100, s=5$}
\subsubsubsection{Relative risk (to null model)}
\includegraphics[scale=0.82]{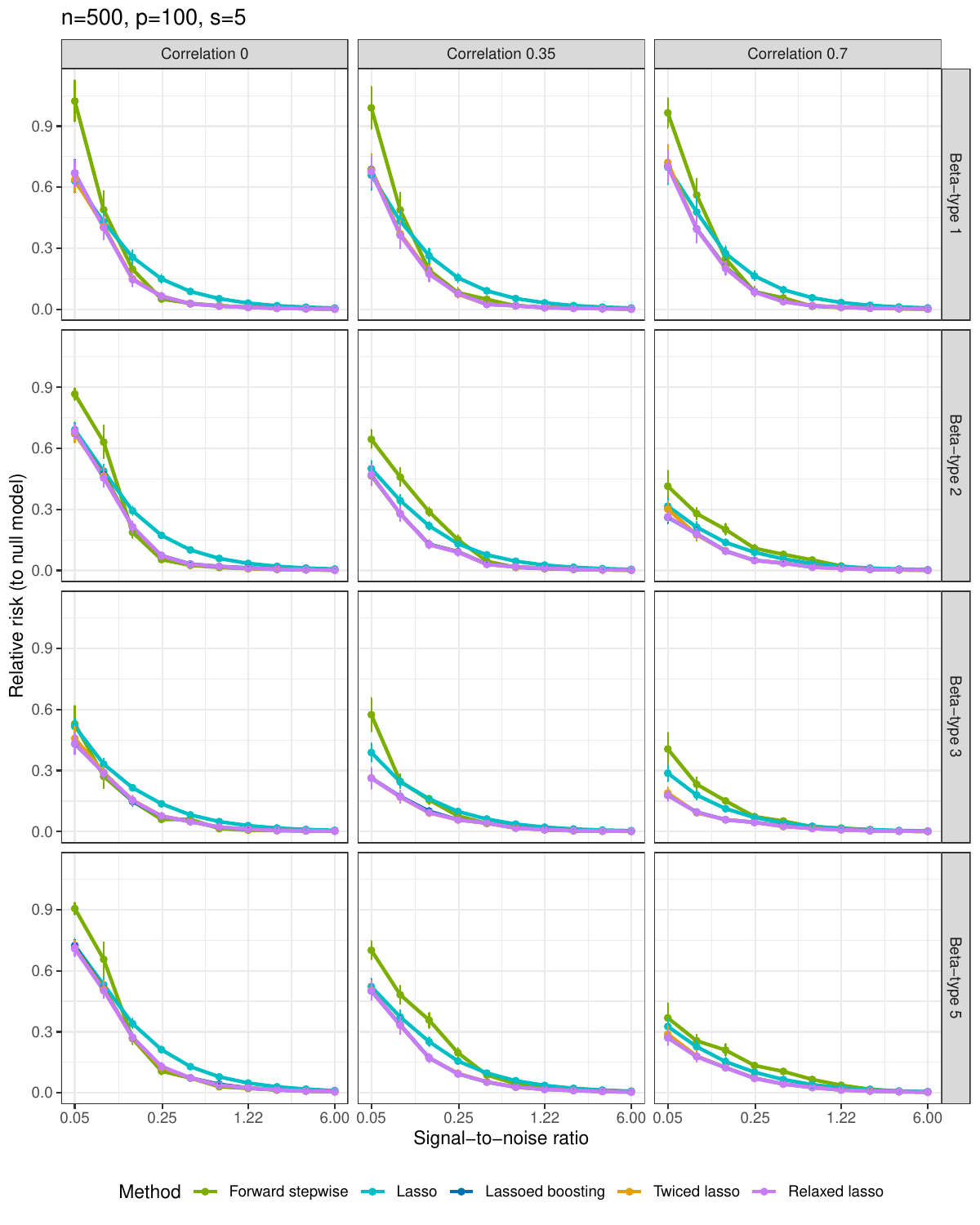}
\subsubsubsection{Relative test error (to Bayes)}
\includegraphics[scale=0.82]{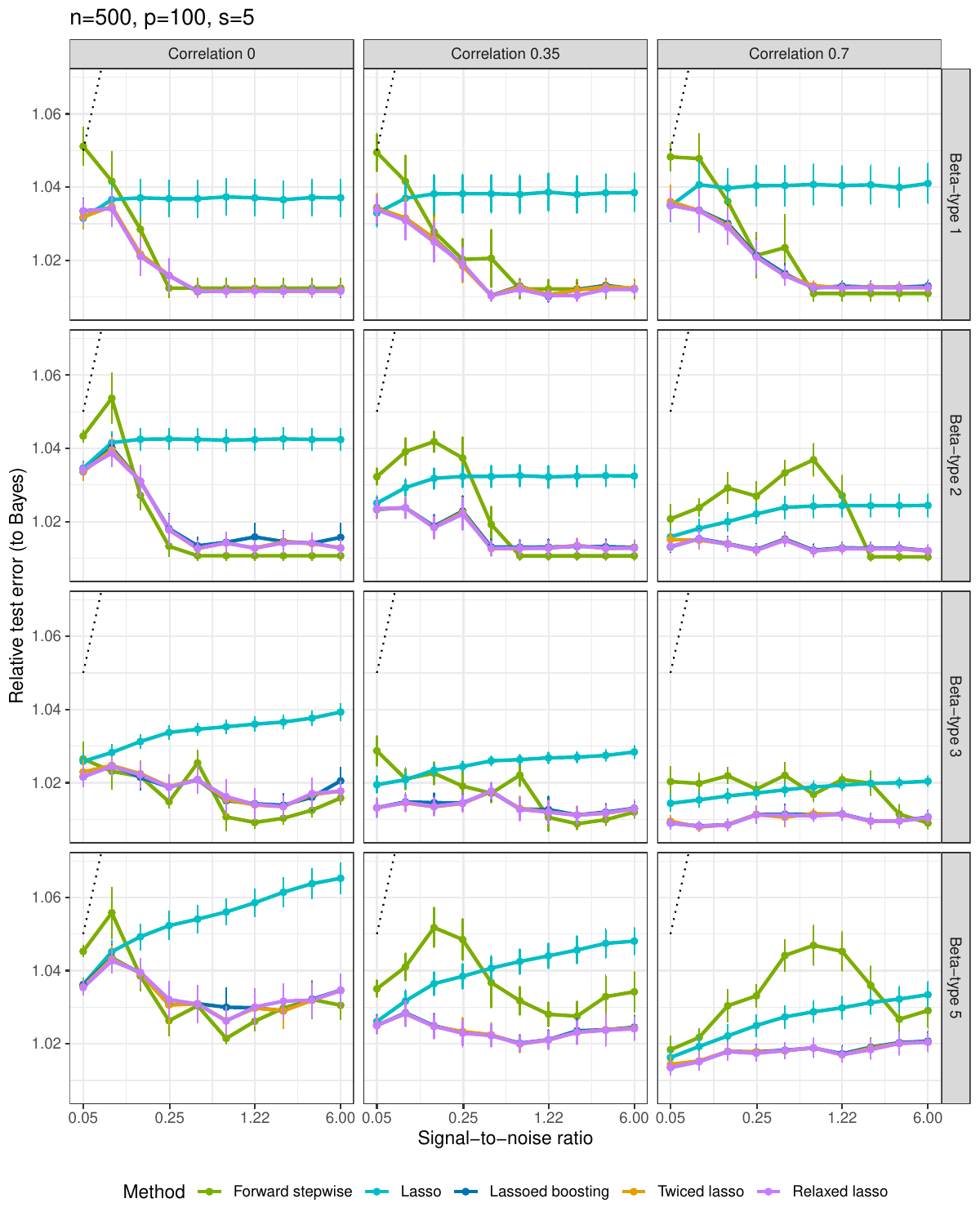}
\subsubsubsection{Proportion of variance explained}
\includegraphics[scale=0.82]{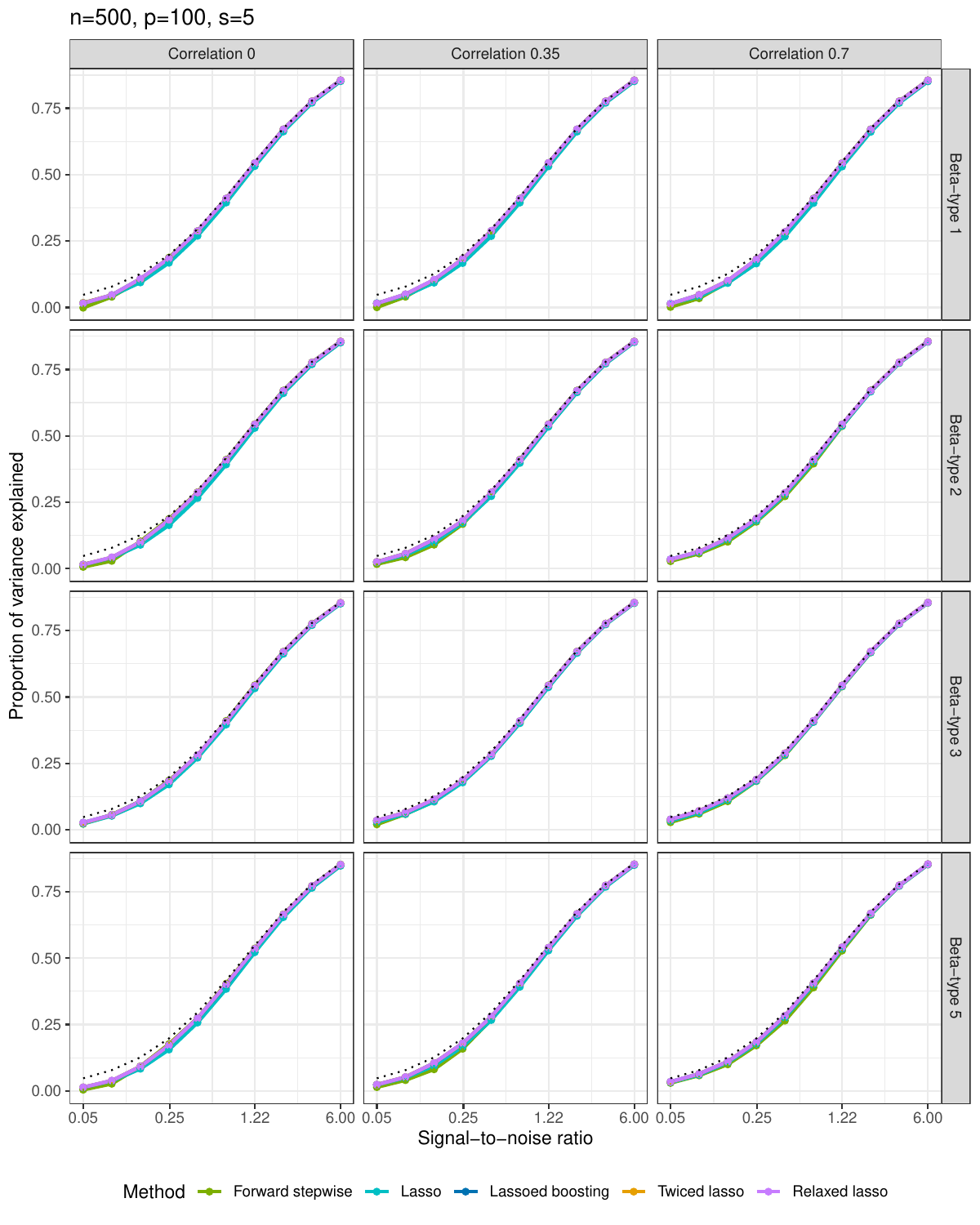}
\subsubsubsection{Number of nonzero coefficients}
\includegraphics[scale=0.82]{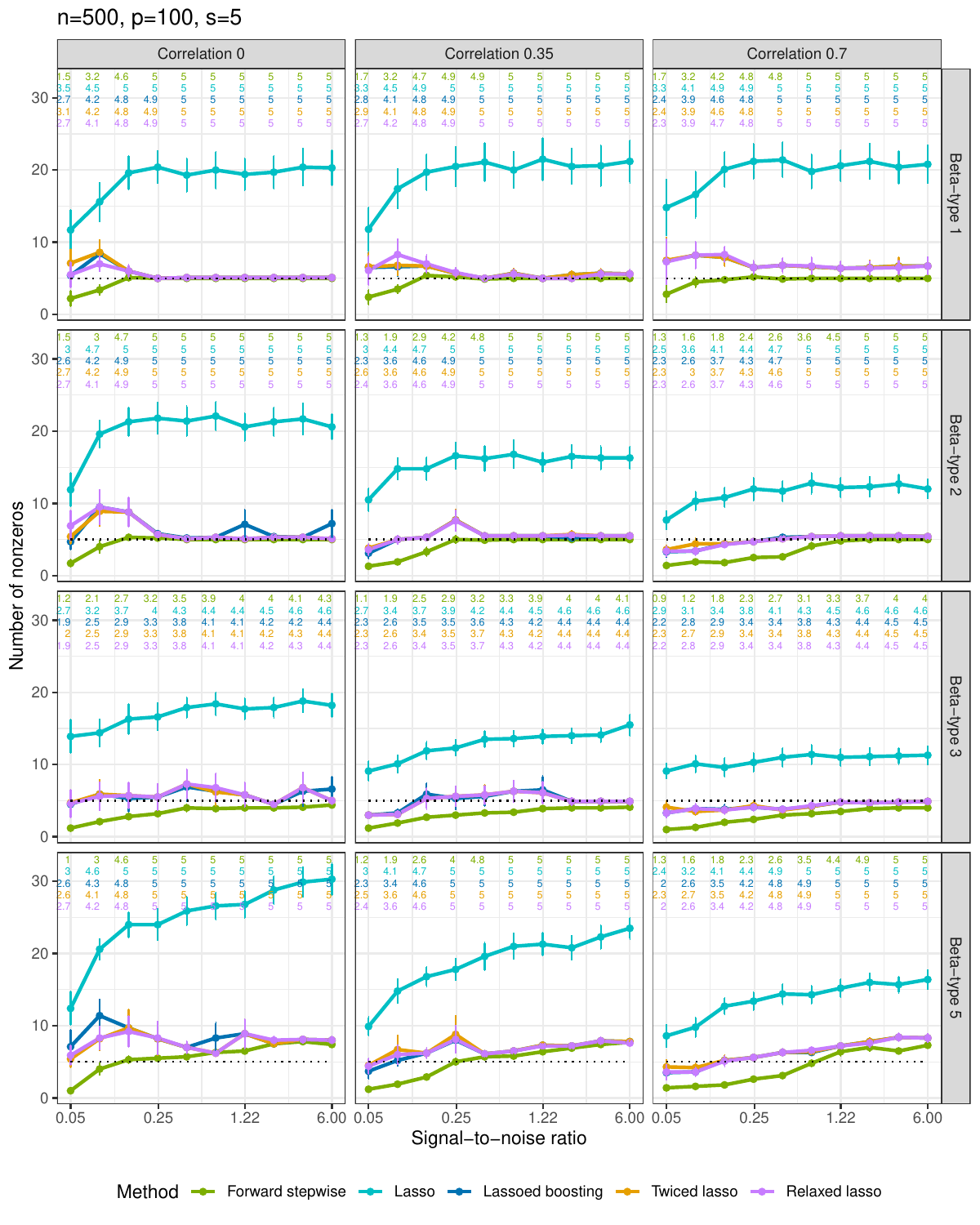}

\subsubsection{High-5 setting: $n=50, p=1000, s=5$}
\subsubsubsection{Relative risk (to null model)}
\includegraphics[scale=0.82]{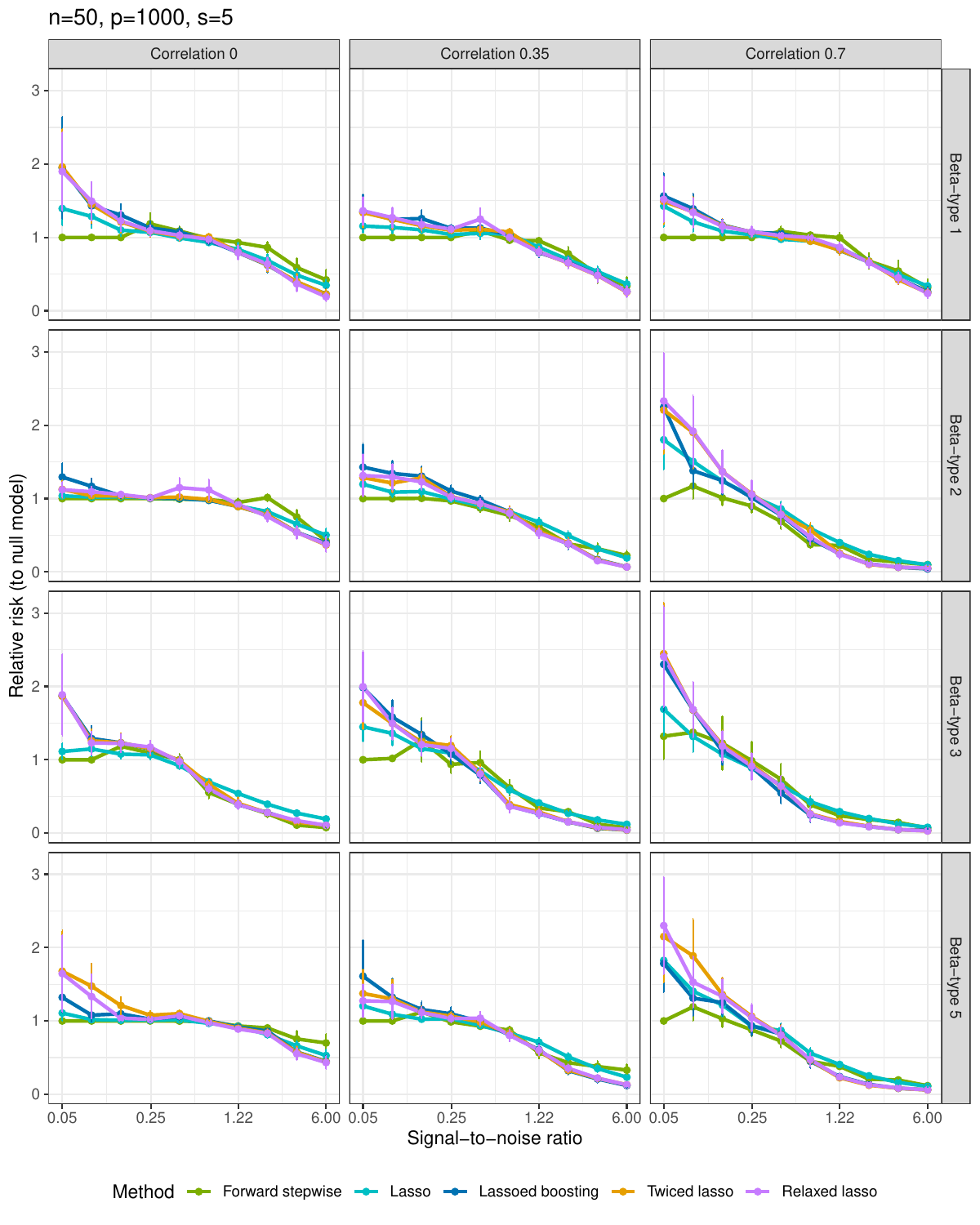}
\subsubsubsection{Relative test error (to Bayes)}
\includegraphics[scale=0.82]{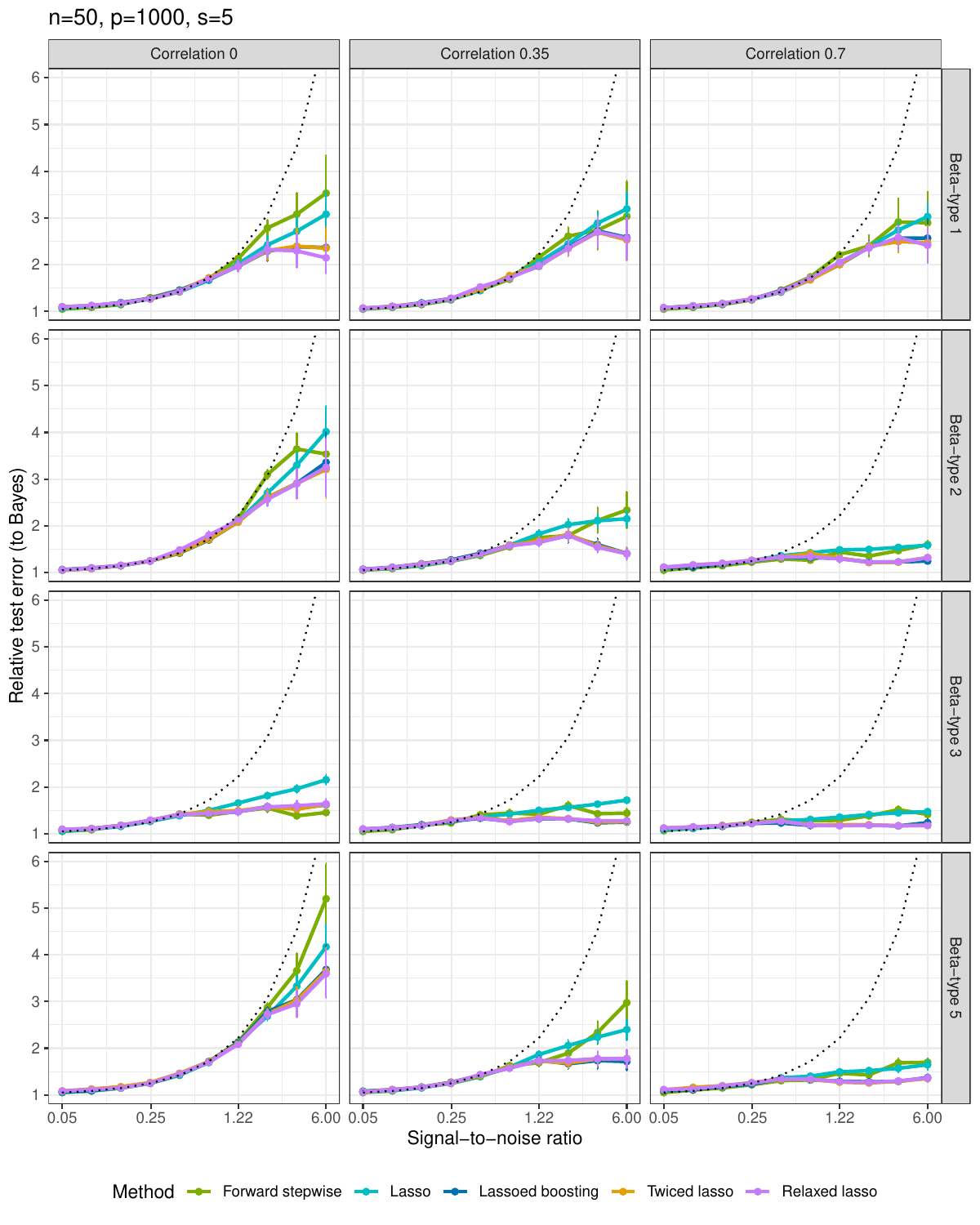}
\subsubsubsection{Proportion of variance explained}
\includegraphics[scale=0.82]{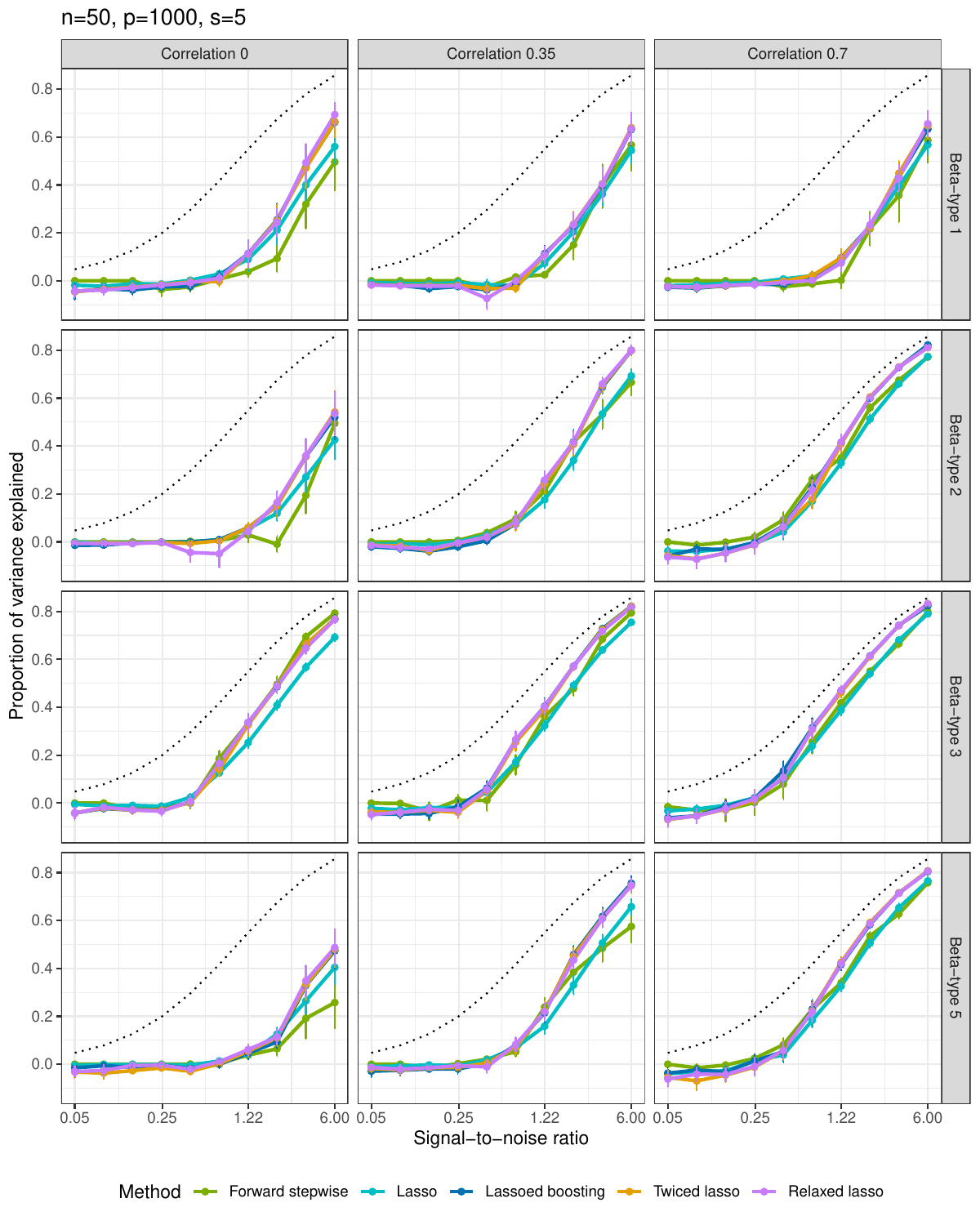}
\subsubsubsection{Number of nonzero coefficients}
\includegraphics[scale=0.82]{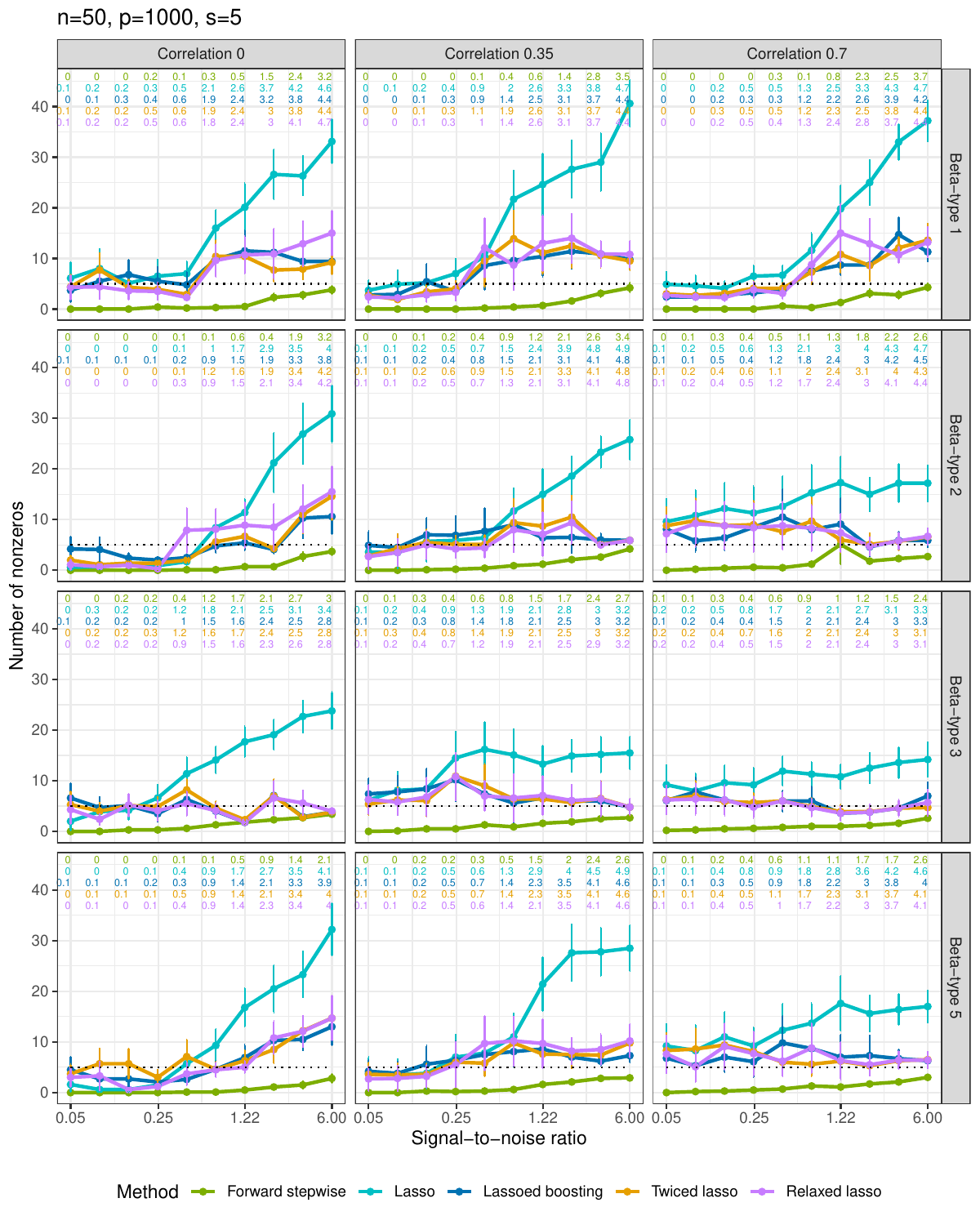}

\subsubsection{High-10 setting: $n=100, p=1000, s=10$}
\subsubsubsection{Relative risk (to null model)}
\includegraphics[scale=0.82]{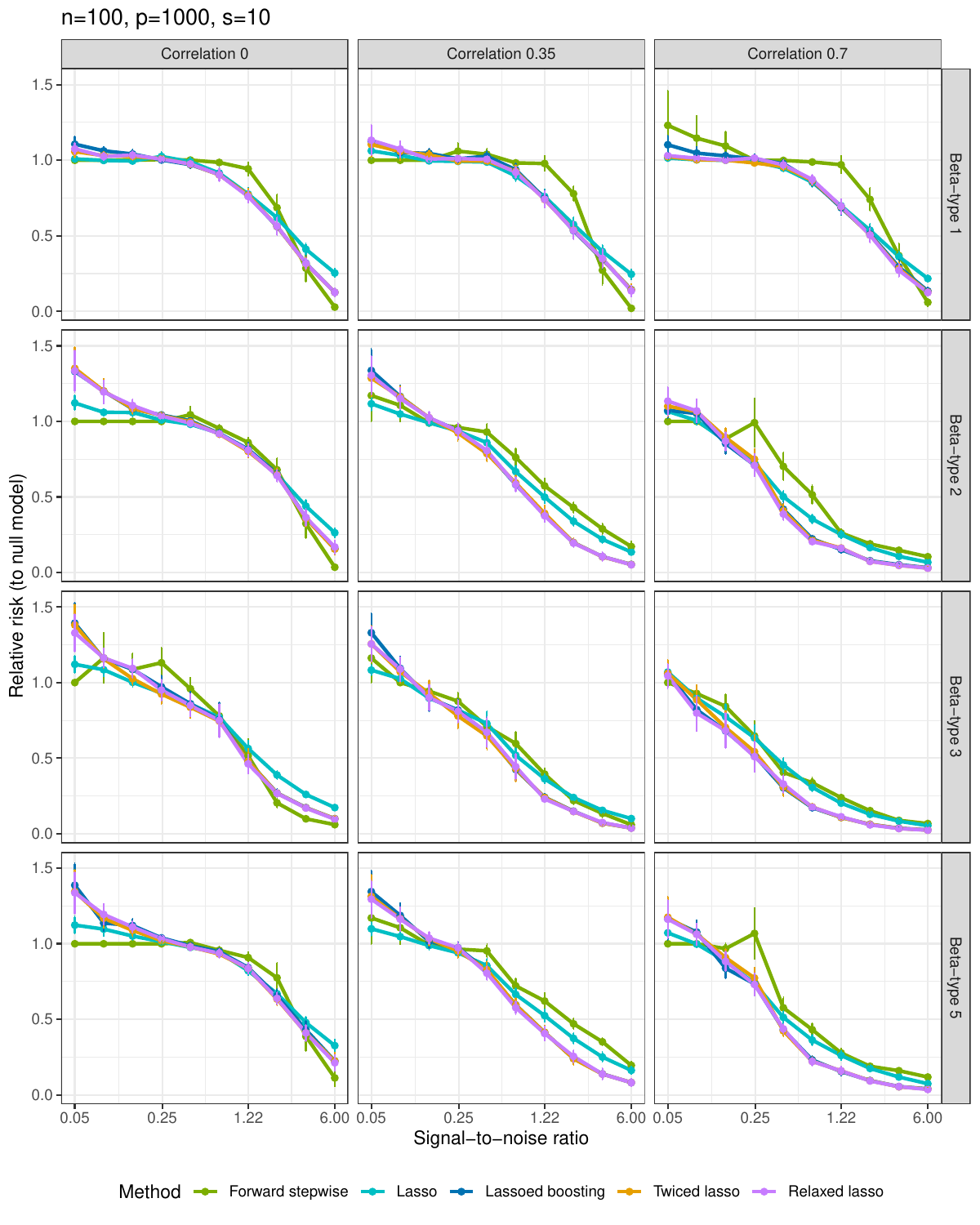}
\subsubsubsection{Relative test error (to Bayes)}
\includegraphics[scale=0.82]{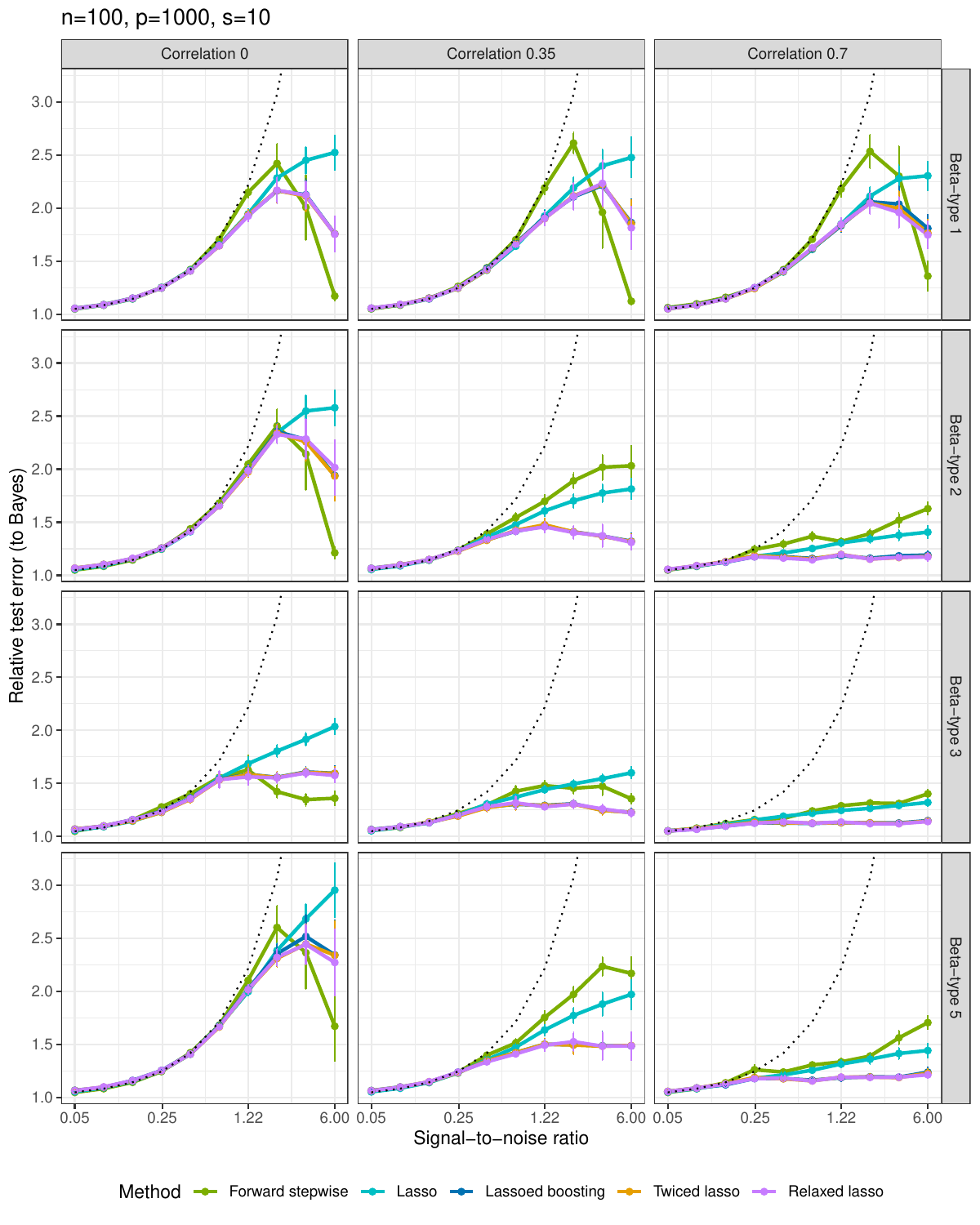}
\subsubsubsection{Proportion of variance explained}
\includegraphics[scale=0.82]{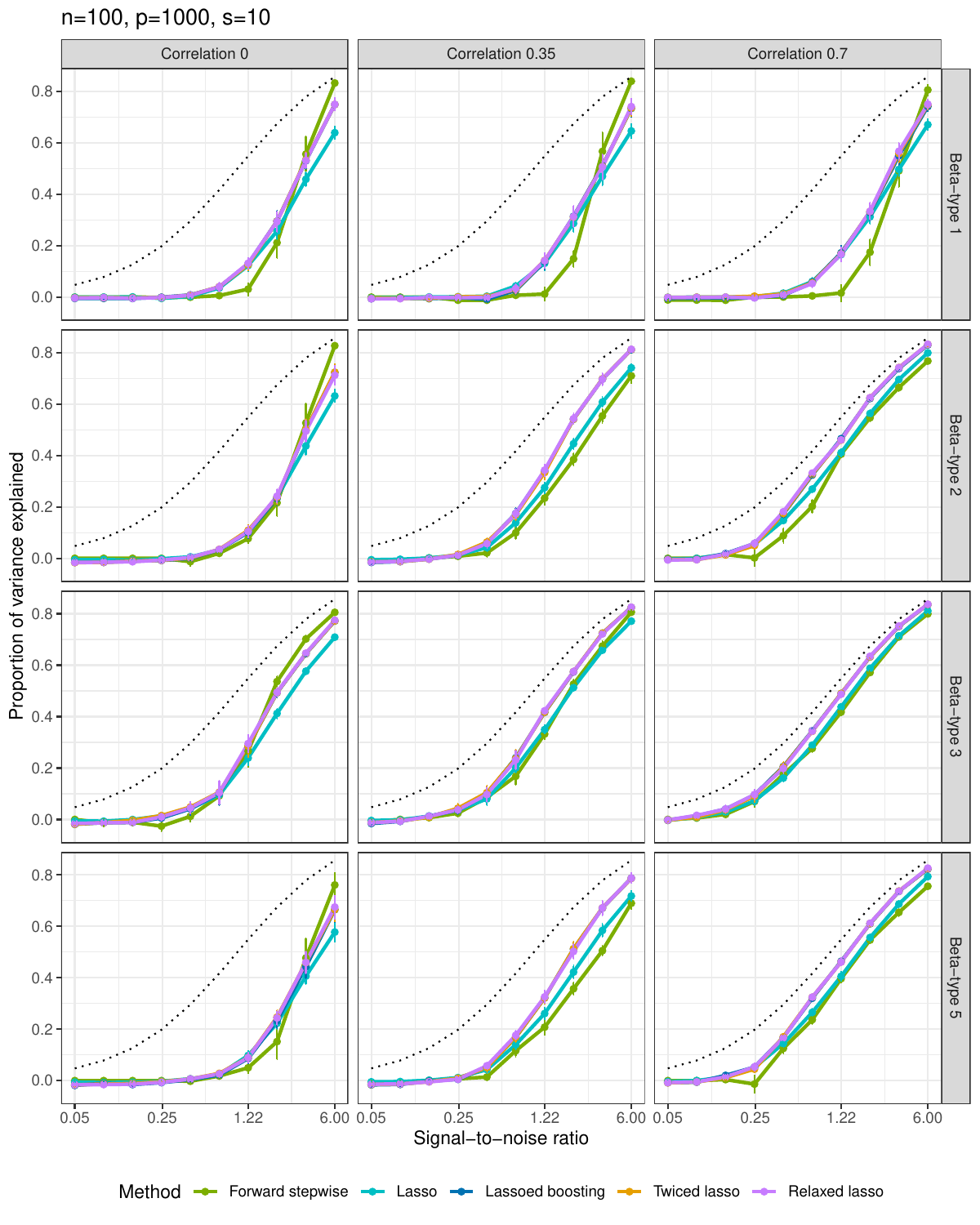}
\subsubsubsection{Number of nonzero coefficients}
\includegraphics[scale=0.82]{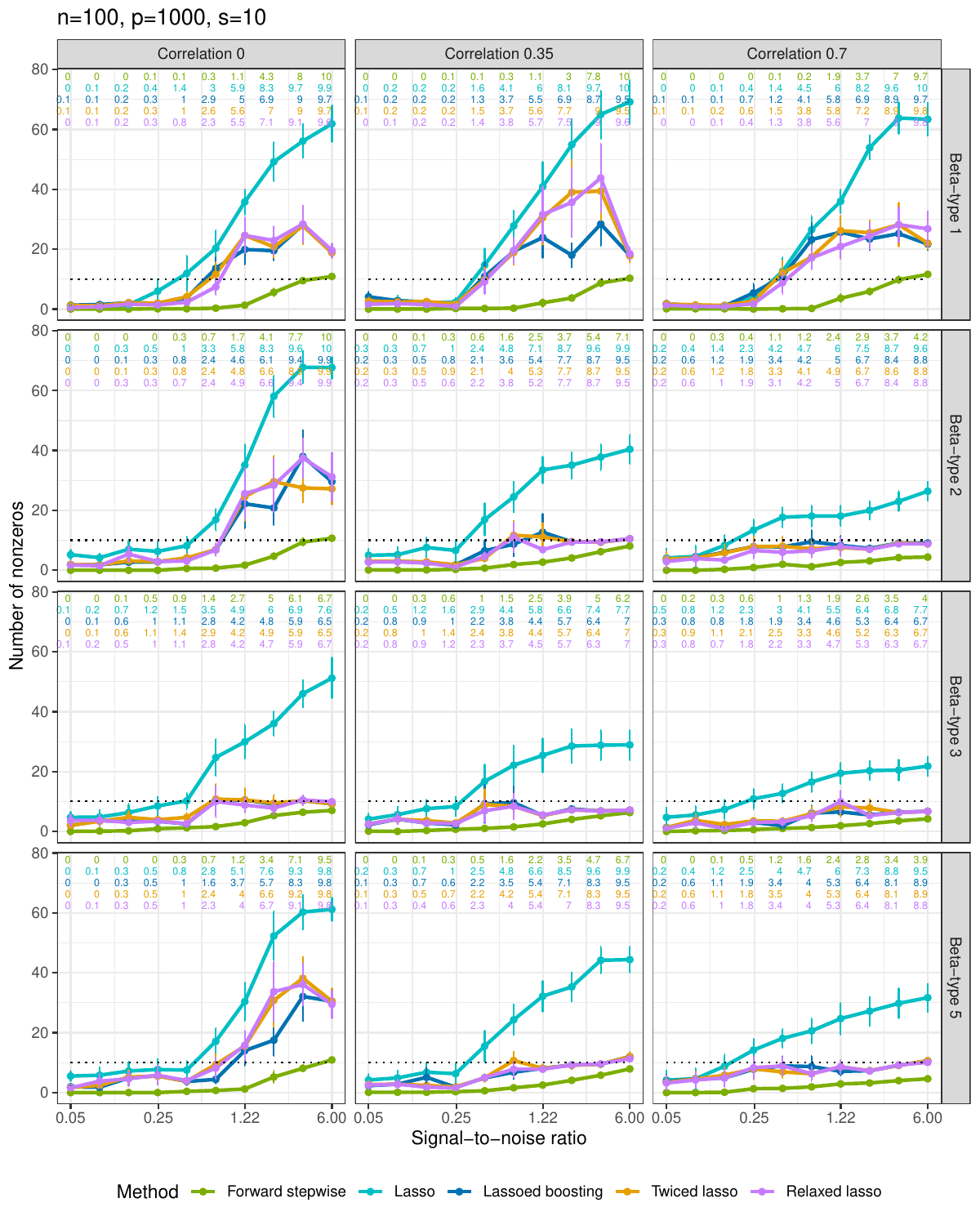}

\subsection{Oracle tuning figures in simulation}
\subsubsection{Low setting: $n=100, p=10, s=5$}
\subsubsubsection{Relative risk (to null model)}
\includegraphics[scale=0.79]{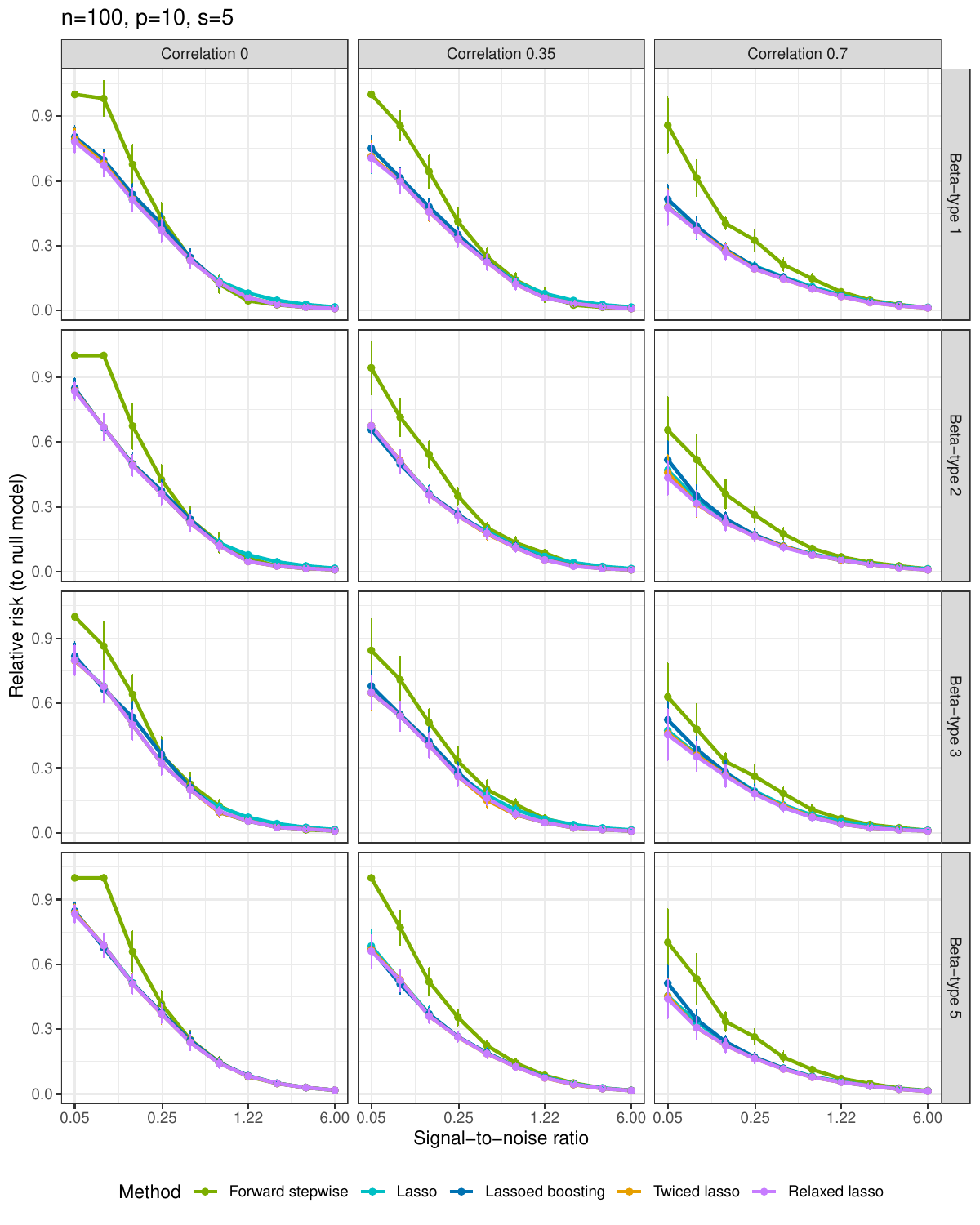}
\subsubsubsection{Relative test error (to Bayes)}
\includegraphics[scale=0.82]{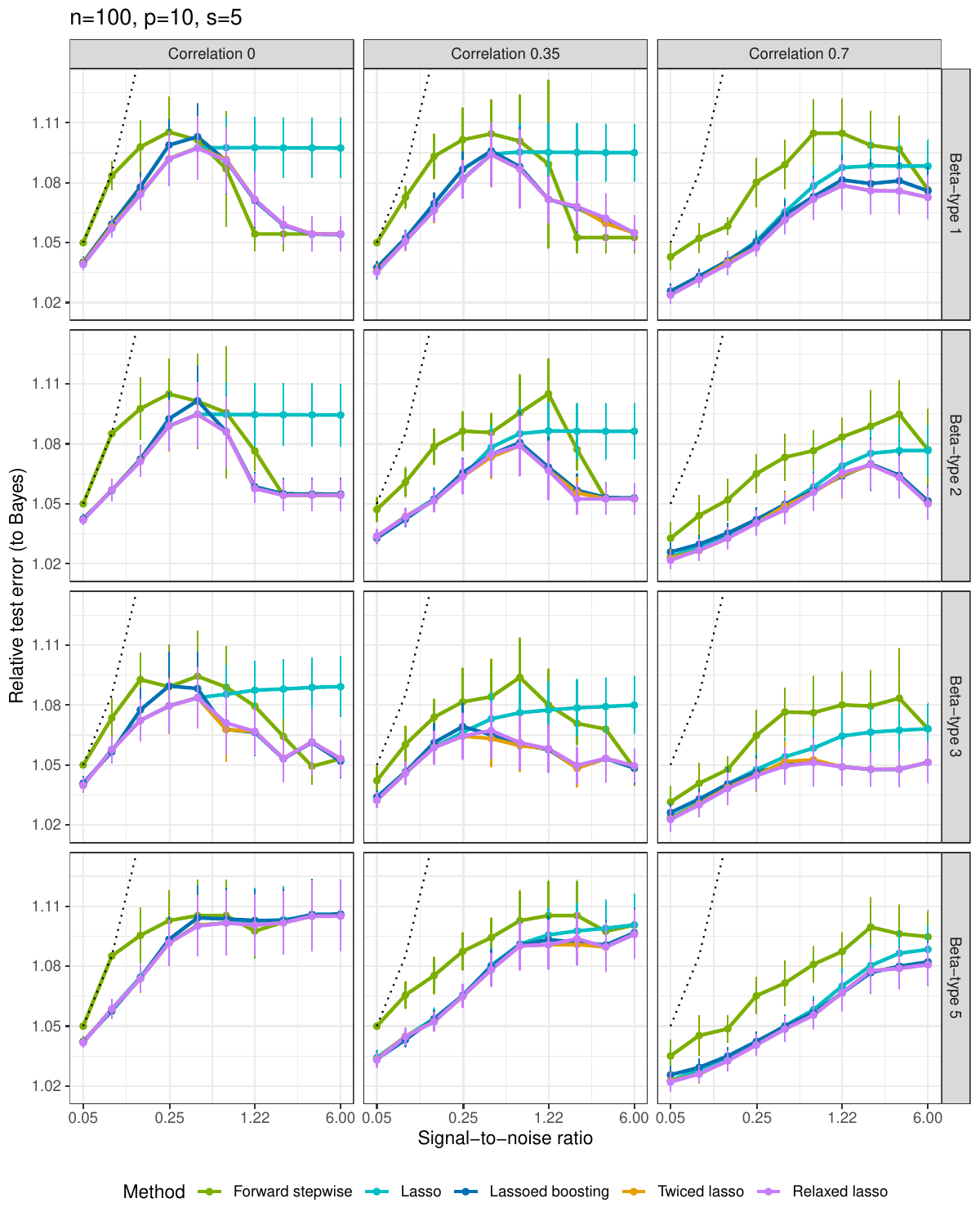}
\subsubsubsection{Proportion of variance explained}
\includegraphics[scale=0.82]{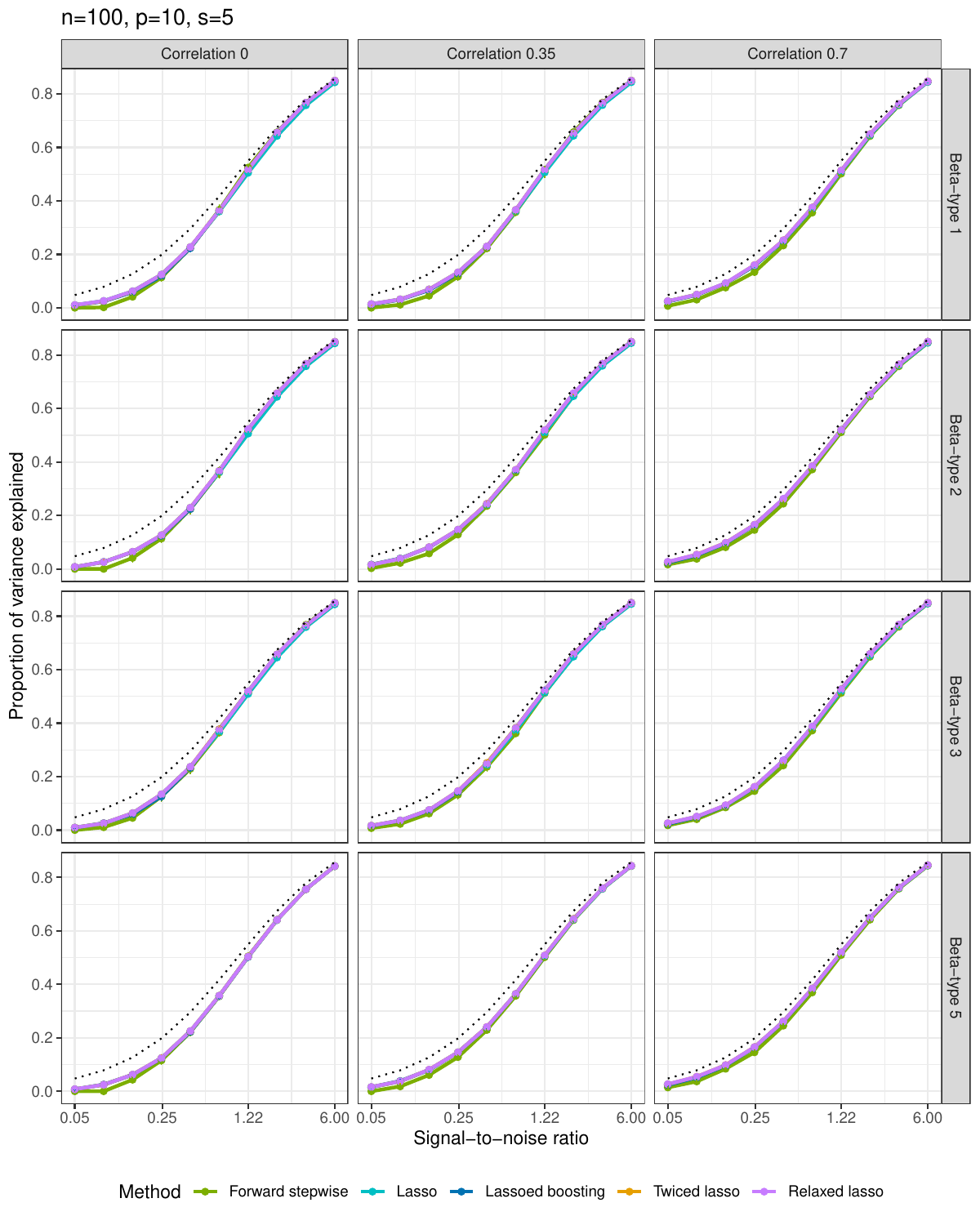}
\subsubsubsection{Number of nonzero coefficients}
\includegraphics[scale=0.82]{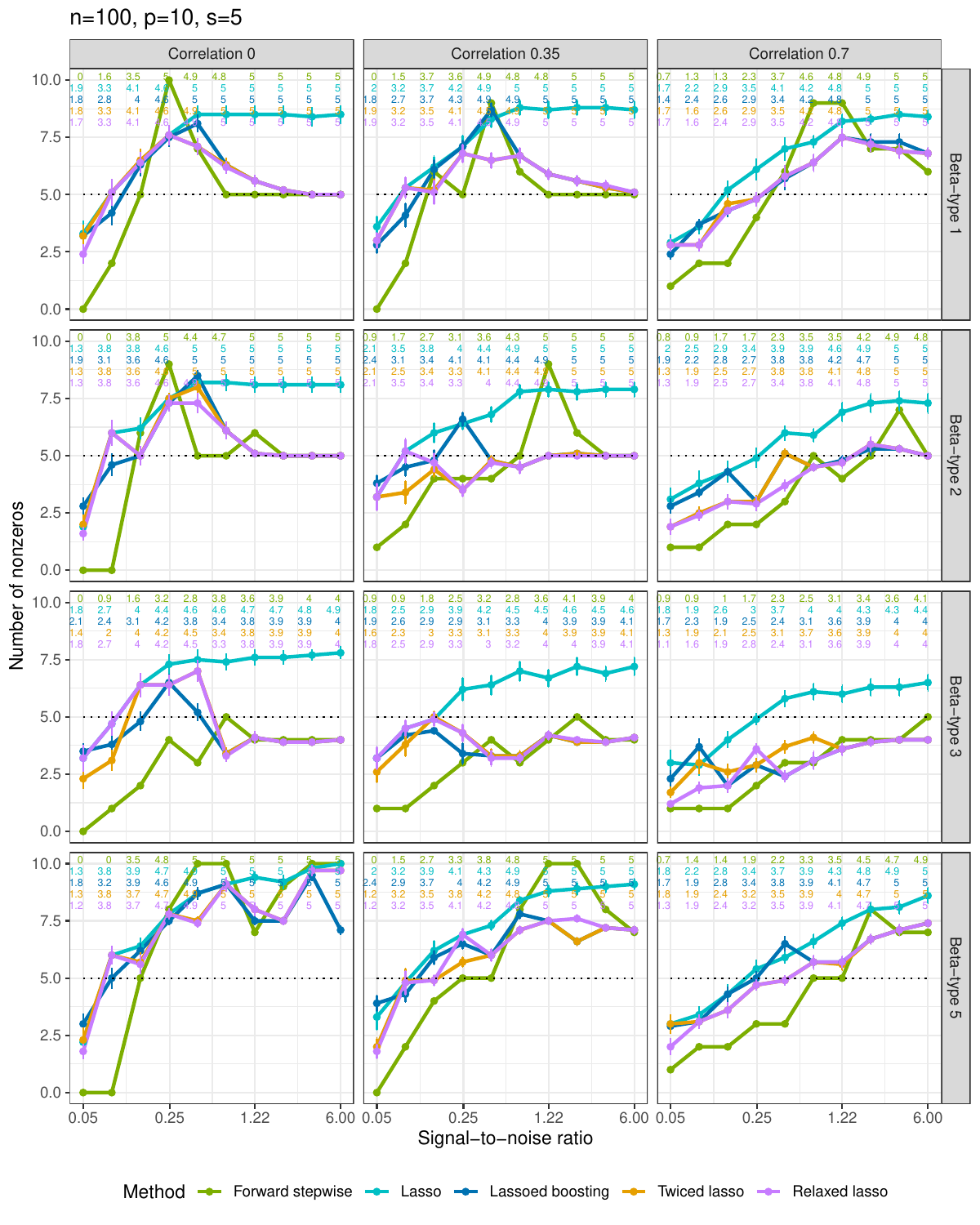}

\subsubsection{Medium setting: $n=500, p=100, s=5$}
\subsubsubsection{Relative risk (to null model)}
\includegraphics[scale=0.82]{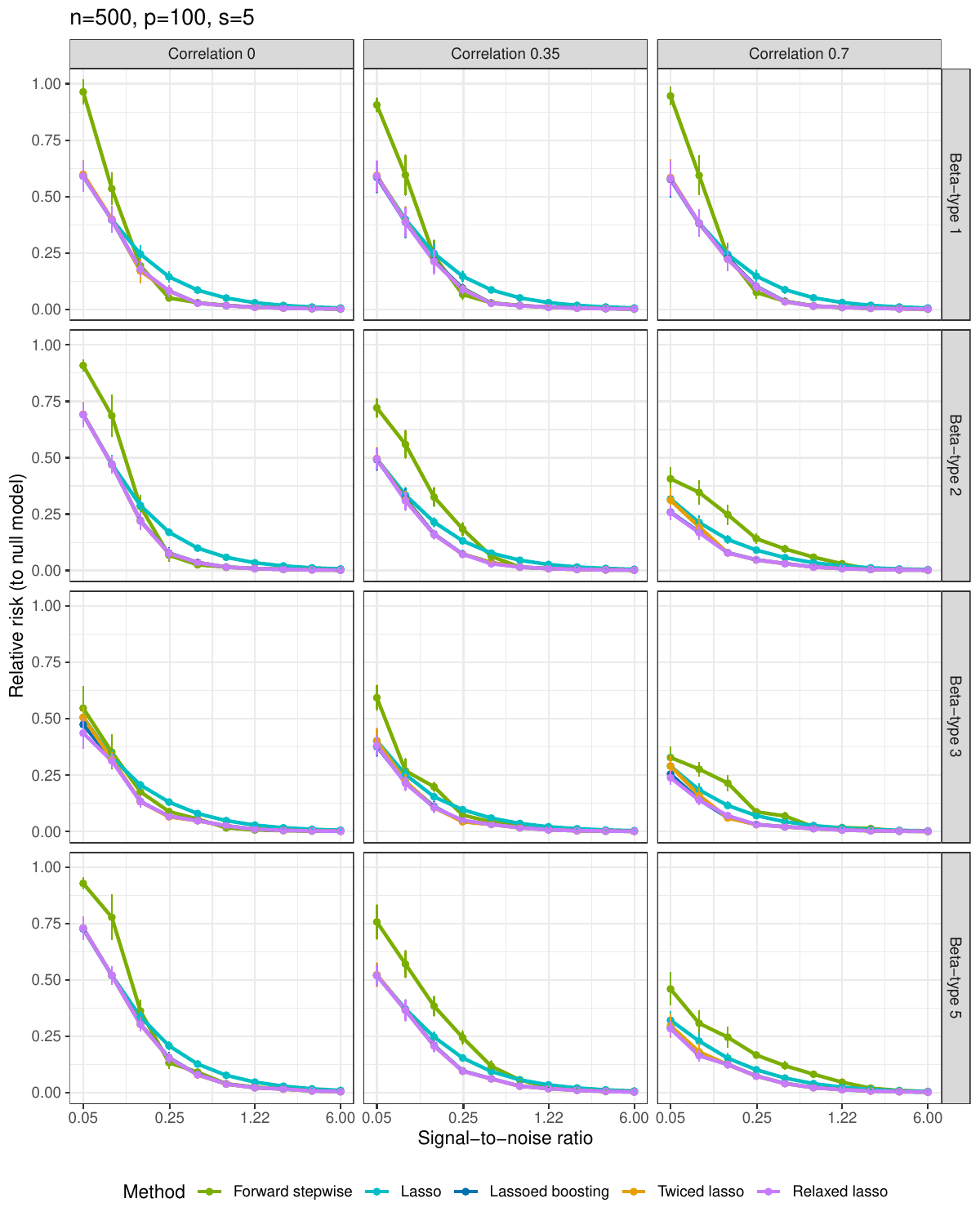}
\subsubsubsection{Relative test error (to Bayes)}
\includegraphics[scale=0.82]{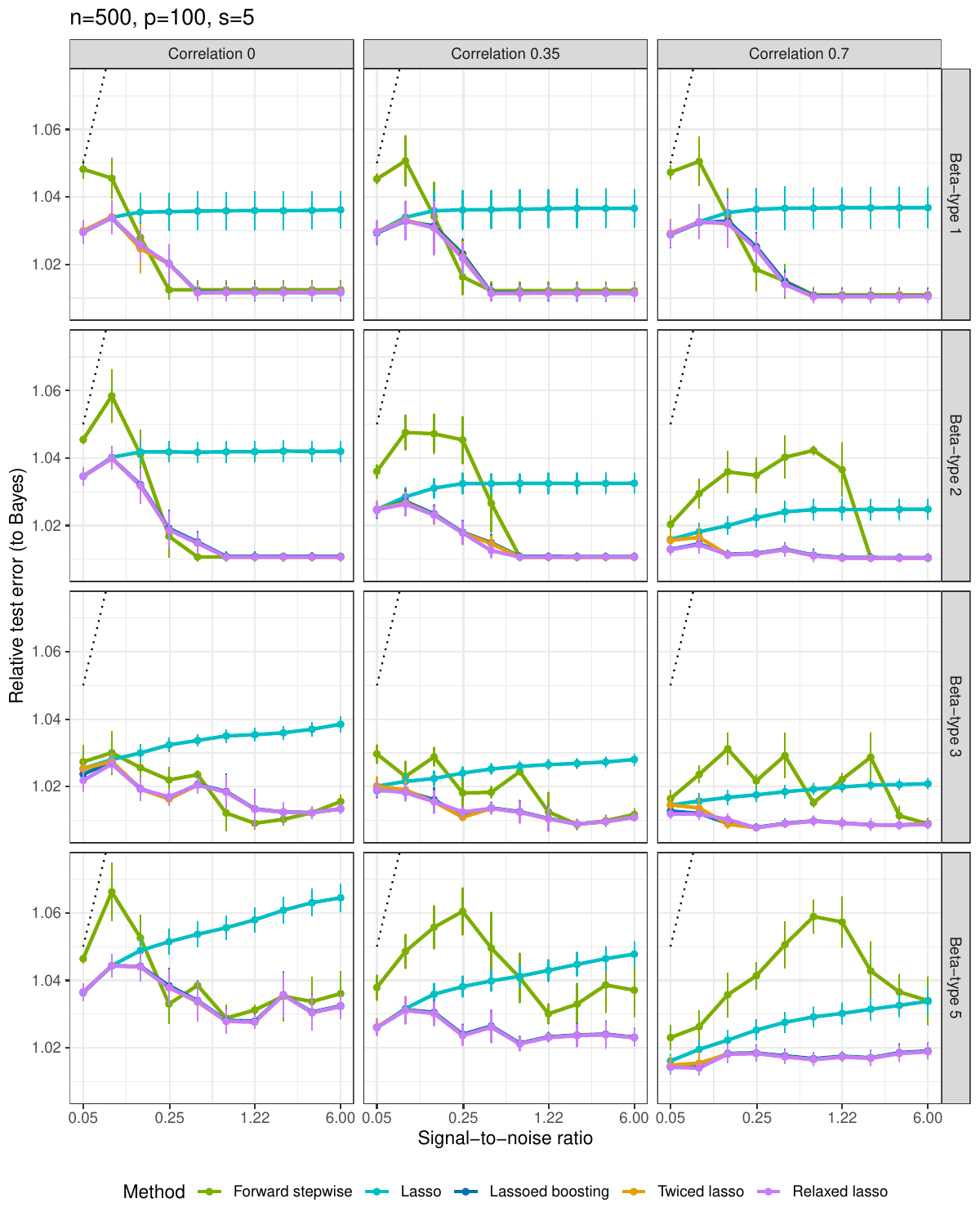}
\subsubsubsection{Proportion of variance explained}
\includegraphics[scale=0.82]{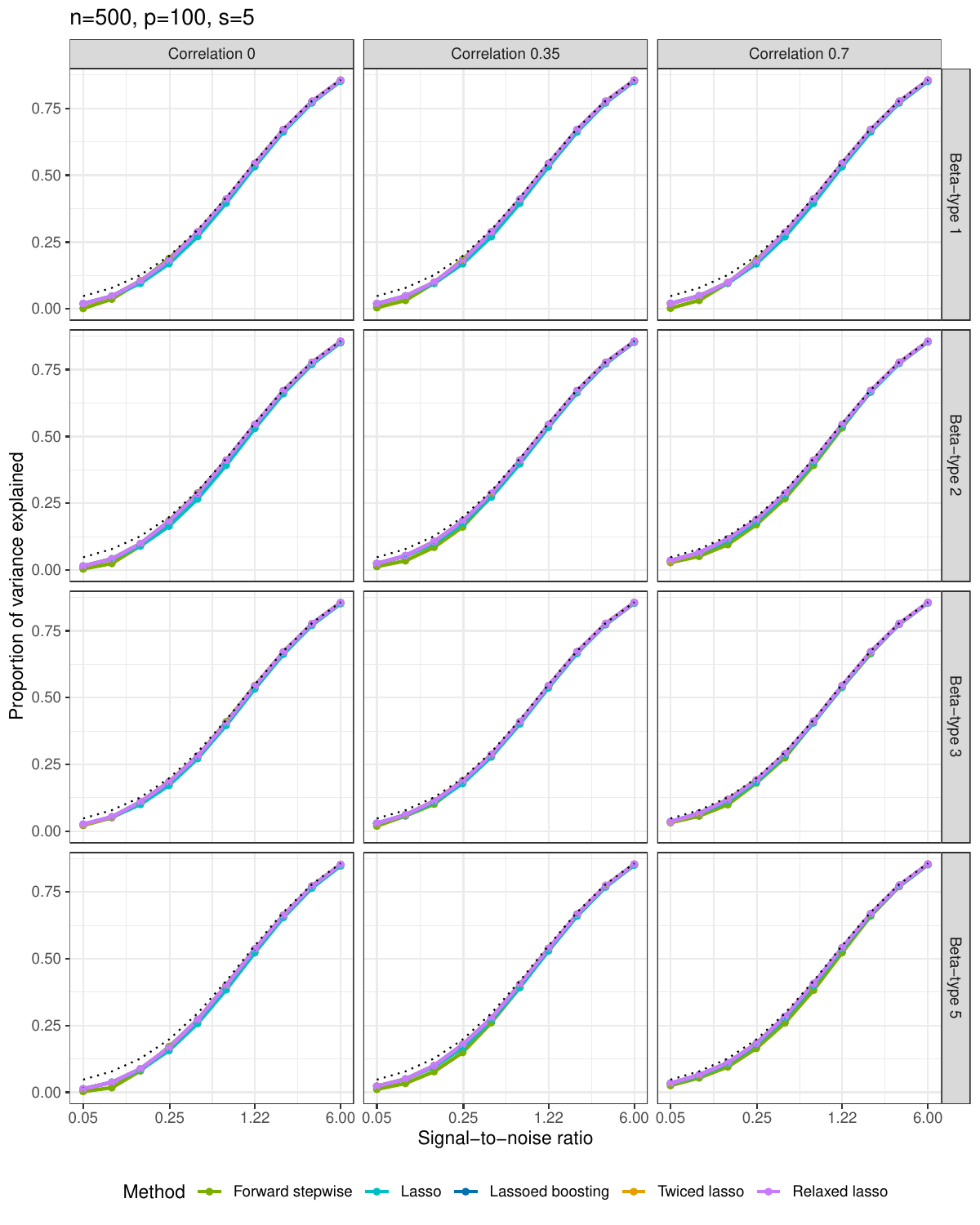}
\subsubsubsection{Number of nonzero coefficients}
\includegraphics[scale=0.82]{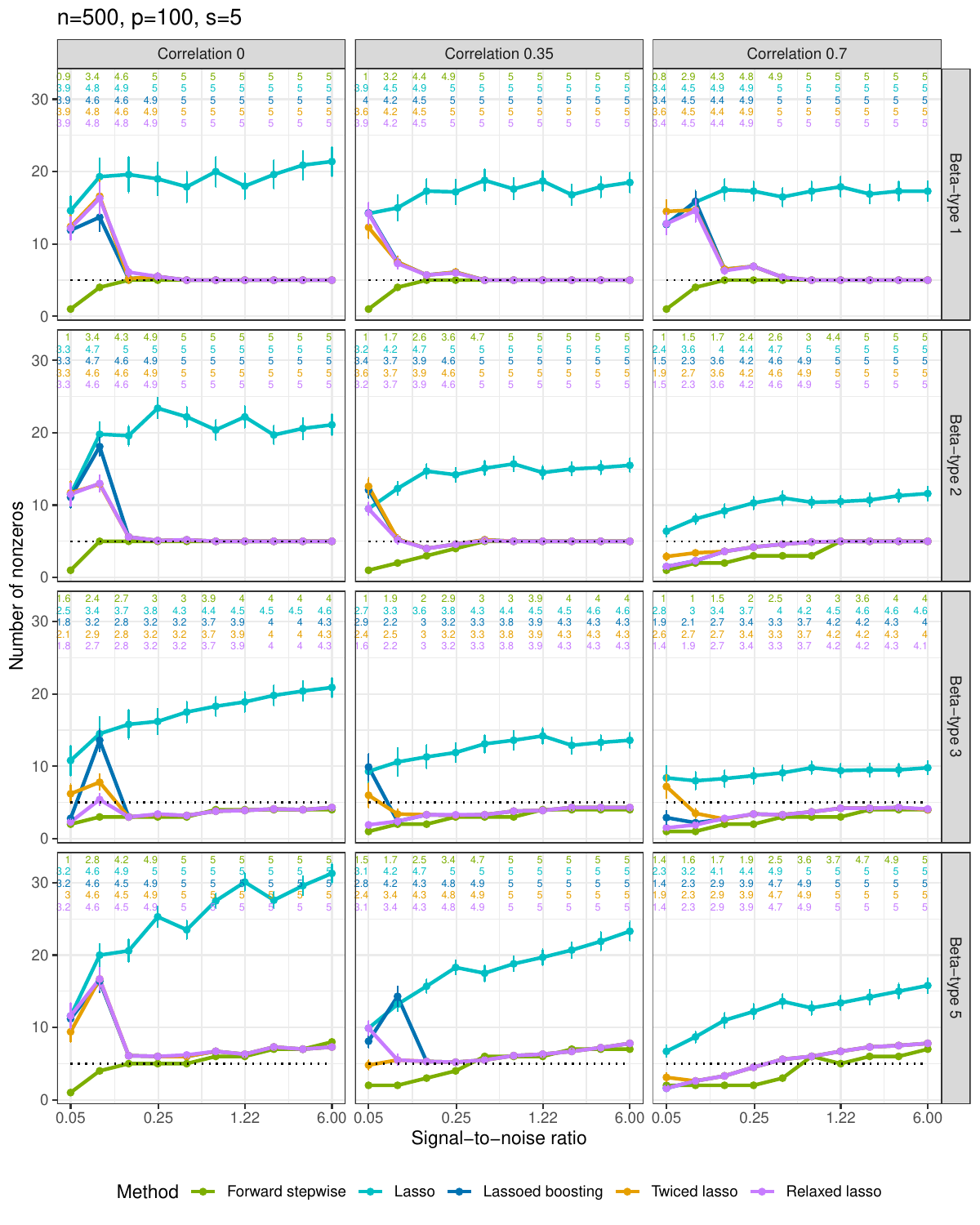}

\subsubsection{High-5 setting: $n=50, p=1000, s=5$}
\subsubsubsection{Relative risk (to null model)}
\includegraphics[scale=0.82]{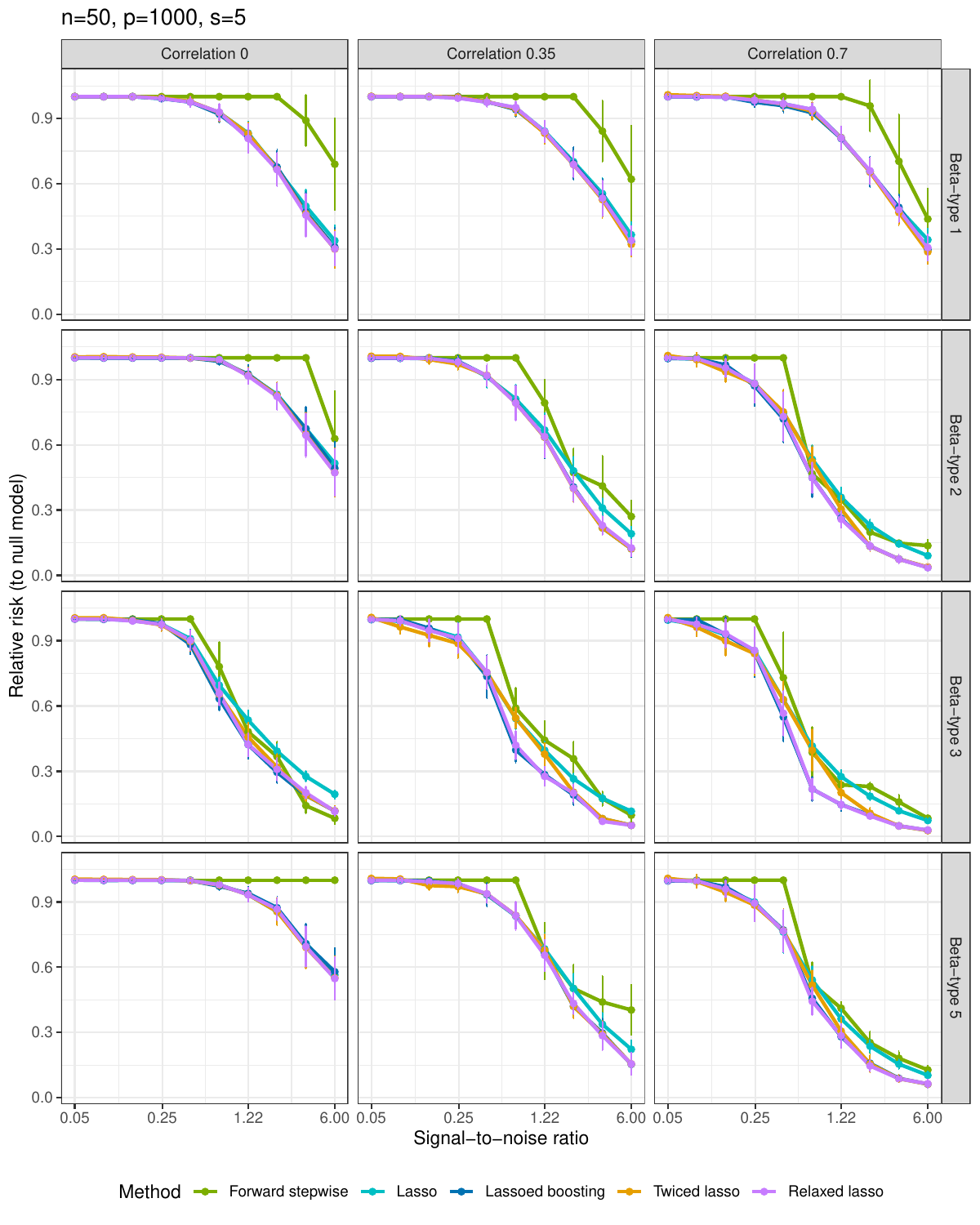}
\subsubsubsection{Relative test error (to Bayes)}
\includegraphics[scale=0.82]{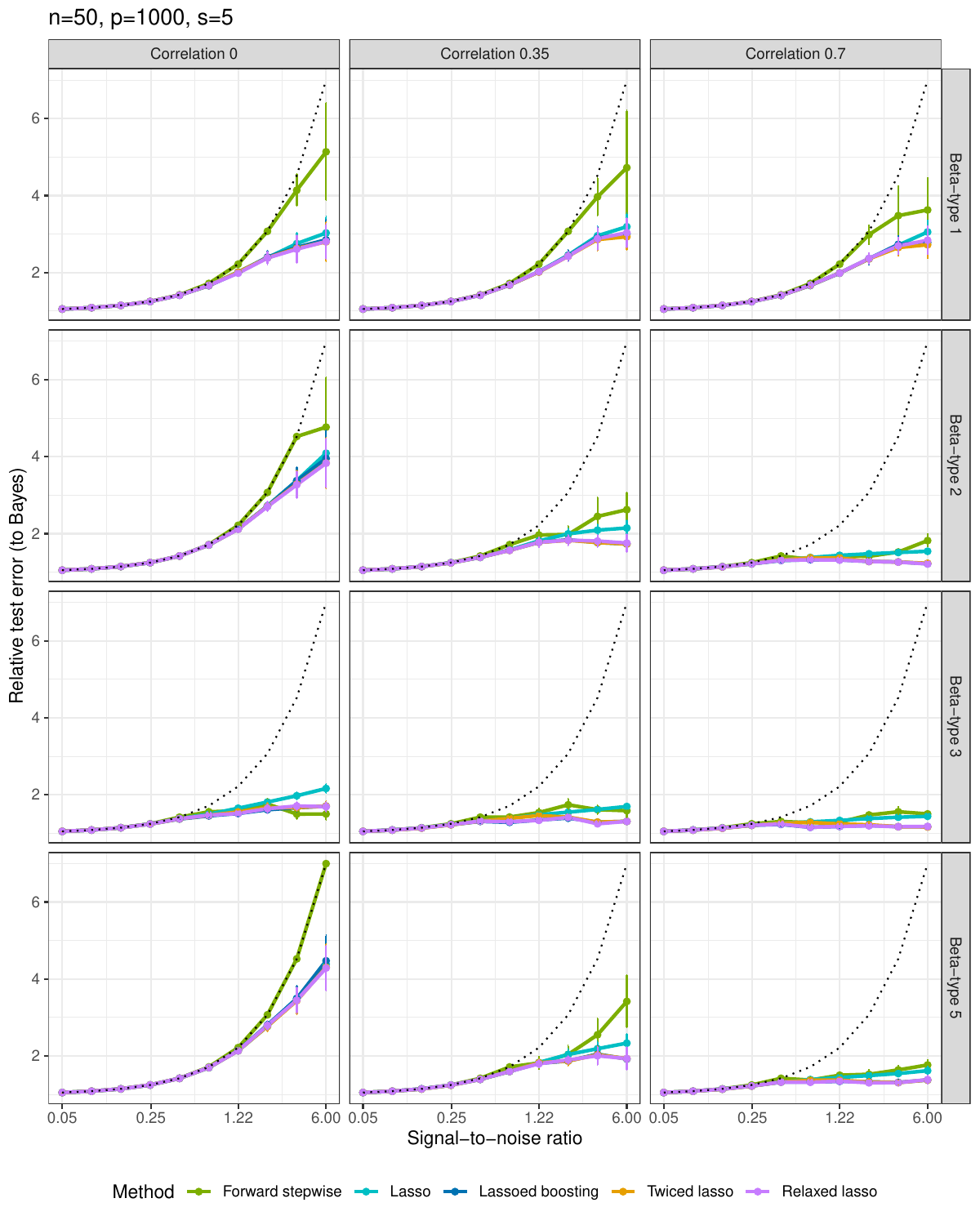}
\subsubsubsection{Proportion of variance explained}
\includegraphics[scale=0.82]{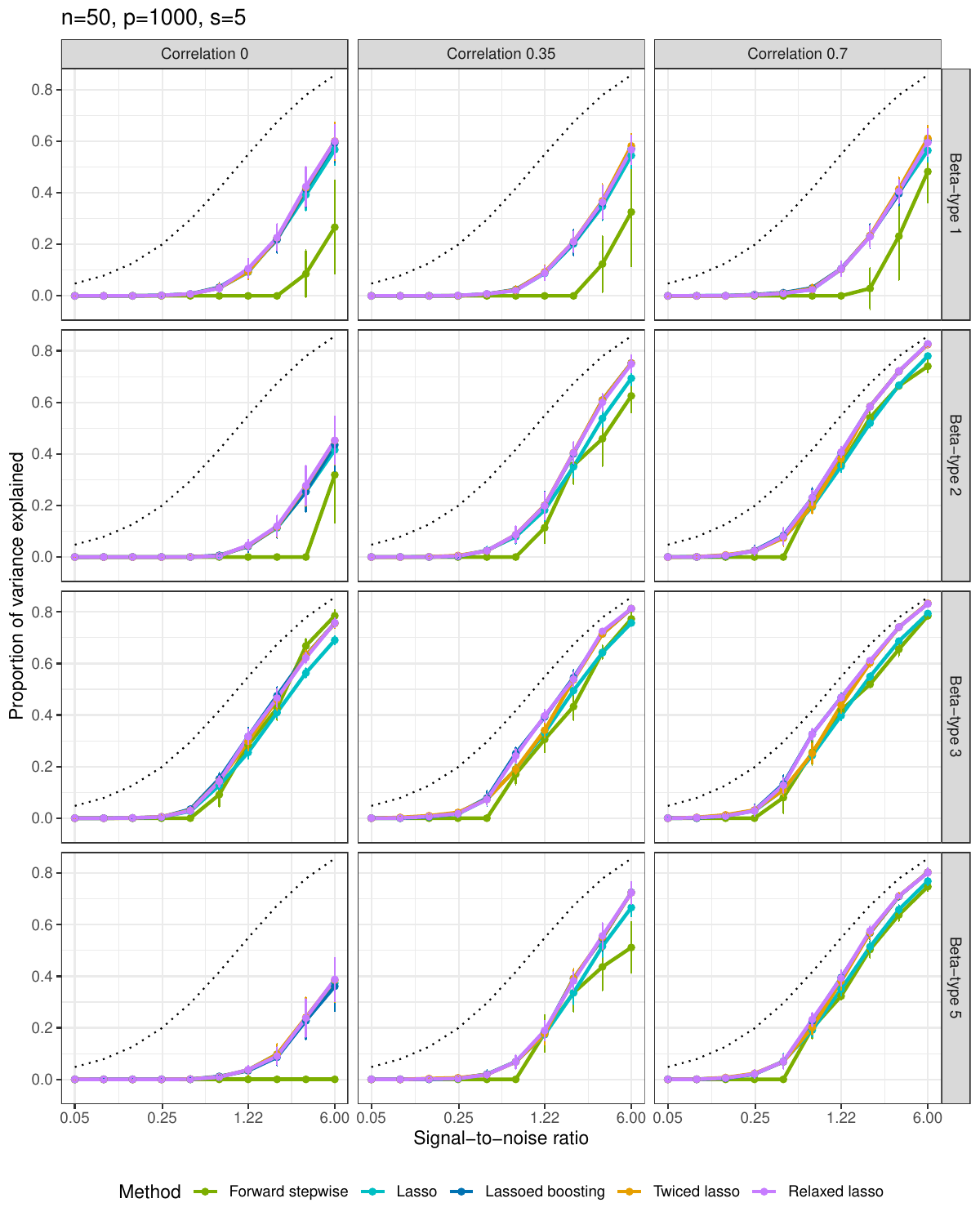}
\subsubsubsection{Number of nonzero coefficients}
\includegraphics[scale=0.82]{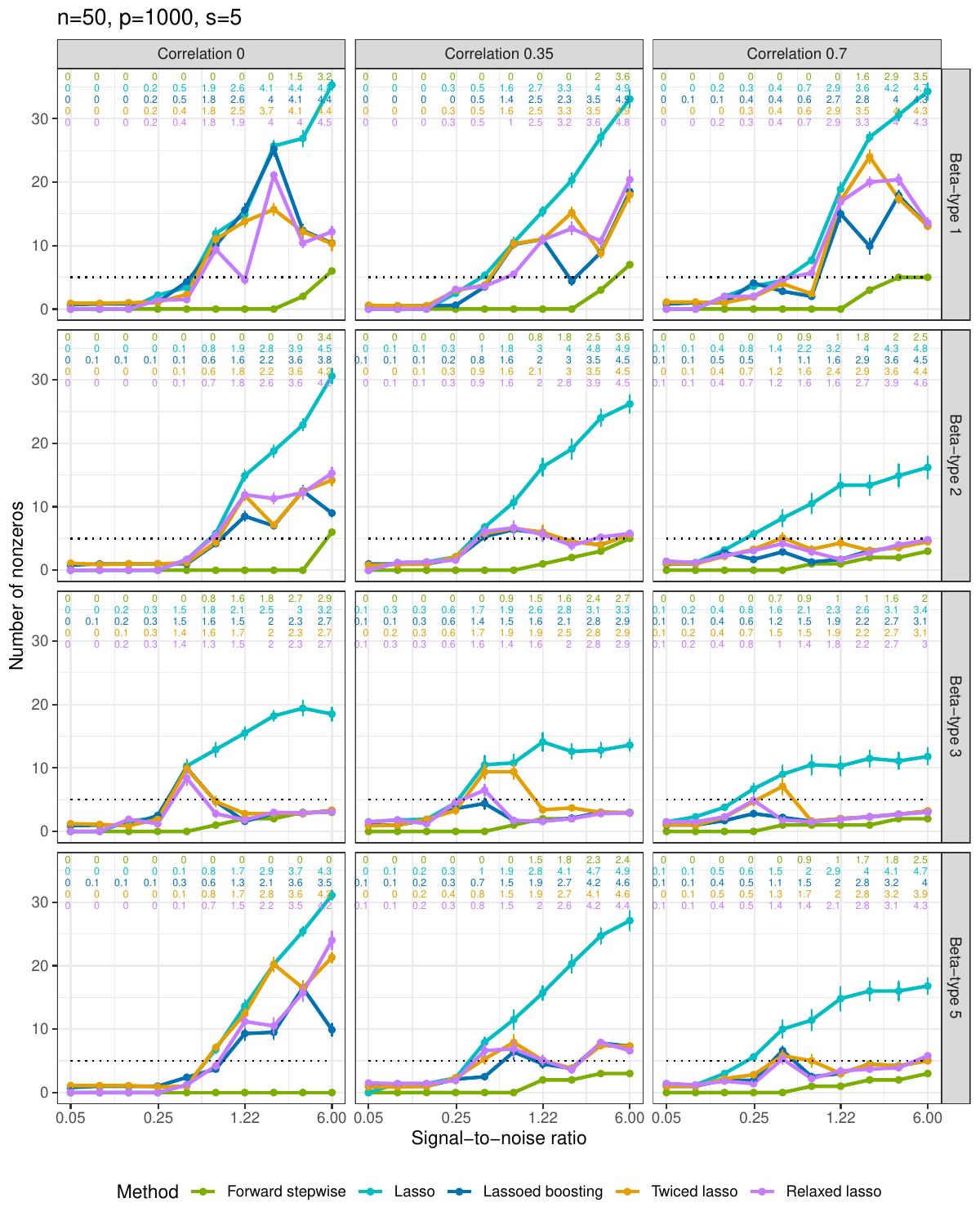}

\subsubsection{High-10 setting: $n=100, p=1000, s=10$}
\subsubsubsection{Relative risk (to null model)}
\includegraphics[scale=0.82]{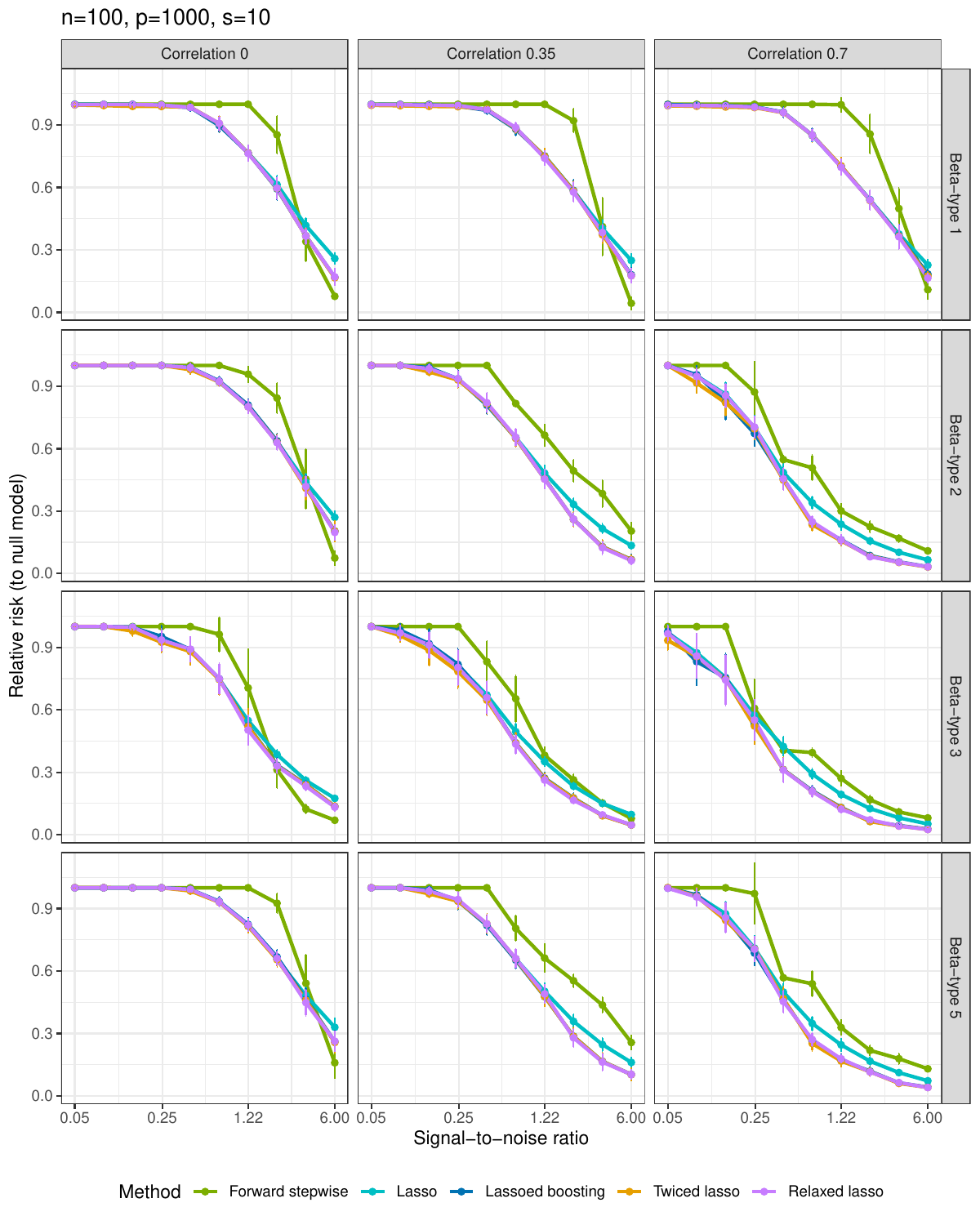}
\subsubsubsection{Relative test error (to Bayes)}
\includegraphics[scale=0.82]{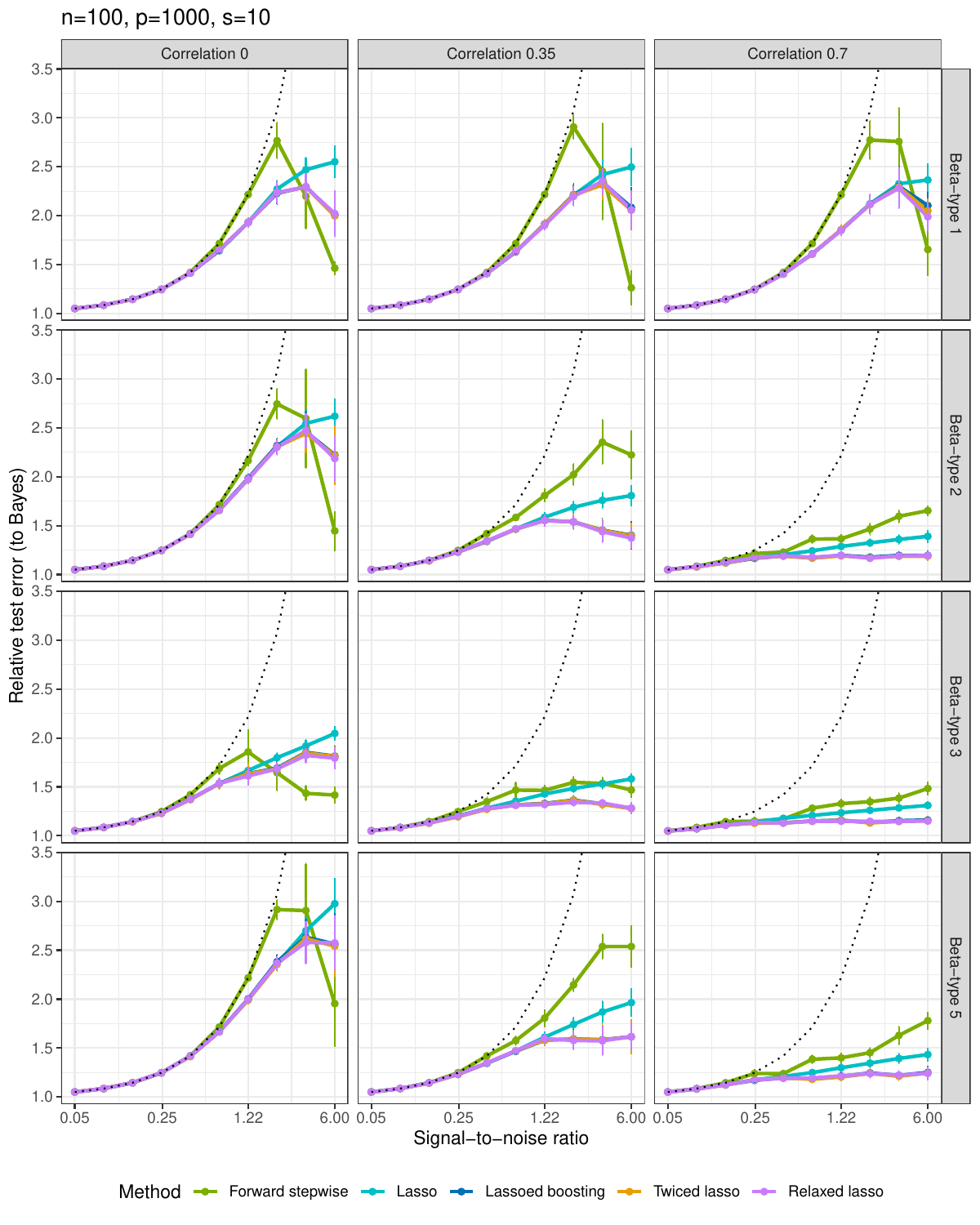}
\subsubsubsection{Proportion of variance explained}
\includegraphics[scale=0.82]{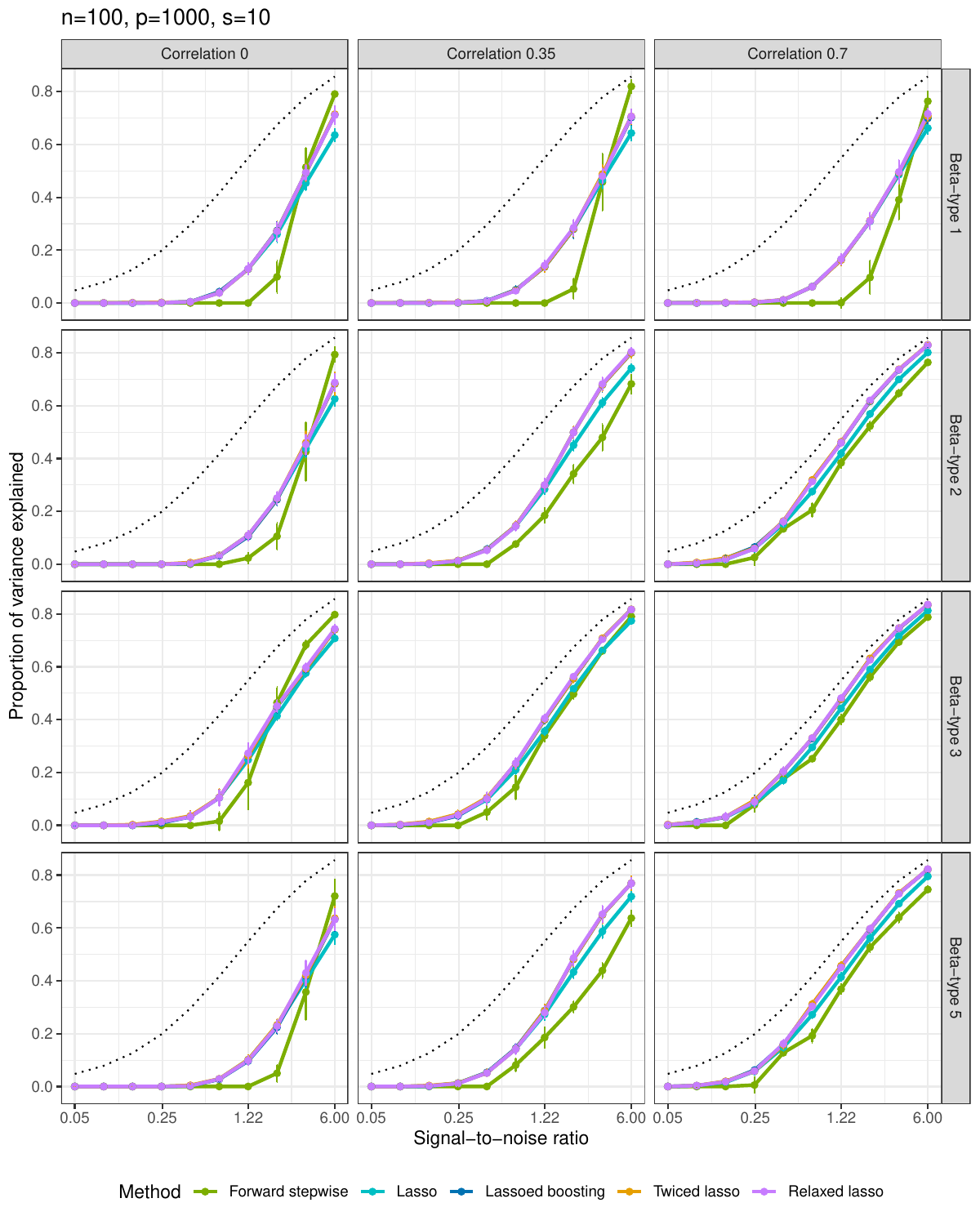}
\subsubsubsection{Number of nonzero coefficients}
\includegraphics[scale=0.82]{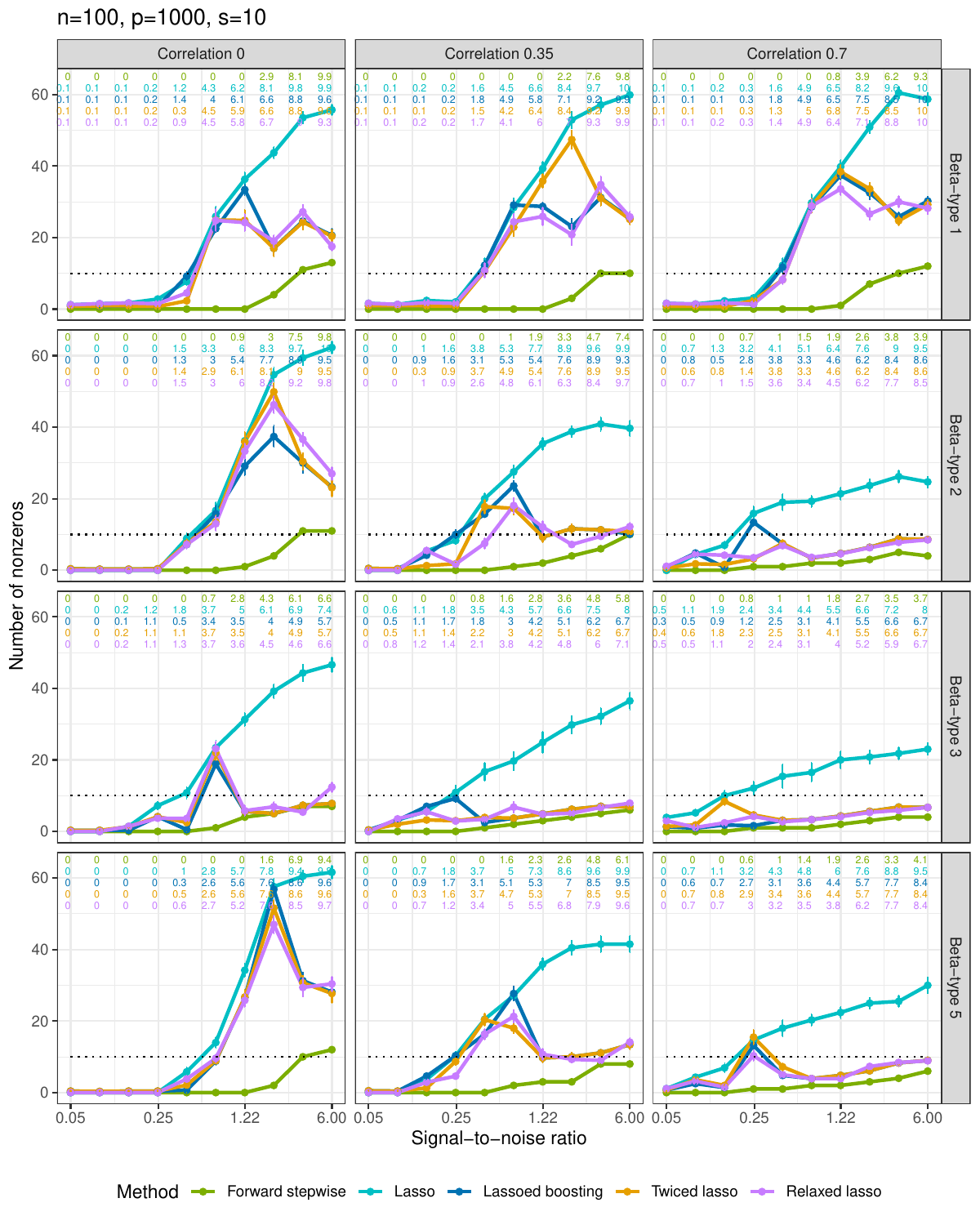}

\newpage
\setstretch{0.8}
\subsection{Variable definitions in the application} \label{supp:variable def}

%	\caption{this is caption}
	\begin{tabularx}{\linewidth}{ p{2.6cm} p{5cm} p{1.9cm} p{5cm}}
		\caption{Variables used in the application (Table 1 in \cite{Greenetal2017stockreturns})}\label{tab:variable list}\\
		%\toprule
		\midrule
		\endfirsthead
		%\toprule
		\midrule
		\endhead
		\midrule
		\multicolumn{4}{r}{\footnotesize( continued )}
		\endfoot
		%\bottomrule
		\endlastfoot
		\text{Acronym} & \text{Firm characteristic} & \text{Acronym} & \text{Firm characteristic}  \\
		\midrule
		\textit{absacc} & Absolute accruals & \textit{divo} & Dividend omission \\
		\textit{acc} & Working capital accruals & \textit{dolvol} & Dollar trading volume \\
		\textit{aeavol} & Abnormal earnings & \textit{dy} & Dividend to price \\
		\multicolumn{1}{r}{} & announcement volume & \multicolumn{1}{r}{} & \multicolumn{1}{r}{} \\
		\textit{age} & \# years since first & \textit{ear} & Earnings announcement \\
		\multicolumn{1}{r}{} & Compustat coverage & \multicolumn{1}{r}{} & return \\
		\textit{agr} & Asset growth & \textit{egr} & Growth in common
		\\
		\multicolumn{1}{r}{} & \multicolumn{1}{r}{} & \multicolumn{1}{r}{} & shareholder equity \\
		\textit{baspread} & Bid-ask spread & \textit{ep} & Earnings to price \\
		\textit{beta} & Beta  & \textit{fgr5yr} & Forecasted growth in \\
		\multicolumn{1}{r}{} & \multicolumn{1}{r}{} & \multicolumn{1}{r}{} & 5-year EPS \\
		\textit{betasq} & Beta squared & \textit{gma} & Gross profitability \\
		\textit{bm} & Book-to-market & \textit{grCAPX} & Growth in capital \\
		\multicolumn{1}{r}{} & \multicolumn{1}{r}{} & \multicolumn{1}{r}{} & expenditures \\
		\textit{bm\_ia} & Industry-adjusted book to & \textit{grltnoa} & Growth in long-term net \\
		\multicolumn{1}{r}{} & market & \multicolumn{1}{r}{} & operating assets \\
		\textit{cash} & Cash holdings & \textit{herf} & Industry sales \\
		\multicolumn{1}{r}{} & \multicolumn{1}{r}{} & \multicolumn{1}{r}{} & concentration \\
		\textit{cashdebt} & Cash flow to debt & \textit{hire} & Employee growth rate \\
		\textit{cashpr} & Cash productivity & \textit{idiovol} & Idiosyncratic return \\
		\multicolumn{1}{r}{} & \multicolumn{1}{r}{} & \multicolumn{1}{r}{} & volatility \\
		\textit{cfp} & Cash-flow-to-price ratio & \textit{ill} & illiquidity \\
		\textit{cfp\_ia} & Industry-adjusted & \textit{indmom} & Industry momentum \\
		\multicolumn{1}{r}{} & cash-flow-to-price ratio & \multicolumn{1}{r}{} & \multicolumn{1}{r}{} \\
		\textit{chatoia} & Industry-adjusted change & \textit{invest} & Capital expenditures and \\
		\multicolumn{1}{r}{} & in asset turnover & \multicolumn{1}{r}{} &  inventory \\
		\textit{chcsho} & Change in shares & \textit{IPO} & New equity issue \\
		\multicolumn{1}{r}{} & outstanding & \multicolumn{1}{r}{} &  \multicolumn{1}{r}{} \\
		\textit{chempia} & Industry-adjusted change & \textit{lev} & Leverage \\
		\multicolumn{1}{r}{} & in employees & \multicolumn{1}{r}{} & \multicolumn{1}{r}{} \\
		\textit{chfeps} & Change in forecasted EPS & \textit{lgr} & Growth in long-term debt \\
		\textit{chinv} & Change in inventory & \textit{maxret} & Maximum daily return \\
		\textit{chmom} & Change in 6-month & \textit{mom12m} & 12-month momentum \\
		\multicolumn{1}{r}{} & momentum & \multicolumn{1}{r}{} & \multicolumn{1}{r}{} \\
		\textit{chnanalyst} & Change in number of & \textit{mom1m} & 1-month momentum \\
		\multicolumn{1}{r}{} & analysts & \multicolumn{1}{r}{} & \multicolumn{1}{r}{} \\
		\textit{chpmia} & Industry-adjusted change in profit margin & \textit{mom36m} & 36-month momentum \\
		\textit{chtx} & Change in tax expense & \textit{mom6m} & 6-month momentum \\
		\textit{cinvest} & Corporate investment & \textit{ms} & Financial statement score \\
		\textit{convind} & Convertible debt indicator & \textit{mve} & Size \\
		\textit{currat} & Current ratio & \textit{mve\_ia} & Industry-adjusted size \\
		\textit{depr} & Depreciation / PP\&E & \textit{nanalyst} & Number of analysts \\
		\multicolumn{1}{r}{} & \multicolumn{1}{r}{} & \multicolumn{1}{r}{} & covering stock \\
		\textit{disp} & Dispersion in forecasted & \textit{nincr} & Number of earnings \\
		\multicolumn{1}{r}{} & EPS & \multicolumn{1}{r}{} & increases \\
		\textit{divi} & Dividend initiation & \textit{operprof} & Operating profitability \\
		\textit{orgcap} & Organizational capital & \textit{roeq} & Return on equity \\
		\textit{pchcapx\_ia} & Industry adjusted \%  & \textit{roic} & Return on invested capital \\
		\multicolumn{1}{r}{} & change in capital expenditures & \multicolumn{1}{r}{} & \multicolumn{1}{r}{} \\
		\textit{pchcurrat} & \% change in current ratio & \textit{rsup} & Revenue surprise \\
		\textit{pchdepr} & \% change in depreciation & \textit{salecash} & Sales to cash \\
		\textit{pchgm\_pchsale} & \% change in gross margin & \textit{saleinv} & Sales to inventory \\
		\multicolumn{1}{r}{} & - \% change in sales & \multicolumn{1}{r}{} & \multicolumn{1}{r}{} \\
		\textit{pchquick} & \% change in quick ratio & \textit{salerec} & Sales to receivables \\
		\textit{pchsale\_pchinvt} & \% change in sales & \textit{secured} & Secured debt \\
		\multicolumn{1}{r}{} & - \% change in inventory & \multicolumn{1}{r}{} & \multicolumn{1}{r}{} \\
		\textit{pchsale\_pchrect} & \% change in sales & \textit{securedind} & Secured debt indicator \\
		\multicolumn{1}{r}{} & - \% change in A/R   & \multicolumn{1}{r}{} & \multicolumn{1}{r}{} \\
		\textit{pchsale\_pchxsga} & \% change in sales & \textit{sfe} & Scaled earnings forecast \\
		\multicolumn{1}{r}{} & - \% change in SG\&A & \multicolumn{1}{r}{} & \multicolumn{1}{r}{} \\
		\textit{pchsaleinv} & \% change & \textit{sgr} & Sales growth \\
		\multicolumn{1}{r}{} & sales-to-inventory & \multicolumn{1}{r}{} & \multicolumn{1}{r}{} \\
		\textit{pctacc} & Percent accruals & \textit{sin} & Sin stocks \\
		\textit{pricedelay} & Price delay & \textit{SP} & Sales to price \\
		\textit{ps} & Financial statements score & \textit{std\_dolvol} & Volatility of liquidity \\
		\multicolumn{1}{r}{} & \multicolumn{1}{r}{} & \multicolumn{1}{r}{} & (dollar trading volume) \\
		\textit{quick} & Quick ratio & \textit{std\_turn} & Volatility of liquidity \\
		\multicolumn{1}{r}{} & \multicolumn{1}{r}{} & \multicolumn{1}{r}{} & (share turnover) \\
		
		\textit{rd} & R\&D increase & \textit{stdacc} & Accrual volatility \\
		\textit{rd\_mve} & R\&D to market & \textit{stdcf} & Cash flow volatility \\
		\multicolumn{1}{r}{} & capitalization & \multicolumn{1}{r}{} & \multicolumn{1}{r}{} \\
		\textit{rd\_sale} & R\&D to sales & \textit{sue} & Unexpected quarterly \\
		\multicolumn{1}{r}{} & \multicolumn{1}{r}{} & \multicolumn{1}{r}{} & earnings \\
		\textit{realestate} & Real estate holdings & \textit{tang} & Debt capacity/firm \\
		\multicolumn{1}{r}{} & \multicolumn{1}{r}{} & \multicolumn{1}{r}{} & tangibility \\
		\textit{retvol} & Return volatility & \textit{tb} & Tax income to book \\
		\multicolumn{1}{r}{} & \multicolumn{1}{r}{} & \multicolumn{1}{r}{} & income \\
		\textit{roaq} & Return on assets & \textit{turn} & Share turnover \\
		\textit{roavol} & Earnings volatility & \textit{zerotrade} & Zero trading days \\
		\bottomrule
	\end{tabularx}

\setstretch{1.5}

\subsection{Path difference and parameter attribution} \label{sec:path difference}

Common methods to compare the difference between the lasso and LS-boost include visual inspection of the solution paths, computing MSPEs, \textit{etc}. We show that the integrated gradient along the solution paths of the lasso and LS-boost can also be used to study the difference between these two methods.

In many applications of network modeling, it is useful to attribute the prediction of a network to each input (variables). \cite{sundararajan2017attribution} propose the idea of integrated gradients for attribution. Consider a function $f:R^p \rightarrow R$. Given a $p \times 1$ input vector $z$, select a baseline vector $z'$, the integrated gradient along the $j$th dimension on the straight line connecting $z$ and $z'$ is defined as:
\begin{equation} \label{eq:IG}
	\text{integrated gradient}_j := (z_j - z_j') \times \int_{\alpha = 0}^{1} \frac{\partial f(z'+\alpha \times (z-z'))}{\partial z_j}d\alpha, \text{ for }j=1,\cdots,p.
\end{equation}
When the path connecting $z'$ and $z$ is not a straight line, we can use the path integrated gradient for attribution.
\begin{equation} \label{eq:path IG}
	\text{path integrated gradient}_j := \int_{\alpha = 0}^{1} \frac{\partial f(\phi(\alpha))}{\partial \phi_j(\alpha)} \frac{\partial \phi_j(\alpha)}{\partial \alpha}d\alpha, \text{ for }j=1,\cdots,p,
\end{equation}
where $\phi = (\phi_1,\cdots,\phi_p):[0,1] \rightarrow R^p$ is a function specifying a path linking $z'$ and $z$ with $\phi(0) = z'$ and $\phi(1) = z$. 

\Cref{eq:IG,eq:path IG} describe a method for variable attribution. The same idea can be used for parameter attribution. To see that, consider the objective function in \cref{eq:ls loss} and let $f = L_n(\cdot)$. Since data are given, $f$ becomes a function of $\beta$. Let the $\hat{\beta}^{(0)} = \boldsymbol{\mathrm{0}}_{p\times1}$ be the starting value in an algorithm (the $z$ in \cref{eq:IG}) and $\hat{\beta}^{(1)}$ be the estimates at the end of the solution path (the $z'$ in \cref{eq:IG}). Since the solution path for the lasso and LS-boost is not a straight line connecting $\hat{\beta}^{(0)}$ and $\hat{\beta}^{(1)}$, we need to accumulate the gradients along the path on which the coefficients travel and use \cref{eq:path IG}. Without specifying a piecewise linear function $\phi$ for the lasso or LS-boost, we opt for the simple method of numerical integration based on the trapezoid rule. This method seems to work well in our case, probably due to the piecewise linear pattern of the solutions.

Use the data in January, 2010 as an example. Both the relaxed lasso and lassoed boost select the same $6$ variables (\textit{lev}, \textit{mve}, \textit{mom1m}, \textit{baspread}, \textit{mom12m}, \textit{retvol}) while the lasso selects $44$ variables. Assume both the lasso and LS-boost start with the same $6$ variables. Let the lasso use $1000$ equally spaced penalty values on $[0,\lambda_0]$ and LS-boost use the learning rate $0.1$ and iteration number $1000$ (the iteration number based on the corrected AIC is $127$.) Let $q$ denote the step in the lasso or LS-boost. We use the following formula for numerical integration. For $j = 1,\cdots,6 \text{ and } q=1,\cdots,1000$,
\begin{equation} \label{eq:trapezoid rule}
	G_{jq} = \frac{\partial L_n(\hat{\beta}_{q+1}) / \partial \beta_j + \partial L_n(\hat{\beta}_{q}) / \partial \beta_j}{2} \times (\hat{\beta}_{q+1} - \hat{\beta}_{q}),
\end{equation}
and we record \cref{eq:trapezoid rule} as the $jq$th element of the matrix $G$.
The approximation to the path integrated gradient in \cref{eq:path IG} becomes the row sums of $G$
\begin{equation} \label{eq:approx path integral}
	G_{j\cdot} = \sum_{q=1}^{1000} G_{jq}, \text{ for }j=1,\cdots,6.
\end{equation} 
To gauge the precision of the numerical integration, we sum the starting value of the loss function $L_n(\hat{\beta}^{(0)})$ and $\sum_{j=1}^{6}G_{j\cdot}$, and compare it to the value of the loss function $L_n(\hat{\beta}^{(1)})$. For the lasso, the two quantities are equal up to the $8$th digit; for LS-boost, they are equal up to the $15$th digit. This simply verifies the fundamental theorem of calculus.  

\begin{table}[htp] \centering
	\begin{center}
		\caption{Path integrated gradient for each $\beta$ in the lasso and LS-boost} 
		\label{tab:path integrated gradients} 
		\begin{threeparttable}
			\begin{tabular}{lcccccc}  
				\toprule
				& $\beta_\text{lev}$ & $\beta_\text{mve}$ & $\beta_\text{mom1m}$ & $\beta_\text{baspread}$ & $\beta_\text{mom12m}$  & $\beta_\text{retvol}$\\
				\midrule
				lasso & $-56.666$ & $-32.765$&$-26.522$&$-9.864$& $-9.653 $ & $-3.661$ \\
				LS-boost &$-56.872$&$-32.803$&$-26.471$&$-9.777$&$-9.646$ & $-3.561$ \\
				\bottomrule
			\end{tabular}
			\begin{tablenotes}[flushleft]
				\setlength\labelsep{0pt}
				\item[] \textit{Notes}: See \Cref{tab:variable list} in the Supplement for the variable definition. The number in each cell is obtained via \cref{eq:approx path integral}. Multiplying the numbers by $10^{-5}$ gives the actual path integrated gradients. The columns are arranged based on the order in which variables enter the model.
			\end{tablenotes}
		\end{threeparttable}
	\end{center} 
\end{table}
\Cref{tab:path integrated gradients} reports the path integrated gradient for both the lasso and LS-boost. All numbers are negative because we consider function minimization. The variables are listed in the order in which they enter the lasso and LS-boost models. Because it is not possible for LS-boost to reach a full LS solution even with a large number of iterations, rigorously speaking, the numerical differences in these numbers also reflect a small numerical difference between the minimized loss of the lasso and LS-boost. In our case, the two minimized loss functions based on demeaned data differ by less than $3 \times 10^{-9}$. Hence, numerical difference in \Cref{tab:path integrated gradients} is largely due to that fact the lasso and LS-boost visit different solution paths. This provides a new perspective on understanding the difference between these two methods.

One should not conclude that the small numerical difference in \Cref{tab:path integrated gradients} indicates the two methods always have similar parameter attribution. This is a simple regression model with only $6$ variables. With many variables, the two methods may select different models, and the attributions in \Cref{tab:path integrated gradients} will be very different. This is likely to happen more frequently when comparing the lasso to lassoed boosting. \Cref{tab:path integrated gradients} does not focus on the relaxed lasso. But it is interesting to note that the attribution method for the relaxed lasso is a hybrid procedure of using both 
\cref{eq:IG,eq:path IG}.

\begin{figure}[htp]
	\centering
	\subfloat[SAPA in the lasso]{\label{fig:attr-lasso}\includegraphics[width=0.48\linewidth,keepaspectratio,scale=1]{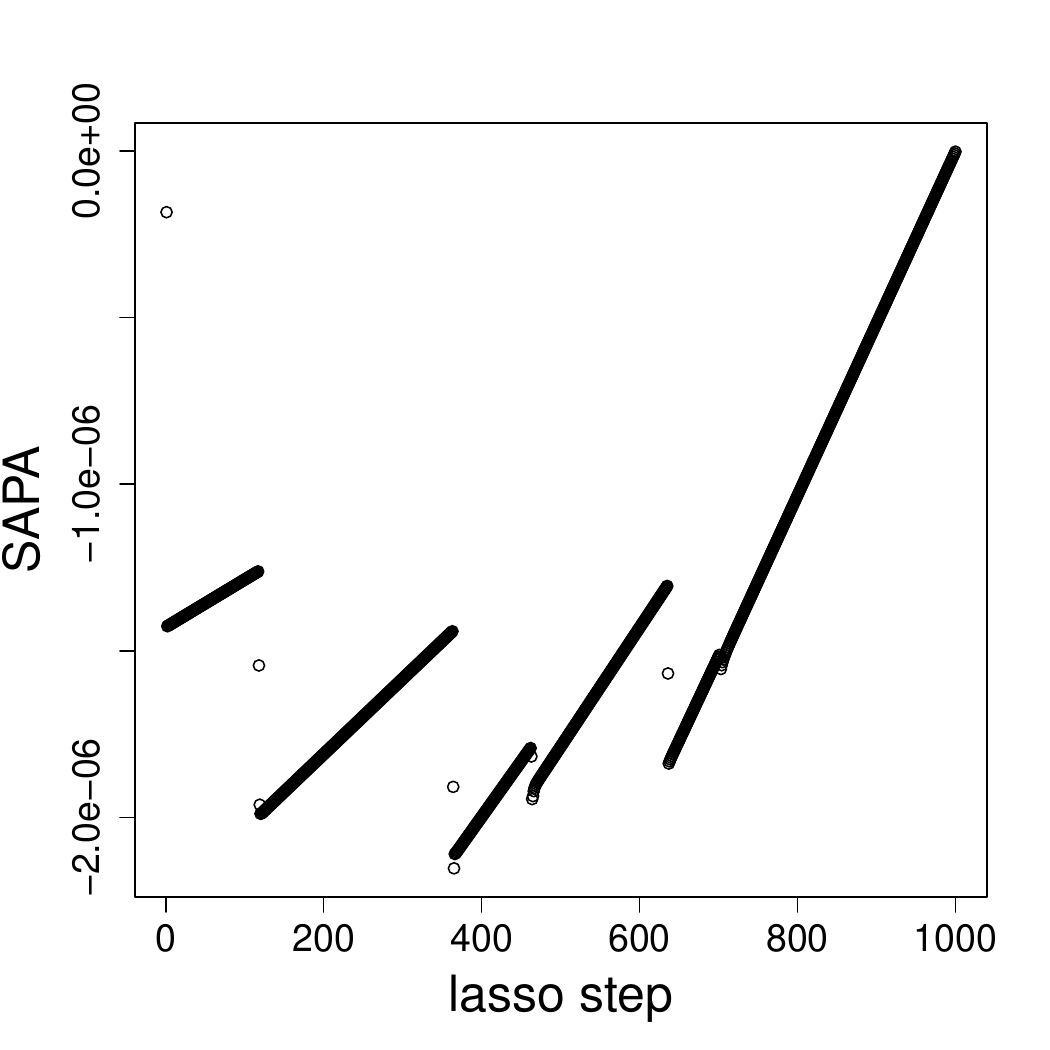} }%
	\subfloat[SAPA in LS-boost, $\varepsilon=0.1$]{\label{fig:attr-boost}\includegraphics[width=0.48\linewidth,keepaspectratio,scale=1]{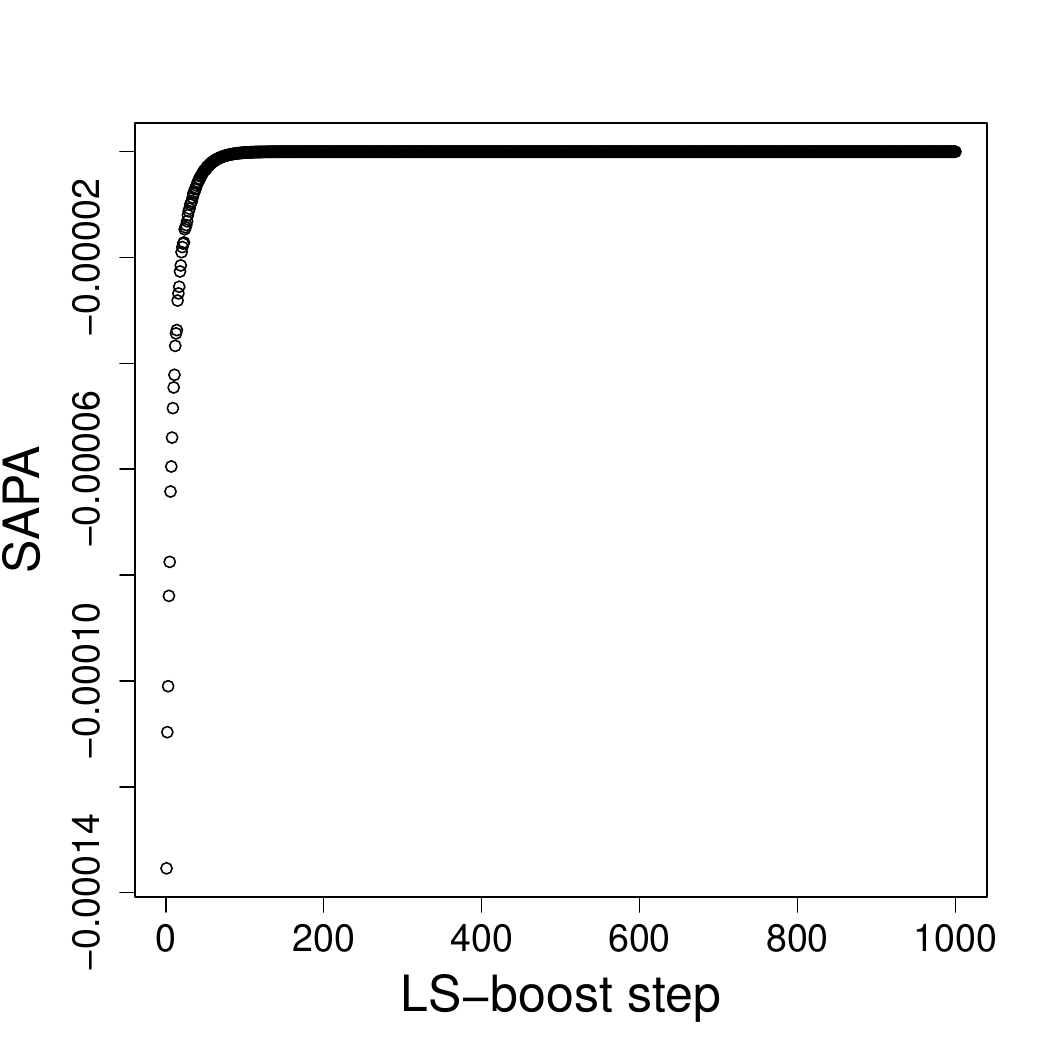} }
	\caption{SAPA in the lasso and LS-boost for $6$ variables (\textit{lev}, \textit{mve}, \textit{mom1m}, \textit{baspread}, \textit{mom12m}, \textit{retvol}) in January, 2010. In \Cref{fig:attr-lasso}, there is a small break at the end of the $5$th segment, resulting a total of $6$ segments.}%
	\label{fig:attribution}%
\end{figure}

We can also compute the column sums of $G$, which measures the stepwise aggregate parameter attribution (SAPA) of a method. SAPA is also the stepwise decrease in the loss function. \Cref{fig:attribution} plots SAPA of both the lasso and LS-boost. We make several observations. The SAPA of the lasso consists of $6$ segments, each of which represent a gradual decrease (in absolute value) in SAPA after a new variable enters the model. Interestingly, for LS-boost, its SAPA exhibits a smoother, continuous pattern. This indicates that the lasso and LS-boost may give completely different descending pattern of the loss function during optimization. For the lasso, SAPA at the beginning of each segment always shows up as a jump. This is due to the fact that when a new variable is just selected, its path integrated gradient at the current step is computed w.r.t. a zero value in its coefficient. After that, its path integrated gradient is computed w.r.t. a nonzero coefficient, which helps smooth the values, and they mostly line up along a straight path until the next variable enters the model. We also observe that these segments of the lasso SAPAs have different length and slope. A long and relatively flat segment indicates lasso is traveling on a solution path that might drive down the loss function considerably. The pattern of the SAPA in \Cref{fig:attr-lasso} is also implicitly connected to the well-known fact that the lasso solution path is piecewise linear. Our analysis shows that path integrated gradient can be a useful tool to study the differences between the lasso and LS-boost. The above analysis also applies to lassoed boosting.

\subsection{Additional figures for parameter attribution in the lasso and LS-boost} \label{sec:path difference additional figures}

\begin{figure}[htp]
	\centering
	\subfloat[SCPA plot for the lasso]{\label{fig:arealasso}\includegraphics[width=0.46\linewidth,keepaspectratio]{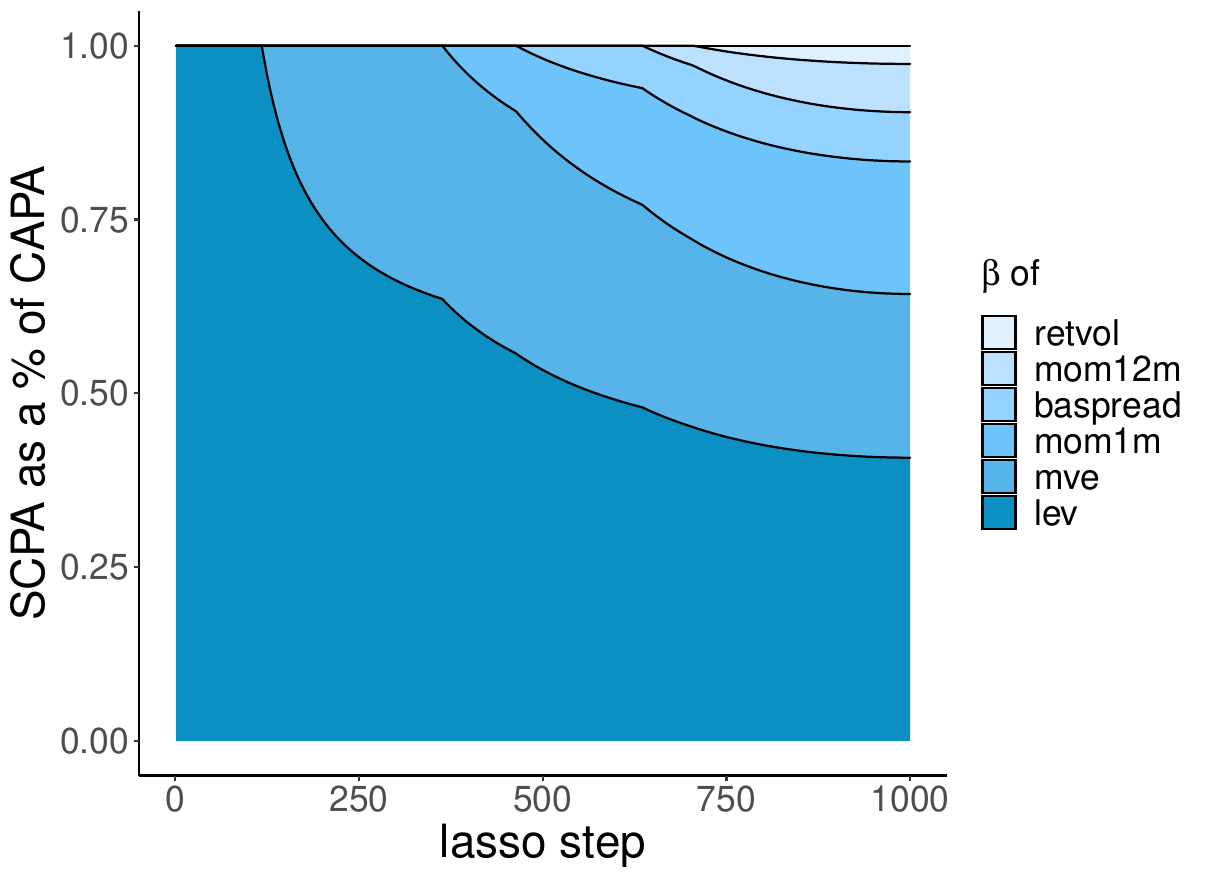} }%
	\subfloat[SCPA plot for LS-boost, $\varepsilon = 0.1$]{\label{fig:areaboost01}\includegraphics[width=0.46\linewidth,keepaspectratio]{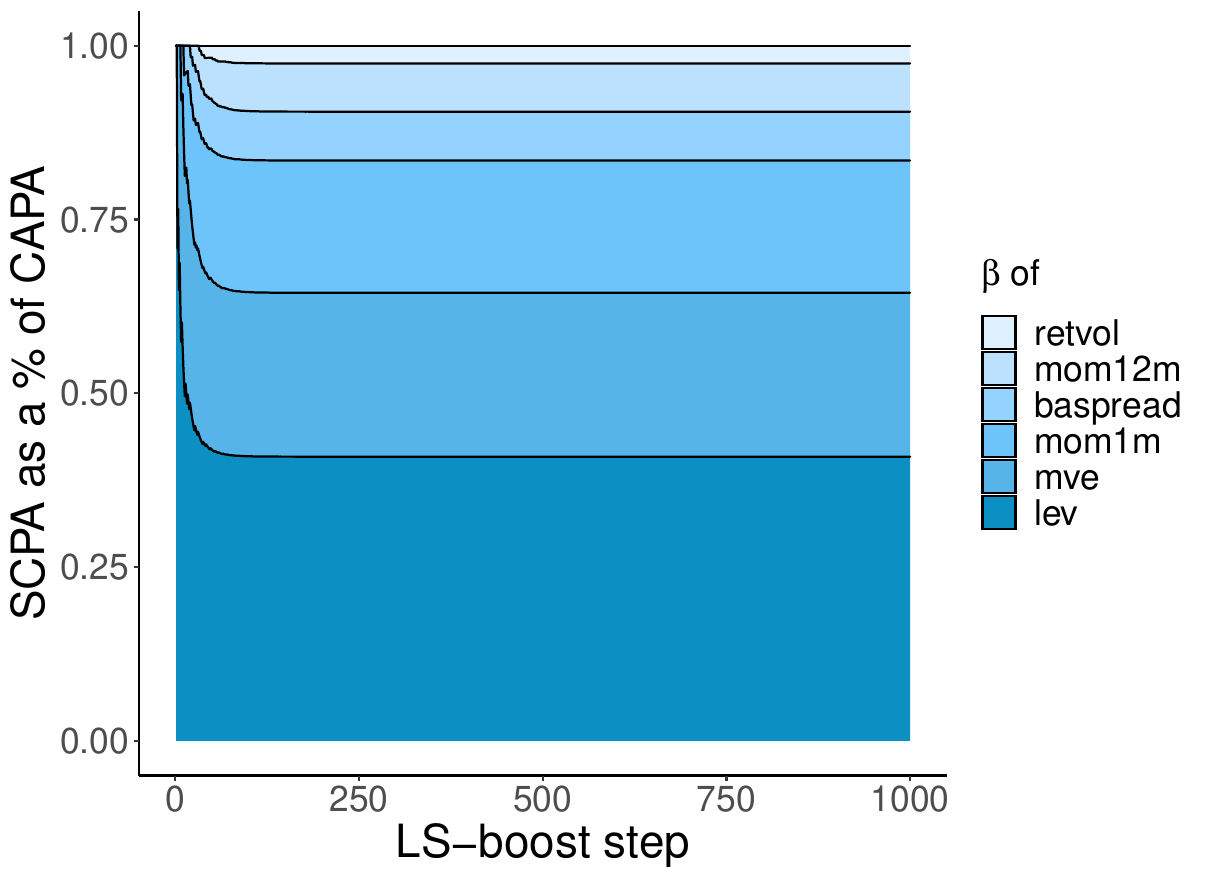} }\\
	\subfloat[SAPA in LS-boost, $\varepsilon = 0.01$]{\label{fig:attr-boost001}\includegraphics[width=0.46\linewidth,keepaspectratio]{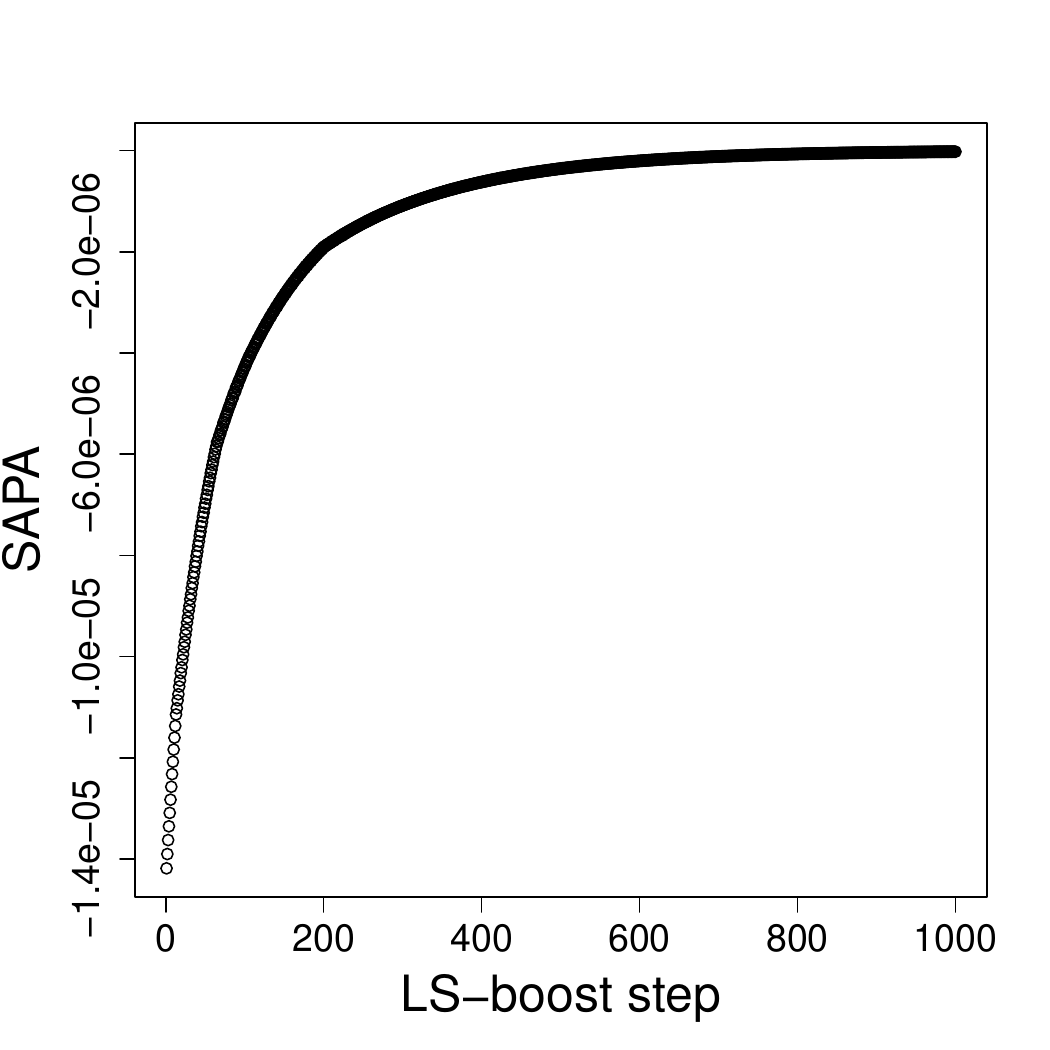} }%
	\subfloat[SCPA plot for LS-boost, $\varepsilon=0.01$]{\label{fig:areaboost001}\includegraphics[width=0.46\linewidth,keepaspectratio]{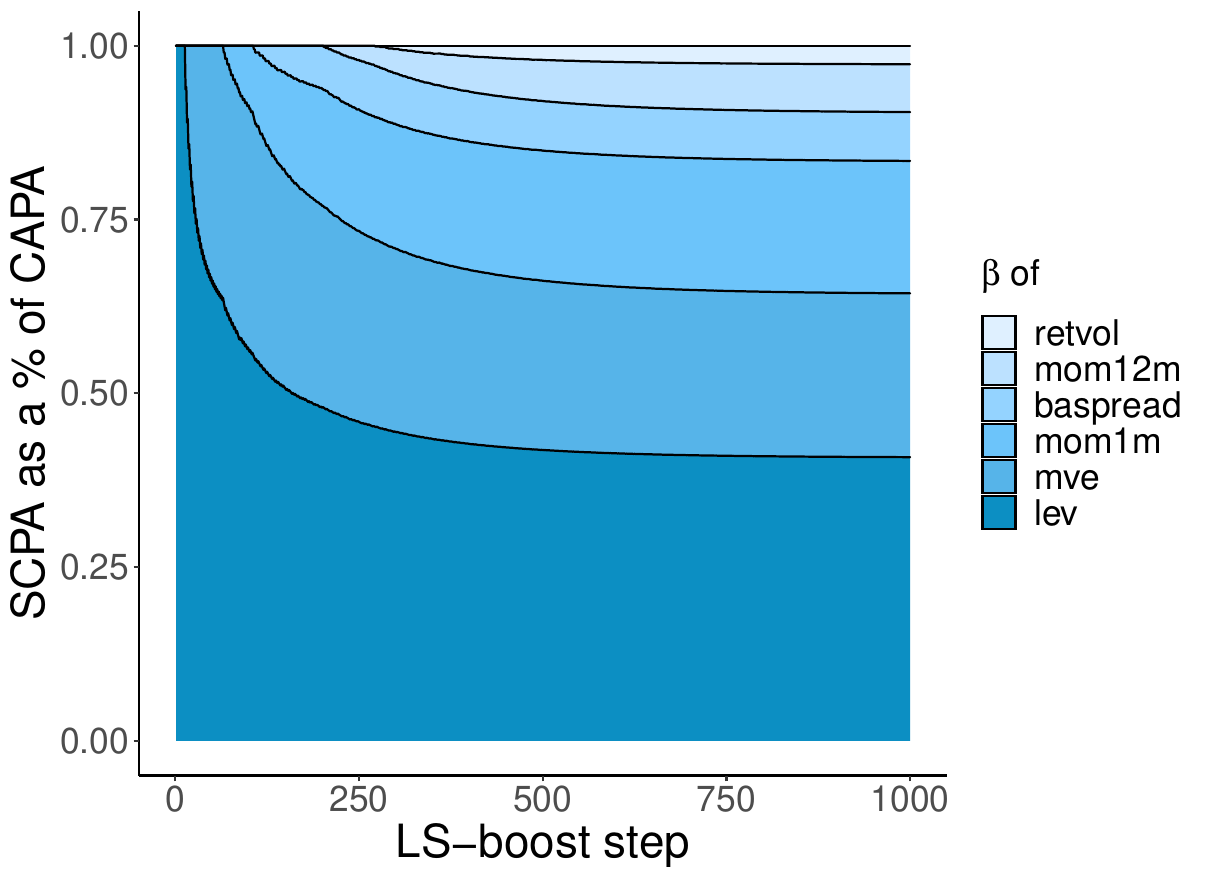} }%
	\caption{\Cref{fig:arealasso,fig:areaboost01,fig:areaboost001} are SCPA plots for lasso and LS-boost. \Cref{fig:attr-boost001} is the SAPA plot for LS-boost with $\varepsilon=0.01$.}%
	\label{fig:paths}%
\end{figure}

\Cref{fig:arealasso,fig:areaboost01,fig:areaboost001} plot, at each step, a parameter estimate $\hat{\beta}_j$'s stepwise cumulative parameter attribution (SCPA) as a percentage of the cumulative aggregate parameter attribution (CAPA) up to each step. These three figures provide an additional way to visualize the difference between the lasso and LS-boost. Compared with \Cref{fig:attr-boost}, \Cref{fig:attr-boost001} illustrates how a different learning rate can alter the pattern of SAPA in LS-boost.

\end{appendices}

\end{document}